\documentclass[twocolumn, notitlepage, 10pt, aps, floatfix, showpacs, prb, citeautoscript, superscriptaddress]{revtex4-1}
\usepackage{graphicx}
\usepackage{amsmath, amssymb}
\usepackage{bm}
\usepackage{xcolor}
\usepackage[colorlinks, citecolor=blue, urlcolor=blue, linkcolor=red]{hyperref}
\usepackage{bookmark}
\usepackage[version=4]{mhchem}
\usepackage{microtype}
\usepackage{float}
\usepackage[load=physical,load=abbr, range-units=single]{siunitx}

\begin{document}
\title{Theory of Coulomb blockaded transport in realistic Majorana nanowires}

\author{Yi-Hua Lai}
\author{Sankar Das Sarma}
\author{Jay D. Sau}
\affiliation{Department of Physics, Condensed Matter Theory Center and the Joint Quantum Institute, University of Maryland, College Park, MD 20742}

\begin{abstract}
Coulomb blockaded transport of topological superconducting nanowires provides an opportunity to probe the localization of states at both ends of the system in a two-terminal geometry. In addition, it provides a way for checking for subgap states away from the leads. At the same time, Coulomb blockade transport is difficult to analyze because of the interacting nature of the problem arising from the nonperturbative Coulomb interaction inherent in the phenomenon. Here we show that the Coulomb blockade transport can be modeled at the same level of complexity as quantum point contact tunneling that has routinely been used in mesoscopic physics to understand nanowire experiments provided we consider the regime where the tunneling rate is below the equilibration rate of the nanowire. This assumption leads us to a generalized Meir-Wingreen formula for the tunnel conductance which we use to study various features of the nanowire such as Andreev bound states, self-energy, and soft gap. We anticipate that our theory will provide a route to interpret Coulomb blockade transport in hybrid Majorana systems as resulting from features of the nanowire, such as Andreev bound states and soft gaps.
\end{abstract}

\maketitle

\section{Introduction}\label{sec:level1_1}
% Remember to attach the ideal case in this section first
Topology has become an intrinsic part of condensed matter physics since 1980 when the quantum Hall state was discovered \cite{Klitzing1980New}. The quantized Hall conductance, which only takes integer values in units of $e^2/h$, is based on its topological robustness arising from the existence of a Chern index characterizing the quantization\cite{Laughlin1981Quantized,Thouless1982Quantized}. Typically, insulators and superconductors are the platforms to manifest topological phenomena because they both have bulk gaps. However, only under strict conductions, i.e., having robust boundary gapless states protected by the bulk gap, can insulators and superconductors become topological. The bulk-boundary correspondence is the key to topological materials. There are certain topological systems where the boundary gapless excitations (often referred to as Majorana modes since these excitations are their own anti-particles) are non-Abelian anyons, manifesting nontrivial braiding statistics, which can be used for fault-tolerant topological quantum computation\cite{Nayak2008NonAbeliana,Kitaev2001Unpaireda,Read2000Paired,DasSarma2005Topologically}. Therefore, some topological systems [e.g. 5/2 fractional quantum Hall system or the Moore-Read pfaffian state, one- and two-dimensional spinless $p$-wave superconductors or topological superconductors] are promising candidates for topological quantum computers\cite{Nayak2008NonAbeliana,Sarma2015Majoranaa,Lutchyn2018Majoranaa}. Among the topological superconductors, there exists one kind of zero-energy quasi-particles, called Majorana bound states (MBSs), which obey anyonic non-Abelian statistics, thus playing an essential role in quantum computation. Microsoft Corporation has chosen MBS-based topological quantum computation in topological superconductors as its preferred quantum computing platform, making MBS a well-known idea in the technical popular press\cite{Castelvecchi2017Quantum}. Several experimental systems can host MBSs, such as at the ends of 1D topological superconductors\cite{Kitaev2001Unpaireda}, or the point defects of 2D topological superconductors \cite{Fu2008Superconducting}. The most studied experimental scheme to realize MBSs is superconductor (SC)-proximitized semiconductor nanowire with spin-orbit coupling and Zeeman spin splitting from the external magnetic field \cite{Lutchyn2010Majoranaa,Oreg2010Helicala,Sau2010NonAbeliana}. Different experimental searches for MBSs using this kind of setup, either with InSb or InAs as semiconductor, combined with NbTiN or Al as superconductor on top of it have been reported \cite{Mourik2012Signaturesa,Das2012Zerobiasa,Deng2012Anomalousa,Finck2013Anomalousa,Churchill2013Superconductornanowirea,Krogstrup2015Epitaxy,Chen2017Experimentala,Gul2018Ballistica}. For convenience and brevity in describing this system, we will call it “Majorana nanowire” in this paper. Recently, the predicted zero-bias conductance peak (ZBCP) above a critical Zeeman field was observed in the experiments, which has been touted as a possible milestone evidence for the existence of MBSs \cite{Mourik2012Signaturesa,Das2012Zerobiasa,Deng2012Anomalousa,Finck2013Anomalousa,Churchill2013Superconductornanowirea,Krogstrup2015Epitaxy,Chen2017Experimentala,Gul2018Ballistica}. This critical Zeeman field is the topological quantum phase transition (TQPT) field, for the emergence of the topological regime. However, the TQPT Zeeman field is unknown in the experiment. In fact, the ZBCP can also be induced by generic low-lying in-gap fermionic bound states in Majorana nanowire, such as impurity disorder \cite{Liu2012ZeroBiasa,Bagrets2012Classa,Pikulin2012zerovoltagea,Sau2013Densitya,Mi2014Xshapeda}, inhomogeneous chemical potential \cite{Kells2012Nearzeroenergya,Prada2012Transporta,Moore2018Quantizeda,Moore2018Twoterminala,Liu2017Andreeva,Vuik2019Reproducing}, or low-lying Andreev bound states (ABSs) \cite{Lee2012ZeroBiasa,Kells2012Nearzeroenergya,Liu2017Andreeva,Liu2018Distinguishinga,Moore2018Twoterminala,Vuik2019Reproducing,Pan2020Physical}. Therefore, observation of ZBCP is not a guarantee for the existence of topological MBS. Many theoretical papers have proposed protocols to distinguish MBS from ABS \cite{Chiu2017Conductance,Setiawan2017Electrona,Liu2018Distinguishinga,Moore2018Quantizeda,Moore2018Twoterminala,Stanescu2019Robust,Lai2019Presence}. The issue is totally open whether MBS have been seen or not and if not, what needs to be done to validate MBS existence, in spite of the large number of theoretical and experimental papers.

The relatively short experimental device length for epitaxially grown superconductor-semiconductor nanowires allows for a measurement of transport in the Coulomb blockade (CB) regime\cite{Albrecht2016Exponentiala,Shen2018Parity,Vaitiekenas2020Fluxinduced} simply by lowering the transmission to either end of the wire. Transport in the CB regime, while somewhat more complicated to theoretically interpret relative to quantum point contact (QPC) tunneling, provides information about the states at both ends and possibly also bulk transport\cite{Albrecht2016Exponentiala,vanHeck2016Conductancea,Chiu2017Conductance}. The theoretical complication arises from the fact that one must take into account both MBS and CB physics on an equal and nonperturbative footing, while at the same time address the nonequilibrium physics of tunneling transport. Direct measurement of tunnel conductance at both ends requires a three-terminal configuration\cite{Menard2020ConductanceMatrix}, which risks generating additional spurious subgap states at the third contact\cite{Huang2018Quasiparticlea}. The CB measurement dispenses with the need for such a third contact leading to a simpler measurement. Additionally, as we will discuss later in this work, the CB conductance is less sensitive to sub-gap states that are localized only at one end or another, with the exception of MBSs, making it a particularly attractive experimental approach. This possibility of exploring both the wire ends and the bulk so that the bulk-boundary topological correspondence could be investigated in a single experiment in a single sample is what compels us to carry out the extensive theoretical analysis presented in this current work.

\begin{figure}
	\includegraphics[scale=0.3]{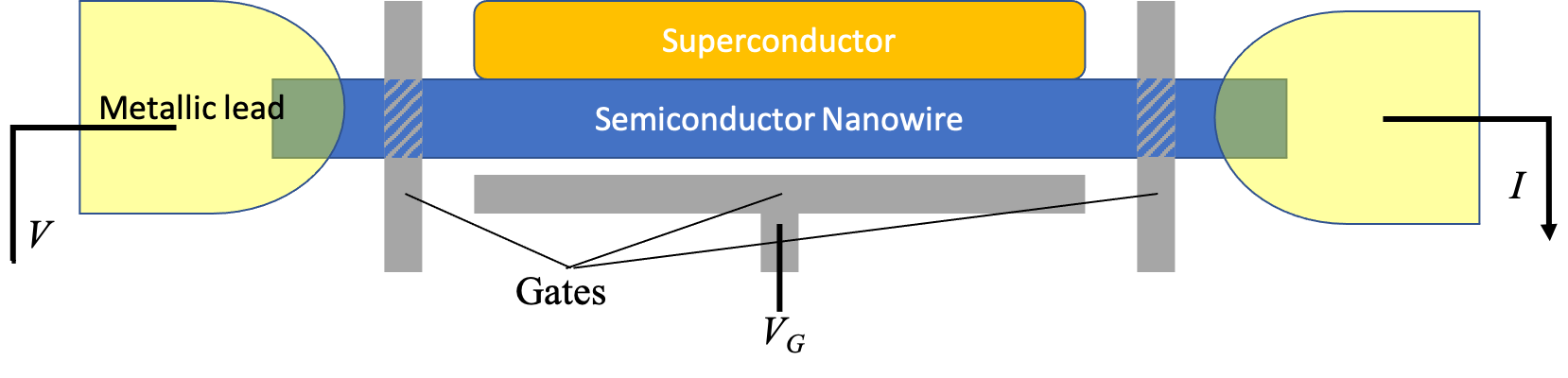}
	\caption{Schematic plot of the Coulomb blockaded Majorana nanowire. This semiconducting nanowire (e.g. InAs or InSb) is proximitized by the parent $s$-wave superconductor (e.g. Al) which only covers part of the nanowire. The gate with voltage $V_G$ is for controlling charge on the nanowire. The Coulomb blockaded transport is measured between the leads shown at the two ends of the Majorana nanowire. The coupling to the leads is pinched off to the required degree by the tunnel gates shown.}
	\label{fig:scheme}
\end{figure}

Despite the advantages of CB measurements, they are more complicated to interpret relative to quantum point contact both because of the involvement of tunneling at both ends as well as Coulomb interaction. As a result, there is no standard formalism to model such transport analogous to the Blonder-Tinkham-Klapwijk (BTK) formalism to model quantum point contact tunneling\cite{Liu2012ZeroBiasa,Bagrets2012Classa,Pikulin2012zerovoltagea,Kells2012Nearzeroenergya,Prada2012Transporta,Sau2013Densitya,Mi2014Xshapeda,Liu2017Andreeva,Moore2018Quantizeda,Moore2018Twoterminala,Vuik2019Reproducing}. In fact, the interplay of Coulomb interactions and low temperature Fermi-liquid correlations can lead to intricate many-body physics such as the Kondo effect\cite{Beri2012Topological,Altland2014Multichannel,Cheng2014Interplay,Bao2017Topological,Michaeli2017Electron}, which can further complicate the interpretation of data. Such complications in understanding can be avoided for these systems for temperatures above the Kondo temperature where transport can be modeled by perturbation theory in tunneling \cite{vanHeck2016Conductancea}. However, CB transport through a complex system such as a semiconductor-superconductor nanowire, which has many low-energy levels, is difficult to treat numerically and is characterized by an exponential complexity of the perturbative rate equations\cite{Chiu2017Conductance}. 

The difficulty of interpreting the CB condcutance manifests itself in terms of some (at best) partially understood measurements on these systems so far\cite{Albrecht2016Exponentiala,Shen2018Parity}. The schematic set-up for such measurements is shown in the set-up in Fig. \ref{fig:scheme} which describes a Coulomb blockaded superconductor/semiconductor island between two leads. The transport in the absence of a Zeeman field is expected to be dominated by tunneling of Cooper pairs with charge $2e$, which manifests as $2e$ periodicity of the CB conductance with the gate voltage $V_g$ (shown in Fig. \ref{fig:scheme})\cite{Albrecht2016Exponentiala,Shen2018Parity,Vaitiekenas2020Fluxinduced}. The observation of such a $2e$ periodic CB conductance establishes a parity gap in these materials, which is not directly accessible by quantum point contact tunneling\cite{Albrecht2016Exponentiala}. The application of a Zeeman potential can reduce the parity gap so that the $2e$ periodic CB peaks split into pairs of resonances\cite{Albrecht2016Exponentiala}. Further increase of the Zeeman field leads to the CB peaks becoming $1e$ periodic\cite{Albrecht2016Exponentiala} or going back to $2e$ periodic with a $1e$ shift i.e., a parity switch\cite{Shen2018Parity}. Some of these features can be understood in terms of an ideal Majorana nanowire. In this scenario, one expects to see a relatively bright $2e$ periodic peak splitting into relatively dark peaks associated with transport of electrons through the bulk states that approach zero energy as the gap is closed by the Zeeman field\cite{vanHeck2016Conductancea}. These peaks then morph into $1e$ periodic bright conductance peaks associated with non-local transport through the pair of end MBSs\cite{Albrecht2016Exponentiala,vanHeck2016Conductancea}. While this bright-dark-bright pattern is understood in terms of this picture of an ideal topological wire, the higher intensity of the $2e$ peaks relative to the MBS peak remains a puzzle\cite{vanHeck2016Conductancea}. Additional information about these states is obtained from converting the peak position as a function of Zeeman field into an anticipated spectrum, also known as the oscillating conductance peak spacing (OCPS)\cite{Albrecht2016Exponentiala}. The OCPS in semiconductor/superconductor systems appears to show oscillations as a function of Zeeman field\cite{Albrecht2016Exponentiala} similar to what is expected from split MBSs\cite{Cheng2009Splittinga,Chiu2017Conductance}. However, the experimentally measured oscillations are found to decrease with Zeeman field\cite{Albrecht2016Exponentiala,Shen2018Parity,Vaitiekenas2020Fluxinduced}, in contrast to what is expected for MBSs\cite{Chiu2017Conductance}. This is expected to be a rather generic consequence of the MBSs becoming delocalized as the Zeeman field suppresses the gap, consequently increasing the Majorana localization length (or equivalently, the effective superconducting coherence length).  

In this work, we will start by combining the rate equation formalism with an equilibration assumption to derive a two-lead generalization of the Meir-Wingreen mesoscopic theoretic formalism for the interacting CB system. We will then apply this formalism to a semi-realistic model for a semiconductor nanowire that has been used to study quantum point contact tunneling\cite{Sau2010NonAbeliana,Liu2012ZeroBiasa,Bagrets2012Classa,Pikulin2012zerovoltagea,Kells2012Nearzeroenergya,Prada2012Transporta,Sau2013Densitya,Mi2014Xshapeda,Liu2017Andreeva,Setiawan2017Electrona,Moore2018Quantizeda,Moore2018Twoterminala,Liu2018Distinguishinga,Vuik2019Reproducing,Stanescu2019Robust,Lai2019Presence} in Majorana nanowires. We will then study various limits of the model to develop a generic correspondence between features seen in CB transport experiments\cite{Albrecht2016Exponentiala,Shen2018Parity,Vaitiekenas2020Fluxinduced} manifesting characteristics of the semiconductor wire model such as self-energy, soft gap etc.

\begin{figure}[htbp]
	\includegraphics[scale=0.35]{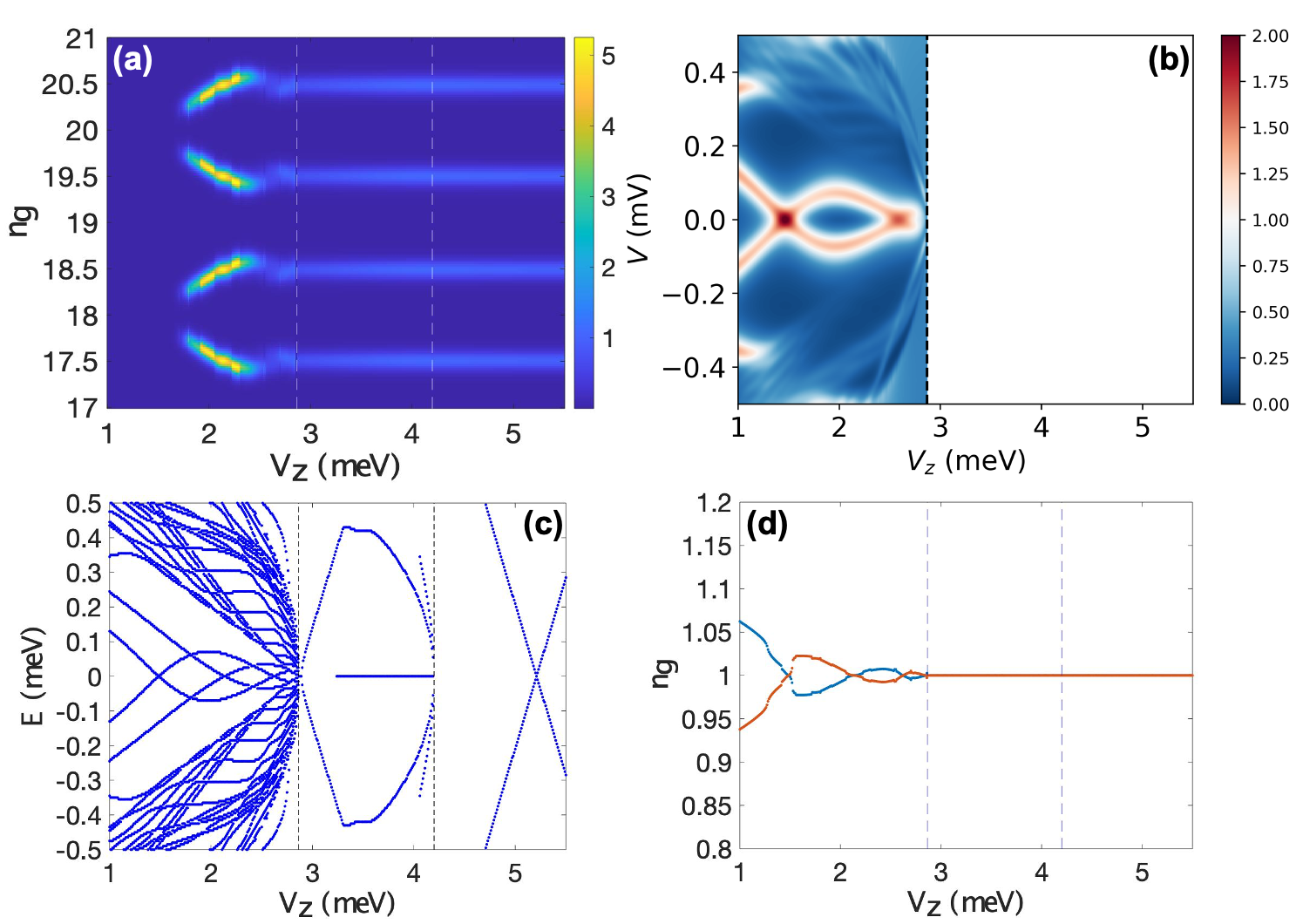}
	\caption{The ideal case results, which show the most similar features as experimental data by fine-tuning. The parameters are: the temperature $T=0.01$ meV, the wire length $L=1.5$ $\micro$m, the SC gap at zero Zeeman field $\Delta_0=0.9$ meV, the SC collapsing field $V_c=4.2$ meV. Other relevant parameters are given in Sec.~\ref{sec:level1_5}. (a) Coulomb blockaded conductance $G$ as a function of the gate-induced charge number $n_g$ and Zeeman field $V_z$ at $E_c=0.13$ meV. We only show two periods in the range of $n_g \in[17,21]$. (b) Non-Coulomb blockaded conductance $G$ from the left lead as a function of bias voltage $V$ and Zeeman field $V_z$. (c) Energy spectrum as a function of Zeeman field $V_z$. (d) OCPS as a function of Zeeman field $V_z$ that is extracted from the vertical peak spacing in panel (a).}
	\label{fig:idealCase}
\end{figure}

One of the main motivations of our work is to provide a qualitative understanding of measured CB transport experiments\cite{Albrecht2016Exponentiala,Shen2018Parity,Vaitiekenas2020Fluxinduced}. Therefore, before presenting the details of the formalism in Secs.~\ref{sec:level1_2},~\ref{sec:level1_3} and ~\ref{sec:level1_4}, we present a preliminary description of our main results to motivate our work. In Fig. \ref{fig:idealCase}(a), we show a representative result that qualitatively resembles some of the recent experimental data\cite{Albrecht2016Exponentiala,Shen2018Parity,Vaitiekenas2020Fluxinduced} for the CB conductance as a function of gate-induced charge number $n_g$ and Zeeman field $V_z$. As discussed in more detail in Sec.~\ref{sec:level1_5}, Figs. \ref{fig:idealCase}(b,c,d) allow us to compare the information from Fig. \ref{fig:idealCase}(a) to other characteristics of the wire such as end conductance [Fig. \ref{fig:idealCase}(b)] that is measured more typically (i.e., QPC)\cite{Mourik2012Signaturesa,Das2012Zerobiasa,Deng2012Anomalousa,Finck2013Anomalousa,Churchill2013Superconductornanowirea,Krogstrup2015Epitaxy,Chen2017Experimentala,Gul2018Ballistica} and the nanowire spectrum [Fig. \ref{fig:idealCase}(c)], which is the information desired from tunneling transport. The OCPS shown in Fig. \ref{fig:idealCase}(d) is directly obtained from the peak spacing along $n_g$ in Fig. \ref{fig:idealCase}(a) and can be compared to parts of the spectrum in Fig. \ref{fig:idealCase}(c). The model used to obtain the results in Fig. \ref{fig:idealCase}(a), which most closely resembles experimental data\cite{Albrecht2016Exponentiala,Shen2018Parity,Vaitiekenas2020Fluxinduced}, is a Majorana nanowire model that includes quantum dots, self-energy effects, suppression of SC and soft SC gap in addition to CB (see Sec.~\ref{sec:level1_4} for details).  

Figure \ref{fig:idealCase}(a) does not show the $2e$ periodic (in $n_g$) part of the Coulomb blockade conductance that appears as the brightest features in the experimental data\cite{Albrecht2016Exponentiala,Shen2018Parity,Vaitiekenas2020Fluxinduced} at the lowest part of the range of Zeeman field. This is because the quasiparticle gap in this range of Zeeman field, which is below that shown in Fig. \ref{fig:idealCase}(a), is larger than the charging energy $E_c$, so that the transport is dominated by Cooper-pair transport\cite{vanHeck2016Conductancea}. The charging energy $E_c$ is defined by the electrostatic energy
\begin{equation}\label{electrostaticE}
	U(N)=E_c(N-n_g)^2
\end{equation}
where $N$ is the total electron number in the Majorana nanowire. The range of Zeeman energy plotted in Fig. \ref{fig:idealCase}(a) is where the quasiparticle gap is below $E_c$, so that the conductance is dominated by single-electron transport. As will be discussed in more detail in Sec.~\ref{sec:level1_2}, the calculated conductance $G$ is $2e$ periodic in the gate charge $n_g$. As a result, the conductance plot shown in Fig. \ref{fig:idealCase}(a) is representative of the conductance over the entire range of gate charge $n_g$. Note that the CB conductance [e.g. Fig. \ref{fig:idealCase}(a)] in this work is shown in units where the normal state conductance peak (i.e., at high $V_z$) is equal to one. This is a natural unit to use in CB situations, and further details of this choice are discussed following Eq. \eqref{tunnel_Lambda}.

The energy spectrum in Fig. \ref{fig:idealCase}(c) shows (as elaborated in Sec.~\ref{sec:level1_5}) that the oscillations seen in the OCPS in Fig. \ref{fig:idealCase}(d) arise from ABSs\cite{Kells2012Nearzeroenergya,Prada2012Transporta,Moore2018Quantizeda,Moore2018Twoterminala,Liu2017Andreeva,Vuik2019Reproducing} (and not MBSs) at the ends of the wire. The dark region of the conductance in Fig. \ref{fig:idealCase}(a) arises from the part of the ABS where the energy difference between the two ABSs at the two ends exceeds the temperature. The structure of this energy spectrum contains multiple subgap states as in Shen \textit{et al}.\cite{Shen2018Parity} As elaborated in Sec.~\ref{sec:level1_5}, we find such oscillations in OCPS with decreasing amplitude only in the case of quantum dot generated ABSs before the TQPT shown by the first dashed line in Fig. \ref{fig:idealCase}(a). One of the puzzling features of the CB conductance data\cite{Albrecht2016Exponentiala,Shen2018Parity,Vaitiekenas2020Fluxinduced} in Majorana nanowires is the intensity of the $2e$ periodic peaks being higher than the $1e$ periodic part, which is the opposite of what the naive expectation is. This occurs despite the $2e$ conductance arising from Cooper-pair transport which is higher-order in the tunneling, while the $1e$ periodic peaks being from electron transport. As a result, as discussed in Sec.~\ref{sec:level1_3}, the conductance of the $2e$ periodic part is theoretically expected to be smaller\cite{vanHeck2016Conductancea}. However, our results in Fig. \ref{fig:idealCase}(a) show a significant suppression of the $1e$ periodic conductance both from the self-energy corrections as well as from the soft-gap compared with systems where these effects are not included. This suppression can explain the discrepancy between the experiment\cite{Albrecht2016Exponentiala,Shen2018Parity,Vaitiekenas2020Fluxinduced} and the simple model of Majorana tunneling for $1e$ periodic conductance\cite{vanHeck2016Conductancea}. A more detailed understanding of the features in Fig. \ref{fig:idealCase} as well as the role of the various contributing features such as soft-gap, quantum dots etc. are provided in Sec.~\ref{sec:level1_5}.

The rest of this paper is organized as follows. In Sec.~\ref{sec:level1_2}, we will derive the two-terminal generalized Meir-Wingreen formula by Fermi's golden rule, steady-state rate equations, and linear-response conductance. In Sec.~\ref{sec:level1_3}, we will analytically reduce our formalism to simple limiting in the strong Coulomb blockade limit, such as the case of single bound state, the case of one sub-gap level at each end, and Cooper-pair transitions, which simultaneously explain the observed phenomenon. In Sec.~\ref{sec:level1_4}, we describe the details of the microscopic model for the semiconductor nanowire to which we apply our generalized Meir-Wingreen formalism. Then we will demonstrate our numerical results in Sec.~\ref{sec:level1_5}, and discuss the effects from temperature, nanowire length, SC collapsing field, chemical potential, and quantum dots. In Sec.~\ref{sec:level1_6}, we discuss the key features that show up our numerical results. Section~\ref{sec:level1_7} presents a summary of our results together with potential experimental directions. Many technical details are relegated to the Appendices, which we refer to as appropriate in the main sections of the text. The relegation of the detailed derivations to the Appendices enables a seamless discussion in the main sections using the relevant formula and equations of the theory taken from the Appendices.

\section{Two-terminal generalized Meir Wingreen formula}\label{sec:level1_2}
\subsection{Setup}\label{sec:level1_2_1}
Let us consider transport through a thermalizing dot $Q$, which has $N$ units of charge on it. Note that this thermalizing dot is our main system, i.e., the nanowire in our case, which has nothing to do with the unintentional quantum dots induced by disorder in our context. It is described by the Hamiltonian
\begin{equation}\label{H_Q}
	H_Q=\sum_j E_j|\psi_j\rangle\langle\psi_j|
\end{equation}
, where $\{E_j\}$ is the set of energies of the system. The dot $Q$ is coupled to two leads $L$ and $R$, with chemical potential $\mu_L$ and $\mu_R$, respectively, through a tunneling Hamiltonian
\begin{equation}\label{H_t}
	H_t=\sum_{k,\alpha=L,R}t_\alpha\left[a_{k\alpha}^\dagger c_\alpha+\text{H.c.}\right]
\end{equation}
, where $c_{\alpha=L,R}$ is an electron annihilation operator on the left and right ends of $Q$. The operators $a_{k,\alpha}^\dagger$ create electrons in the leads $\alpha=L,R$. $t_\alpha$ is the tunneling matrix element at the $\alpha$ end.

We treat the tunneling to the lead perturbatively as in Ref. \cite{Mallayya2019Prethermalization}. The tunnel coupling is assumed to be weak enough so that the quantum dot $Q$ can reach thermal equilibrium between successive tunneling events that change the conserved change $N$. The assumption here being that the tunneling is slow compared with the equilibration time, which can always be ensured by tuning the tunnel barrier. Within this framework, the states of dot $Q$ within a specific charge sector $N$, labeled $i$, have a conditional probability given by the Boltzmann distribution
\begin{equation}\label{P_N}
	P_N(E_i)=Z_N^{-1}e^{-\beta E_i},
\end{equation}
where $E_i$ is the energy of state $i$, $\beta$ is the inverse of temperature, and 
\begin{equation}\label{Z_N}
	Z_N=\sum_i e^{-\beta E_i}
\end{equation}
is the normalized partition function for each charge $N$, i.e., state $i$ has $N$ electrons. Note that
\begin{equation}\label{CondP}
	P_N(E_i)=\frac{P(E_i)}{P_{0,N}},
\end{equation}
where $P(E_i)$ is the probability of the state $i$ with energy $E_i$ and $P_{0,N}$ is the probability of having $N$ electrons. It is assumed that the state $i$ has $N$ electrons.

The probability distribution $P_{0,N}$ of having $N$ electrons is determined by the balance of two processes where the system $Q$ either gains or loses the electron from the leads via a tunneling process. The tunneling rate of electrons from the leads into $Q$, which is assumed to be in a charge state $N$, can be computed using Fermi's golden rule to be
\begin{equation}\label{FermiGolden}
	\Gamma_N^\alpha=\tau_\alpha\sum_{i,j}P_N(E_i)\int d\epsilon f(\epsilon-\mu_\alpha)\delta(E_j-E_i-\epsilon)|\langle\psi_j|c_\alpha^\dagger|\psi_i\rangle|^2
\end{equation}
where $f(\epsilon)=(1+e^{\beta\epsilon})^{-1}$ is the Fermi function, $c_\alpha^\dagger$ is the electron creation operator at the end $\alpha=L,R$ of the system, and $\tau_\alpha=t_\alpha^2\rho_\alpha$ is the basic tunneling rate into the lead $\alpha$ with $\rho_\alpha$ being the density of states in lead $\alpha=L,R$. Considering the current from tunneling at voltages large compared to the SC gap, the tunneling rate $\tau_\alpha$ can be written as $\tau_\alpha=g_\alpha/2\pi \nu_{1D,\alpha}$\cite{vanHeck2016Conductancea}, where $g_\alpha$ is the normal state dimensionless conductance at the end $\alpha$ and $\nu_{1D,\alpha}$ is the density of states at the end of the Majorana nanowire. Following the derivation in Appendix~\ref{sec:levelA_1}, one can show that the reverse tunneling rate of electrons $\Lambda_N^\alpha$, i.e., from dot $Q$ into the leads, is related to the rate $\Gamma_N^\alpha$ as
\begin{equation}\label{eEmission}
	\Lambda_N^\alpha=e^{-\beta\mu_\alpha}\frac{Z_{N-1}}{Z_N}\Gamma_{N-1}^\alpha.
\end{equation} 
Such a relation is consistent with the requirement of satisfying the correct number distribution in $Q$ when the system is decoupled from one of the leads.

More generally, let us consider the process of transferring $j$ electrons into the dot as $\Gamma_N^{\alpha,j}$. Equilibrium with the lead $\alpha$ requires the rate of the reverse process to be
\begin{equation}\label{eEmission_j}
	\Lambda_N^{\alpha,j}=e^{-j\beta\mu_\alpha}\frac{Z_{N-j}}{Z_N}\Gamma_{N-j}^{\alpha,j}.
\end{equation}

\subsection{Steady-state rate equations}\label{sec:level1_2_2}
The steady-state probability distribution $P_{0,N}$ can be determined from the rates $\Gamma_N^\alpha$ and $\Lambda_N^\alpha$ by equating the rate of electrons transitioning from having $N$ electrons to $N\pm 1$ electrons to the rate of electrons making the reverse transition. This equation, following the definitions of the rates $\Gamma_N^\alpha$ and $\Lambda_N^\alpha$, can be written to be:
\begin{equation}\label{detailedBalance}
	P_{0,N}\sum_{\alpha,j}(\Gamma_N^{\alpha,j}+\Lambda_N^{\alpha,j})=\sum_{\alpha,j}\left[P_{0,N-j}\Gamma_{N-j}^{\alpha,j}+P_{0,N+j}\Lambda_{N+j}^{\alpha,j}\right]
\end{equation}
Substituting $\Lambda_N^{\alpha,j}$ from Eq. \eqref{eEmission_j} and rearranging the terms, the above condition is given by
\begin{equation}\label{tildeBalance}
	\begin{aligned}
	&\sum_{\alpha,j}\tilde{\Gamma}_N^{\alpha,j}\{\tilde{P}_{0,N}-\tilde{P}_{0,N+j}e^{-j\beta\mu_\alpha}\}\\
	=&\sum_{\alpha,j}\tilde{\Gamma}_{N-j}^{\alpha,j}\{\tilde{P}_{0,N-j}-\tilde{P}_{0,N}e^{-j\beta\mu_\alpha}\}
	\end{aligned}
\end{equation}
where $\tilde{\Gamma}_N^{\alpha,j}=\Gamma_N^{\alpha,j}Z_N$ and $\tilde{P}_{0,N}=P_{0,N}/Z_N$. The above equation is solved by the detailed balance condition, where both sides of the above equation vanish so that
\begin{equation}\label{sol_tildeP}
	\tilde{P}_{0,N}=\left(\frac{\sum_{\alpha,j}\tilde{P}_{0,N+j}\tilde{\Gamma}_N^{\alpha,j}e^{-j\beta\mu_\alpha}}{\sum_{\alpha,j}\tilde{\Gamma}_N^{\alpha,j}}\right).
\end{equation}
Redefining $P'_{0,N}=\tilde{P}_{0,N}e^{-N\beta\mu_L}$ to simplify the equilibrium solution and defining $\mu_\alpha=V_\alpha+\mu_L$ as the voltages for linear response, the above equation becomes
\begin{equation}\label{linearExpP}
    \begin{aligned}
    P'_{0,N}&=\left(\frac{\sum_{\alpha,j}P'_{0,N+j}\tilde{\Gamma}_N^{\alpha,j}e^{-j\beta V_\alpha}}{\sum_{\alpha,j}\tilde{\Gamma}_N^{\alpha,j}}\right)\\
    &\approx\left(\frac{\sum_{\alpha,j}P'_{0,N+j}\tilde{\Gamma}_N^{\alpha,j}}{\sum_{\alpha,j}\tilde{\Gamma}_N^{\alpha,j}}\right)-\beta\left(\frac{\sum_{\alpha,j}jP'_{0,N+j}\tilde{\Gamma}_N^{\alpha,j}V_\alpha}{\sum_{\alpha,j}\tilde{\Gamma}_N^{\alpha,j}}\right)
    \end{aligned}
\end{equation}
There is a trivial solution for equilibrium with $V_\alpha=0$, which is constant with $P'_{0,N}=P_{0,eq}+v_N$ so that
\begin{equation}\label{v_N}
	v_N\approx\left(\frac{\sum_{\alpha,j}v_{N+j}\tilde{\Gamma}_N^{\alpha,j}}{\sum_{\alpha,j}\tilde{\Gamma}_N^{\alpha,j}}\right)-\beta P_{0,eq}\zeta_N
\end{equation}
\begin{equation}\label{zeta_N}
	\zeta_N=\left(\frac{\sum_{\alpha,j}j\tilde{\Gamma}_N^{\alpha,j}V_\alpha}{\sum_{\alpha,j}\tilde{\Gamma}_N^{\alpha,j}}\right).
\end{equation}
This solution is invariant under a constant shift, which in principle is fixed by normalization. In the case of a two-terminal case, only $V_R=V$ is non-zero, so we can expand $\zeta_N=\rho_N V$ where
\begin{equation}\label{rho_N}
	\rho_N=\left(\frac{\sum_j j\tilde{\Gamma}_N^{R,j}}{\sum_{\alpha,j}\tilde{\Gamma}_N^{\alpha,j}}\right).
\end{equation}
The fluctuations $v_N$ can now be expanded $v_N=\beta P_{0,eq}V\nu_N$ which satisfies
\begin{equation}\label{nu_N}
	\nu_N\approx\left(\frac{\sum_{\alpha,j}\nu_{N+j}\tilde{\Gamma}_N^{\alpha,j}}{\sum_{\alpha,j}\tilde{\Gamma}_N^{\alpha,j}}\right)-\rho_N.
\end{equation}

\subsection{Linear response: conductance}\label{sec:level1_2_3}
% Remember to write the final one-end Meir-Wingreen formula in the end of this section.
The current at the left lead $L$ is determined by the balance of electrons tunneling in and out of $L$. Using the definitions of the tunneling rate $\Gamma_N^{\alpha,j}$ and $\Lambda_N^{\alpha,j}$, this current can be written as
\begin{equation}\label{current}
	I=\sum_{N,j}jP_{0,N}\left(\Gamma_N^{L,j}-\Lambda_N^{L,j}\right).
\end{equation}
Using the rescaled variables in Eq.\eqref{tildeBalance}, the current is re-written as
\begin{equation}\label{current_rescaled}
	I=\sum_{N,j}j\tilde{\Gamma}_N^{L,j}e^{\beta N\mu_L}\{P'_{0,N}-P'_{0,N+j}\}.
\end{equation}
Substituting the current to linear order is
\begin{equation}\label{linearCurrent}
	I\approx\sum_{N,j}j\tilde{\Gamma}_N^{L,j}e^{\beta N\mu_L}\{v_N-v_{N+j}\}.
\end{equation}
Fortunately, the current is not affected by the constant shift ambiguity.

Divided by $V$, the conductance is found to be
\begin{equation}\label{conductance}
	\begin{aligned}
	G&=\sum_{N,j}j\beta P_{0,eq}\tilde{\Gamma}_N^{L,j}e^{\beta N\mu_L}\{\nu_N-\nu_{N+j}\}\\
	&=\sum_{N,j}j\gamma_N^{L,j}\{\nu_N-\nu_{N+j}\}
	\end{aligned}
\end{equation}
, where $\gamma_N^{\alpha,j}=-\beta P_{0,N}\Gamma_N^{\alpha,j}$. Since the redefinition is a scaling that depends only on $N$ and the equation for $\nu_N$ only involves ratios of $\tilde{\Gamma}$, the equation for $\nu$ can be re-written in terms of $\gamma_N^{\alpha,j}$ as
\begin{equation}\label{rho_N_2}
	\rho_N=\left(\frac{\sum_j j\gamma_N^{R,j}}{\sum_{\alpha,j}\gamma_N^{\alpha,j}}\right).
\end{equation}
The fluctuations $v_N$ can now be expanded as $v_N=\beta P_{0,eq}V\nu_N$, which satisfies
\begin{equation}\label{nu_N_2}
	\nu_N\approx\left(\frac{\sum_{\alpha,j}\nu_{N+j}\gamma_N^{\alpha,j}}{\sum_{\alpha,j}\gamma_N^{\alpha,j}}\right)-\rho_N.
\end{equation}

\subsection{Genralized Meir-Wingreen formula}\label{sec:level1_2_4}
In this case, we limit to one-electron processes that should dominate in the strict tunneling limit. If necessary, the generalization to multi-electron processes is straightforward, albeit quite cumbersome, but multielectron transport should be negligible in the tunneling limit of interest here.

The zero-bias conductance $G$ can be calculated by expanding the current $I$ to the lowest order in the bias voltage $\mu_R-\mu_L=V$. Refer to the Appendix~\ref{sec:levelA_3},
\begin{equation}\label{dIdV}
	G=\frac{dI}{dV}|_{V=0}=-\beta\sum_N\tilde{P}_{0,N}\frac{\tilde{\Gamma}_N^R\tilde{\Gamma}_N^L}{\tilde{\Gamma}_N^R+\tilde{\Gamma}_N^L}.
\end{equation}
Restoring the variable change from Eq.\eqref{tildeBalance}, the conductance can be re-written in terms of a re-scaled lead conductance $\gamma_N^\alpha=-\beta P_{0,N}\Gamma_N^\alpha$ as
\begin{equation}\label{conductance_gamma}
	G=\sum_N\frac{\gamma_N^R\gamma_N^L}{\gamma_N^R+\gamma_N^L}.
\end{equation}
Incidentally (as detailed in the Appendix), applying all the variable transformation to Eq.\eqref{FermiGolden}, the rescaled transition rate $\gamma_N^\alpha$ is given by
\begin{equation}\label{gamma}
	\gamma_N^\alpha=\tau_\alpha\sum_{i,j}\{P(E_i)+P(E_j)\}f'(E_j-E_i-\mu)\left|\langle\psi_j|c_\alpha^\dagger|\psi_i\rangle\right|^2
\end{equation}
, which is very similar to the effective one-terminal conductance in the Meir-Wingreen's paper\cite{Meir1992Landauer}.

In the strong CB limit, where only two charge states $N$ and $N-1$ participate in transport, $\gamma_N$ can be assumed to vanish except for one value of $N$, so
\begin{equation}\label{G_singleN}
    G=\frac{\gamma_N^R\gamma_N^L}{\gamma_N^R+\gamma_N^L}
\end{equation}
, which is physically the series formula for the conductance at each end. This equation is the same as Eq.(176) of the Aleiner \textit{et al}. review\cite{Aleiner2002Quantum}, except that the matrix elements of Eq.(130) are replaced by the Meir-Wingreen formula.

To proceed further, we assume the system Hamiltonian [i.e., Eq.\eqref{H_Q}] to be of the form
\begin{align}\label{H_Q_v2}
    H_Q=\sum_p \epsilon_p d_p^\dagger d_p + U(N),
\end{align}
where $\epsilon_p$ is the eigen-energy of the quasi-state $p$. The electron number variable $N$ is in general different from the total occupation of Bogoliubov quasiparticles $d_p^\dagger$ and instead is equivalent to the parity of the number of quasiparticles in this case. Specifically, in the limit of strong Coulomb blockade where only two consecutive values of electron number $N_0$, $N_0-1$ are allowed, the electron number $N\in\{N_0,N_0-1\}$ can be uniquely fixed by the relation $(-1)^N=Q_0\cdot(-1)^{\sum_p d_p^\dagger d_p}$, where $Q_0$ is the ground-state fermion parity of the first part (i.e., BdG) of the Hamiltonian $H_Q$, written as
\begin{equation}\label{Q_0}
    Q_0=\text{Pf}\{H_{Q,\text{BdG}}(E=0)\},
\end{equation}
where $H_{Q,\text{BdG}}$ is the first part of $H_Q$ written in  a Majorana basis. Applying this relation to Eq.\eqref{electrostaticE}, we can show that $U(N)=(\Delta U/2)\cdot Q_0\cdot(-1)^{\sum_p d_p^\dagger d_p}$, where $\Delta U=(-1)^{N_0}[U(N_0)-U(N_0-1)]$ is the electrostatic energy difference between the two transition-allowed charge states $N_0,N_0-1$. Substituting the energy eigenvalues $E_i$ and wave-functions $|\psi_i\rangle$ for $H_Q$, the coefficients $\gamma_N^\alpha$ in Eq.\eqref{gamma} can be written in a more explicit form (details are in Appendix~\ref{sec:levelA_4}):
\begin{equation}\label{gamma_simple}
	\gamma_N^\alpha=\beta\frac{e^2}{\hbar}\sum_p\sum_{n=0,1}\sum_{Q=\pm1}\tilde{F}_p(n,Q)\left[(1-n)\Gamma_{p}^\alpha+n\Lambda_{p}^\alpha\right],
\end{equation}
where
\begin{equation}\label{tilde_F}
	\tilde{F}_p(n,Q)=f_{eq}\left[(1-2n)\epsilon_p-Q\Delta U\right]\cdot F_p(n,Q)
\end{equation}
with
\begin{equation}\label{F}
	\begin{aligned}
	&F_p(n,Q)\equiv\\
	&\frac{e^{-\beta\left(\frac{Q\Delta U}{2}+n\epsilon_p\right)}\left[1+Q\cdot Q_0\cdot(-1)^n\prod_{s\neq p}\tanh\left(\frac{\beta\epsilon_s}{2}\right)\right]}{\sum_{Q=\pm 1}\sum_{n=0,1}e^{-\beta\left(\frac{Q\Delta U}{2}+n\epsilon_p\right)}\left[1+Q\cdot Q_0\prod_s\tanh\left(\frac{\beta\epsilon_s}{2}\right)\right]}.
	\end{aligned}
\end{equation}
Note that $f_{eq}(\epsilon)=(1+e^{\beta\epsilon})^{-1}$ is the Fermi distribution at equilibrium and, following Eqs. \eqref{FermiGolden} and \eqref{eEmission},
\begin{equation}\label{tunnel_Gamma}
	\Gamma_{p}^\alpha=\sum_{\sigma=\uparrow,\downarrow}\left|u_{p,\alpha\sigma}\right|^2
\end{equation}
is the tunneling rate for the electron from the lead at $\alpha$ end to the nanowire (same for the opposite direction), and
\begin{equation}\label{tunnel_Lambda}
    \Lambda_{p}^\alpha=\sum_{\sigma=\uparrow,\downarrow}\left|v_{p,\alpha\sigma}\right|^2
\end{equation}
is the tunneling rate for the hole from the lead at $\alpha$ end to the nanowire (same for the opposite direction), where $\tau_\alpha$ are assumed to be the same at both ends $\alpha$. Within the tunneling limit considered here, the value of the tunneling amplitudes $\tau_\alpha$ determines the overall scale of the conductance $G$. In our calculation, the value of $\tau_\alpha$ has been chosen so that the peak height in the normal metal (i.e., large $V_z$) regime is equal to 1. This should be considered a choice of units for our calculation. Comparison to experimental data can be made by scaling the experimental data in a similar way by the normal state conductance. We note that this choice of unit is the natural one in the tunneling limit under consideration here. Note that $u_{p,\alpha\sigma}$ and $v_{p,\alpha\sigma}$ are coefficients of electron and hole relation with quasiparticle and hole that are discussed in more detail in Sec.~\ref{sec:level1_4}.

The equation for $G$ in Eq. \eqref{conductance_gamma}, together with the definitions \eqref{Q_0}-\eqref{tunnel_Lambda}, is the central formalism used in this work to compute the conductance of a system $Q$ coupled to separate leads $L$ and $R$. Since the only constraint in equations for the conductance [Eqs. \eqref{Q_0}-\eqref{tunnel_Lambda}] connecting the number of electrons $N$ to the quasiparticle degrees of freedom is through the parity, the results are invariant as long as $N$ changes by $2$. Using Eq.\eqref{electrostaticE}, this also implies that the conductance $G$ is periodic in $n_g$ with period 2.  Because of this, in this paper, we will only plot the gate charge $n_g$ over two periods i.e., a range of length 4. This formalism reduces, in the case where leads $L$ and $R$ coincide in space, to the well-known conductance derived by Meir and Wingreen\cite{Meir1992Landauer} for interacting systems. Our work generalizes the formalism to the situation with arbitrarily spatially separated $L$ and $R$ leads as appropriate for Majorana nanowire experiments.

\section{Conductance for few-level systems}\label{sec:level1_3}
The evaluation of the conductance $G$ using Eq. \eqref{conductance_gamma} for a realistic Majorana system, which has a complicated spectrum, requires a rigorous numerical treatment. In this section, we analytically evaluate Eq. \eqref{conductance_gamma} in cases where the system $Q$ has one or two levels in the low-energy spectrum. We will find that the conductance $G$ can be written analytically in these cases. The results in these cases will help understand the numerical results for the more complex Majorana wire system, which in certain parameter regimes contains only a few low-energy levels relevant for these analytical results.

\subsection{Rates for few electron process}\label{sec:level1_3_1}
In addition to electron tunneling processes, transport through the system $Q$ also occurs through Cooper-pair tunneling because of the proximity-induced superconductivity. In order to place these two processes on a comparable footing, we rewrite the equation for the scaled transition rate for tunneling of electrons [Eq. \eqref{gamma}] as  
\begin{equation}\label{gamma_1}
	\begin{aligned}
	\gamma_N^{\alpha,1}&=\beta\tau_\alpha^{(1)}\sum_{i,j}P(E_i)\int d\epsilon f(\epsilon-\mu)\delta(E_j-E_i-\epsilon)M_{ij}^{(1)}\\
	&=\beta\tau_\alpha^{(1)}Z_{tot}^{-1}\sum_{i,j}e^{-\beta(E_i-N\mu)}f(E_j-E_i-\mu)M_{ij}^{(1)}
	\end{aligned}
\end{equation}
where $M_{ij}^{(1)}= \left|\langle |c^\dagger_\alpha| \rangle\right|^2$ is the transition matrix element for transferring one electron from the leads to the dot, and $\tau_\alpha^{(1)}$ is the one-electron tunneling rate, which was referred to as $\tau_\alpha$ in Eq. \eqref{FermiGolden}. Similarly, we can write the rate for two-electron (or Cooper-pair) transfer
\begin{equation}\label{gamma_2}
    \begin{aligned}
    \gamma_N^{\alpha,2}=&\beta\tau_\alpha^{(2)}Z_{tot}^{-1}\sum_{i,j}e^{-\beta(E_i-N\mu)}\\
    &\times\int d\epsilon f(\epsilon-\mu)f(E_j-E_i-\epsilon-\mu)M_{ij}^{(2)}\\
    =&\tau_\alpha^{(2)}Z_{tot}^{-1}\sum_{i,j}e^{-\beta(E_i-N\mu)}f^{(2)}(E_j-E_i-2\mu)M_{ij}^{(2)}
    \end{aligned}
\end{equation}
where $f^{(2)}(\epsilon)=(\beta\epsilon)/(e^{\beta\epsilon}-1)$ and $\sqrt{M_{ij}^{(2)}}$ is the matrix element of transferring a Cooper pair into $Q$. The charge of the system changes by $2$, preserving parity under this tunneling process. The parameter $\tau_\alpha^{(2)}$ sets the scale of the Cooper-pair tunneling rate analogous to the one-electron tunneling rate $\tau_{\alpha}^{(1)}$.

Using the fact that $\mu$ and the gate voltage entering $E_j$ play equivalent roles, we can set $\mu$ to zero. In that case, we can write the rates in a more symmetric form
\begin{equation}\label{gamma_1_sym}
	\gamma_N^{\alpha,1}=\beta\tau_\alpha^{(1)}Z_{tot}^{-1}\sum_{i,j}\frac{1}{e^{\beta E_i}+e^{\beta E_j}}M_{ij}^{(1)}
\end{equation}
\begin{equation}\label{gamma_2_sym}
	\gamma_N^{\alpha,2}=\beta\tau_\alpha^{(2)}Z_{tot}^{-1}\sum_{i,j}\frac{E_j-E_i}{e^{\beta E_j}-e^{\beta E_i}}M_{ij}^{(2)}
\end{equation}
where $E_i$ is understood to be replaced by $E_i\rightarrow E_i- \min_i E_i$. The latter can be done since only ratios of $E_i$ enter any formula. In this form, it is clear that any $\gamma_N$ rate is significant if  both energies are less than $\beta^{-1}$.

The tunneling matrix elements $\tau_{\alpha}^{(i=1,2)}$ can be estimated by considering the limits of transport without a superconducting gap and without sub-gap states respectively. In the case without SC, we can rewrite Eq.\eqref{gamma_1_sym} as $\gamma^1\approx \beta\tau^{(1)}\nu_{Q}\int_{\omega>0} d\omega \langle M^{(1)}(\omega)(1+e^{\beta\omega})\rangle$, where $\nu_Q$ is the normal state density of states (DOS) in the system $Q$. Ignoring the frequency dependence of $M^{(1)}(\omega)$ on the frequency on the scale of the temperature $T$ so that $\nu_Q M^{(1)}(\omega)\approx \nu_{1D,\alpha}$, we can write the end conductance $\gamma^{(1)}\equiv P_{0,N}g_\alpha\sim P_{0,N}\tau^{(1)}_\alpha\nu_{1D,\alpha}$ so that $g_\alpha\sim \tau_\alpha^{(1)}\nu_{1D,\alpha}$, which is similar  to the normal state conductance discussed below Eq.\eqref{FermiGolden}.  In the limit of large conductance at the opposite end, which maintains the equilibrium distribution for the number $N$, the normal state conductance is $G\sim g_\alpha$. In the case of Cooper-pair transport with no sub-gap quasiparticle state and negligible charging energy, we can assume $E_j\sim E_i$ so that $\gamma^{(2)}\sim \tau_\alpha^{(2)} M^{(2)}$. The parameter in this approximation is the single-end N-S conductance $g^{(SC)}$ calculated from the BTK formalism\cite{Blonder1982Transition} so that $\tau^{(2)}M^{(2)}\sim g^{(SC)}$. The Beenakker formula\cite{Beenakker1992Quantum} suggests that the gapped SC conductance $g^{\text{(SC)}}\sim g_\alpha^2$ in the limit where $g_\alpha$ is the conductance in units of the quantum of conductance and is assumed to be much smaller than unity. Therefore, in the tunneling limit $\gamma^{(2)}\ll \gamma^{(1)}$, leading to the expectation that the conductance from Cooper-pair transport processes should be much smaller than arising from electron transport\cite{Sau2015Proposal}.

The constraint $\gamma_\alpha^{(2)}\ll\gamma_\alpha^{(1)}$ may be alleviated by enhancement of Cooper-pair transport in the presence of ABSs. To understand this, we note that the Cooper-pair tunneling amplitude $\tau_{\alpha}^{(2)}$ is generated by elastic co-tunneling of two electrons into the superconductor through virtual states 
\begin{equation}\label{tau_alpha2}
	\tau_\alpha^{(2)}=t^2\sum_n \frac{u_n v_n^*}{E_n-\delta}=t^2\int_{\omega>\Delta U} d\omega Tr[\rho(\omega)\tau_+](\omega-\Delta U)^{-1},
\end{equation}
where $u_n$, $v_n$ and $E_n$ are the particle and hole components of the wave-functions of states with Bogoliubov-de Gennes (BdG) eigenvalue $E_n>0$.  Here $\rho(\omega)$ is the local density of the superconducting wire in Nambu space with a particle-hole matrix $\tau_+=\tau_x+i\tau_y$. Considering a simplified superconducting model where we apply a uniform pair potential to the states of a normal metal so that the superconducting density matrix is given by $Tr[\rho(\sqrt{\omega^2+\Delta^2})\tau_+]=\rho_0(\omega)\frac{\Delta}{\sqrt{\Delta^2+\omega^2}}$. Within this approxmation, $\tau_\alpha^{(2)}=t^2\int_{\omega>\Delta U} d\omega \rho_0(\omega)\frac{\Delta}{\sqrt{\omega^2+\Delta^2}(\sqrt{\omega^2+\Delta^2}-\Delta U)}$. In the limit of a uniform density of states, we can scale the integration variable $\omega\rightarrow\omega \Delta$ so that $\tau_\alpha^{(2)}\approx t^2\rho_0$. The conductance $\gamma_2\sim \tau_\alpha^{(2)2}\sim t^4\rho_0^2\sim \gamma_1^2\ll \gamma_\alpha^{(1)}$. Alternatively, if we consider a scenario that may be realistic for a semiconductor/superconductor structure where the local density of states in the semiconductor is suppressed near the Fermi level but enhanced above energy $\omega\gtrsim\Delta$, $\tau_\alpha^{(1)}\sim\rho_0(\omega\sim 0)$ may be suppressed without changing $\tau_\alpha^{(2)}$. This allows a situation where the 2e conductance peaks with height $\gamma_2$ may exceed the normal CB conductance peaks at high magnetic fields.

\subsection{Single-bound-state induced electron transport}\label{sec:level1_3_2}
In the case of one ``active" level, i.e., within the range of thermal activation, there are only two states: one with electron number $N$ and another with $N+1$, where the quasiparticle energy $\epsilon=E_{N+1}-E_N$ is the energy difference between the two states. Substituting the quasiparticle energy into Eq. \eqref{gamma_1_sym} leads to
\begin{equation}\label{gamma_one_level}
	\gamma_N^{\alpha,1}=\frac{\beta\tilde{\tau}_\alpha^{(1)}}{4}\text{sech}^2\left(\frac{\beta\epsilon}{2}\right)
\end{equation}
, where $\tilde{\tau}_\alpha^{(1)}=\tau_\alpha^{(1)}M_{ij}^{\alpha,(1)}$. In this case, the conductance becomes
\begin{equation}\label{G_one-level}
	G=\frac{\beta}{4}\frac{\tilde{\tau}_R^{(1)}\tilde{\tau}_L^{(1)}}{\tilde{\tau}_R^{(1)}+\tilde{\tau}_L^{(1)}}\text{sech}^2\left(\frac{\beta\epsilon}{2}\right)
\end{equation}
, which is consistent with Chiu \textit{et al.} \cite{Chiu2017Conductance}. The single-level case is also consistent with Meir-Wingreen's original formula\cite{Meir1992Landauer}. Note here that the conductance of a state that is localized at one of the ends of the system is substantially suppressed since one of $\tau_{R,L}^{(1)}$ is small. From the last paragraph, $\tilde{\tau}_\alpha\sim g_\alpha \nu_{1D}$ so that $G\sim \beta\nu_{1D}\frac{g_R g_L}{g_R+g_L}\text{sech}^2(\beta\epsilon/2)$. This conductance is enhanced compared to the non-Coulomb blockaded conductance.

One can use Eq. \eqref{G_one-level} to estimate the conductance in the case of a large number of levels with similar transmissions $\tilde{\tau}$. Assuming that the conductance is split among $N$ levels with conductance $\tilde{\tau}/N$ spread out over a range $\Delta$, the resulting conductance can be approximated by
\begin{equation}\label{G_delocalised}
	\begin{aligned}
	G&\simeq \frac{\beta}{4}\frac{\tilde{\tau}_R^{(1)}\tilde{\tau}_L^{(1)}}{\tilde{\tau}_R^{(1)}+\tilde{\tau}_L^{(1)}}\Delta^{-1}\int \text{sech}^2\left(\frac{\beta\epsilon}{2}\right)d\epsilon\\
	&\approx\frac{\beta}{4}\frac{\tilde{\tau}_R^{(1)}\tilde{\tau}_L^{(1)}}{\tilde{\tau}_R^{(1)}+\tilde{\tau}_L^{(1)}}\frac{2T}{\Delta}.
	\end{aligned}
\end{equation} 
We note that the conductance $G$ in this case is suppressed relative to Eq. \eqref{G_one-level} by a temperature-dependent factor of $(T/\Delta)$. This factor cancels the factor $\beta\nu_{1D}$ so that the conductance is now temperature independent and comparable with the conductance of the non-Coulomb blockaded case\cite{Aleiner2002Quantum,Kouwenhoven1997Electron}.

\subsection{One sub-gap level at each end}\label{sec:level1_3_3}
Let us consider the case of a long wire with a pair of levels, one at each of the left and right ends:
\begin{equation}\label{gamma_1_sym_copy}
\gamma_N^{\alpha,1}=\beta\tau_\alpha^{(1)}Z_{tot}^{-1}\sum_{i,j}\frac{1}{e^{\beta E_i}+e^{\beta E_j}}M_{ij}^{(1)}
\end{equation}

We assume that there are levels at the two ends of a wire with energy $\epsilon_\alpha$. Generalizing Eq. \eqref{gamma_one_level} to this case, the left and right conductances would be given by
\begin{equation}\label{generalized_gamma}
	\gamma_N^{\alpha,1}=\left(2Z_{tot}\right)^{-1}\beta\tilde{\tau}_\alpha^{(1)}e^{-\beta\epsilon_\alpha/2}\text{sech}\left(\frac{\beta\epsilon_\alpha}{2}\right).
\end{equation}
Using Eq. \eqref{G_singleN}, the conductance can be written as
\begin{equation}\label{G_twoLevel}
	G=\frac{\beta}{4}\frac{\tilde{\tau}_R^{(1)}\cdot\tilde{\tau}_L^{(1)}}{\tilde{\tau}_0^{(1)}\cosh^2\left(\frac{\beta\epsilon_1}{2}\right)+\tilde{\tau}_1^{(1)}\cosh^2\left(\frac{\beta\epsilon_0}{2}\right)},
\end{equation}
where $\epsilon_0=\min_\alpha(\epsilon_\alpha)$ and $\epsilon_1=\max_\alpha(\epsilon_\alpha)$. Note that at the CB resonance, $\epsilon_0\rightarrow 0$ while $\epsilon_1$ stays positive. This means that the conductance $G$ is exponentially suppressed if a pair of levels near the left end and right end have different energies.

If the energy levels of the ABSs on the left and the right are nearly degenerate, i.e., $\epsilon_L\approx\epsilon_R=\epsilon$, the above equation reduces to the result for a single level, i.e., Eq. \eqref{G_one-level} and the conductance suppression is eliminated. This can be seen in the short nanowire case, considering that the ABSs on both ends are delocalized so that one bound state occupies both ends. This also means the exponential suppression in Eq. \eqref{G_twoLevel} only applies to the long nanowire, where the bound states are localized enough. The conductance for the two-state system in Eq. \eqref{G_twoLevel}, can also be suppressed even in the case of nearly degenerate level $\epsilon_\alpha=\epsilon$ in the presence of gapless states in the bulk of the superconductor that are generated by a magnetic field, as is assumed for the results in Fig. \ref{fig:idealCase}. This suppression can be understood as a suppression of the tunneling matrix elements $\tau_\alpha$ resulting from hybridization between the bound states and the bulk states. As will be elaborated in the discussion, this suppression will play a role in understanding the suppression of conductance relative to that from the $2e$ periodic Cooper-pair transport.

\subsection{Conductance near $N$ and $(N+2)$ degeneracy}\label{sec:level1_3_4}
In this case, transport is dominated by Cooper-pair transfer processes and Eq. \eqref{G_singleN} can be generalized to 
\begin{equation}\label{G_Cooper_1}
	G= \frac{\gamma^{L,2}\gamma^{R,2}}{\gamma^{L,2}+\gamma^{R,2}}
\end{equation}

Using Eq. \eqref{gamma_2_sym}, so that $\gamma^{\alpha,2}=\beta\tau^{\alpha,2} \frac{\epsilon}{\sinh(\beta\epsilon)}$, where $\epsilon=E_{N+2}-E_N$, the Cooper-pair conductance is written as 

\begin{equation}\label{G_Cooper_2}
	G= \frac{\tau^{L,2}\tau^{R,2}}{\tau^{L,2}+\tau^{R,2}}\frac{\beta\epsilon}{\sinh(\beta\epsilon)}.
\end{equation}
We notice that this reaches a maximum value comparable to $g_{\text{SC}}$ that is independent of temperature as $\epsilon$ approaches 0. This is different from the suppression factor for the single-level case in Sec.~\ref{sec:level1_3_2} and the maximum value in this case is simply the non-CB conductance $g_{\text{SC}}$.

For an ideal superconductor, the number of electrons in the system $N$ is even. However, 
sub-gap bound states that are not directly coupled to the leads can play an important role in the CB transport. For example, applied Zeeman fields can drive a state to cross zero energy changing the ground-state parity of $N$. This leads to a shift in the periodicity of the CB conductance of the system\cite{Shen2018Parity}. Another possibility is where the system has states on the order of or lower than the temperature of the system. In this case, the $N$ and $N+2$ degeneracy cannot be reached, because this gate voltage would correspond to a ground state of $N+1$, which has no degeneracy. 

\section{Nanowire Hamiltonians}\label{sec:level1_4}
\subsection{The Hamiltonian for 1D superconducting nanowire}\label{sec:level1_4_1}
The 1D superconductor-proximitized semiconductor nanowire with spin-orbit coupling in the presence of a field-induced Zeeman spin splitting can be described in the following form
\begin{equation}\label{H_BdG}
	\begin{aligned}
	\hat{H}_{\text{BdG}}(\epsilon)=&\sum_x\{C_x^\dagger\left[(2t-\mu)\tau_z\sigma_0+V_z\tau_0\sigma_x+\Sigma(\epsilon)\right]C_x\\
	&+\left[C_{x+a}^\dagger(-t\tau_z\sigma_0+i\alpha\tau_z\sigma_y)C_x+\text{H.c.}\right]\}
	\end{aligned}
\end{equation}
where $C_x=(c_{x\uparrow},c_{x\downarrow},c_{x\uparrow}^\dagger,c_{x\downarrow}^\dagger)$ is the electron operator at position $x$, and
\begin{equation}\label{selfE}
	\Sigma(\epsilon)=-\lambda\frac{\epsilon\tau_0\sigma_0+\Delta\tau_x\sigma_0}{\sqrt{\Delta^2-\epsilon^2}}
\end{equation}
is the self-energy\cite{Stanescu2010Proximity}, with the Zeeman-field-varying superconducting gap
\begin{equation}\label{SCgap}
	\Delta(V_z)=\Delta_0\sqrt{1-(V_z/V_c)^2}
\end{equation}
, where $V_c$ is the Zeeman field that the bulk superconducting gap of the parent superconductor collapses. We can also add the quantum dot (QD) into the nanowire. As an example, the potential confining quantum dots for our numerical results is of this form:
\begin{equation}\label{V_dot}
	V_{dot}(x)=V_D\cos\left(\frac{3\pi x}{2l_D}\right)
\end{equation}
at the left and right ends of the nanowire, but the potential depth $V_D$ value is different on both sides. $l_D$ is the QD length. The whole Hamiltonian with the QD is
\begin{equation}\label{H_QD}
	\begin{aligned}
	\hat{H}_{\text{QD}}(\epsilon)=&\sum_x\{C_x^\dagger\left[(2t-\mu+V_{dot}(x))\tau_z\sigma_0+V_z\tau_0\sigma_x\right]C_x\\
	&+\left[C_{x+a}^\dagger(-t\tau_z\sigma_0+i\alpha\tau_z\sigma_y)C_x+\text{H.c.}\right]\}
	\end{aligned}
\end{equation}

The Hamiltonian of the leads is described by
\begin{equation}\label{H_lead}
    \begin{aligned}
    \hat{H}_{lead}(\epsilon)=&\sum_x\{C_x^\dagger\left[(2t-\mu+E_{lead})\tau_z\sigma_0+V_z\tau_0\sigma_x\right]C_x\\
    &+\left[C_{x+a}^\dagger(-t\tau_z\sigma_0+i\alpha\tau_z\sigma_y)C_x+\text{H.c.}\right]\},
\end{aligned}
\end{equation}
where $E_{lead}$ is the gate voltage on the lead. There is also a normal metal-semiconductor tunnel barrier at the junction between the leads and the nanowire. The Hamiltonian in the area with the tunnel barrier is described by
\begin{equation}\label{H_barrier}
	\begin{aligned}
	\hat{H}_{barrier}&(\epsilon)=\\
	&\sum_x\{C_x^\dagger\left[(2t-\mu+V_{barrier}(x))\tau_z\sigma_0+V_z\tau_0\sigma_x\right]C_x\\
	+&\left[C_{x+a}^\dagger(-t\tau_z\sigma_0+i\alpha\tau_z\sigma_y)C_x+h.c.\right]\}
	\end{aligned}
\end{equation}
where $V_{barrier}=E_{barrier}\Pi_{l_{barrier}}(x)$ is a square potential with potential strength $E_{barrier}$ and width $l_{barrier}$.

The electron creation and annihilation operators are written in terms of quasi-particles and quasi-holes
\begin{equation}\label{e_create}
	c_{x\sigma}^\dagger=\sum_p\left(u_{p,x\sigma}^* d_p^\dagger+v_{p,x\sigma}d_p\right),
\end{equation}
\begin{equation}\label{e_annihilation}
    c_{x\sigma}=\sum_p\left(v_{p,x\sigma}^* d_p^\dagger+u_{p,x\sigma}d_p\right).
\end{equation}
The normalization leads to $\sum_{x,\sigma}\left(|u_{p,x\sigma}|^2+|v_{p,x\sigma}|^2\right)=1$. The quasi-particle and quasi-hole for the energy level $p$ are given by
\begin{equation}\label{quasi_create}
	d_p^\dagger=\sum_{x,\sigma=\uparrow,\downarrow}\left(u_{p,x\sigma}c_{x\sigma}^\dagger+v_{p,x\sigma}c_{x\sigma}\right)
\end{equation}
\begin{equation}\label{quasi_annihilation}
	d_p=\sum_{x,\sigma=\uparrow,\downarrow}\left(v_{p,x\sigma}^*c_{x\sigma}^\dagger+u_{p,x\sigma}^*c_{x\sigma}\right)
\end{equation}

\subsection{Tunneling rate from the density matrix}\label{sec:level1_4_2}
From Eqs. \eqref{tunnel_Gamma} and \eqref{tunnel_Lambda}, the tunneling rate can be expressed as the square of wave function coefficients, based on Eqs. \eqref{quasi_create} and \eqref{quasi_annihilation}. So, the expression in Eq.\eqref{tunnel_Gamma} and \eqref{tunnel_Lambda} is technically $\left|\langle\psi_{\epsilon_p}^e|x\rangle\right|^2$ and $\left|\langle\psi_{\epsilon_p}^h|x\rangle\right|^2$, respectively. Note that $\psi_{\epsilon_p}^e$ ($\psi_{\epsilon_p}^h$) is the electron (hole)-part of wave function with eigen-energy $\epsilon_p$. Our first trial to calculate the tunneling rates of the SC states (above SC gap) and metallic states (above SC collapses) would be to use the local and total density of states (LDOS/DOS). See Appendix~\ref{sec:levelA_6} for more details. However, this approach does not fit in when the degeneracy happens, which is a common situation in our system. Therefore, we have to switch to use the density matrix approach to find the degeneracy and the degenerate wave functions for the tunneling rates. 

We start with the density matrix obtained by taking the anti-Hermitian part of the Green's function, i.e.,
\begin{equation}\label{rho_wire_1}
    \rho_{wire}(\omega)=\left(\frac{G(\omega)-G^\dagger(\omega)}{2 i\pi}\right)
\end{equation}
or
\begin{equation}\label{rho_wire_2}
    \begin{aligned}
    \rho_{wire}(\omega)&\equiv\rho_{wire}(x\sigma\tau;x'\sigma'\tau';\omega)\\
    &=\sum_{n,m}\psi_{n,m}(x\sigma\tau)\psi_{n,m}^\dagger(x'\sigma'\tau')\delta(\omega-\epsilon_n)
    \end{aligned}
\end{equation}
($n$ is the energy-level index and $m$ is the degeneracy index of an energy level), while we express the Green's function in the basis of eigenstates, i.e.,
\begin{equation}\label{G_wire}
    \begin{aligned}
    G(\omega)\equiv&G_{wire}(x\sigma\tau;x'\sigma'\tau';\omega)\\
    =&\left(\hat{H}_{\text{BdG}}(\omega)-\omega+i\eta\right)^{-1}\\
    =&\sum_p\frac{\psi_p(x\sigma\tau)\psi_p^\dagger(x'\sigma'\tau')}{\epsilon_p-\omega+i\eta}\\
    =&\underbrace{\sum_p\psi_p(x\sigma\tau)\psi_p^\dagger(x'\sigma'\tau')P(\frac{1}{\epsilon_p-\omega})}_{\text{Hermitian}}\\
    &+\underbrace{i\pi\sum_p\delta(\omega-\epsilon_p)\psi_p(x\sigma\tau)\psi_p^\dagger(x'\sigma'\tau')}_{\text{anti-Hermitian}}.
    \end{aligned}
\end{equation}
, where the linewidth $\eta$ is assumed to be infinitesimal.\\
We can pick out the sub-density matrix at energy level $\epsilon_n$ by integrating the density matrix over the bound state range $[\epsilon_n-a,\epsilon_n+b]$, i.e.,
\begin{equation}\label{sub_rho_wire_1}
    \begin{aligned}
    \rho_{wire}^{(n)}(x\sigma\tau;x'\sigma'\tau')&\equiv\int_{\epsilon_n-a}^{\epsilon_n+b}\rho_{wire}(\omega)d\omega\\
    &=\sum_m\psi_{n,m}(x\sigma\tau)\psi_{n,m}^\dagger(x'\sigma'\tau').
    \end{aligned}
\end{equation}
Note that the integration grid needs to be much finer than the linewidth $\eta$, so the numerical discrete integration can be close to the continuous integration. The benefit of using density matrix is that we can find the degenerate wave functions of energy $\epsilon_n$ by diagonalizing the sub-density matrix $\rho_{wire}^{(n)}$ and get eigenvalues $\lambda_m$ and eigenwave function $\phi_{n,m}$. Then we can express the sub-density matrix as
\begin{equation}\label{sub_rho_wire_2}
    \rho_{wire}^{(n)}(x\sigma\tau;x'\sigma'\tau')=\sum_m\lambda_m\phi_{n,m}(x\sigma\tau)\phi_{n,m}^\dagger(x'\sigma'\tau').
\end{equation}
Comparing Eqs. \eqref{sub_rho_wire_1} and \eqref{sub_rho_wire_2}, we can get the effective wave function of degeneracy label $m$ of energy $\epsilon_n$:
\begin{equation}\label{eff_psi}
    \psi_{n,m}(x\sigma\tau)=(\lambda_m)^{1/2}\phi_{n,m}(x\sigma\tau).
\end{equation}
By default, we will get $(4\cdot N_{tot})$ of $\lambda_m$ by diagonalizing $\rho_{wire}^{(n)}(x\sigma\tau;x'\sigma'\tau')$ with the dimension $(4\cdot N_{tot})\times(4\cdot N_{tot})$, if $N_{tot}$ is the number of lattice site. In order to get the correct degeneracy for the bound states, we need to technically set some threshold for $\lambda_m$, i.e., only those degenerate states with $\lambda_m$ larger than the threshold can be picked as the degenerate states we are going to include into the calculations. This threshold is also kind of constrained by the linewidth of the Green's function $\eta$ in Eq.\eqref{G_wire}. When $\eta$ is small, the peak of DOS is very sharp and narrow, even a small threshold can select the eigenvalue $\lambda_m$ precisely. On the other hand, when $\eta$ is larger, the peak of DOS becomes broadened, then we need a higher threshold to select out the positions of the central peaks. This threshold for $\lambda_m$ should be universally the same over the whole parameter space, in order to make sure the conductance depends only on the LDOS at both ends of the nanowire near zero energy.

The above formula \eqref{eff_psi} applies to the bound states below the gap. For SC states (above the gap before the gap collapses) and metallic states (after the gap collapses), the integration range in Eq. \eqref{sub_rho_wire_1} will be a bit different, i.e.,
\begin{equation}\label{sub_rho_wire_3}
    \begin{aligned}
    \rho_{wire}^{(n)}(x\sigma\tau;x'\sigma'\tau')&\equiv\int_{\epsilon_n-\Delta\epsilon_{n-1}/2}^{\epsilon_n+\Delta\epsilon_n/2}\rho_{wire}(\omega)d\omega\\
    &\approx\rho_{wire}(\epsilon_n)\left(\frac{\Delta\epsilon_n+\Delta\epsilon_{n-1}}{2}\right)\\
    &\approx\rho_{wire}(\epsilon_n)\cdot\Delta\epsilon_n\\
    &=\sum_m\lambda_m\phi_{n,m}(x\sigma\tau)\phi_{n,m}^\dagger(x'\sigma'\tau')\cdot\Delta\epsilon_n
    \end{aligned}
\end{equation}
with the energy spacing picked as
\begin{equation}\label{E_spacing}
    \Delta\epsilon_n\equiv\epsilon_{n}-\epsilon_{n-1}\approx\frac{D_n}{\rho_{\text{SC}}(\epsilon_n)V_{\text{SC}}}
\end{equation}
where the presumed degeneracy $D_n$ is the size of the density matrix, the bulk BCS DOS of the superconductor $\rho_{\text{SC}}(\epsilon_n)$ is
\begin{equation}\label{rho_SC}
	\rho_{\text{SC}}(\epsilon)=\frac{2\rho_F\epsilon}{\sqrt{\epsilon^2-\Delta(V_z)^2}}\theta\left[\epsilon-\Delta(V_z)\right],
\end{equation}
where $\rho_F$ is the DOS at Fermi level. $V_{\text{SC}}$ is defined from the total DOS above the SC gap as
\begin{equation}\label{rho_tot}
    \rho_{tot}(\epsilon)=V_{\text{SC}}\rho_{\text{SC}}(\epsilon).
\end{equation}
This total DOS is under the assumption that the superconductor is much larger than the nanowire so that the component of the wave function in the nanowire is negligible. After the superconducting gap collapses, i.e., $\Delta(V_z>V_c)=0$, Eq. \eqref{rho_SC} reduces to
\begin{equation}\label{rho_metal}
    \rho_{metal}(\epsilon)=2\rho_F
\end{equation}
, a constant. Note that the factor 2 here is due to the spins. 

To avoid the singularity at $\rho_{\text{SC}}(\epsilon_n=\Delta)$ [check out Eq. \eqref{rho_SC}], we define each energy level (for SC states and metallic states) as follows. First, we define the states density (states existing per unit volume) as
\begin{equation}\label{StatesDensity_1}
    F(\epsilon)\equiv\int_{0}^{\epsilon}\rho_{\text{SC}}(x)dx=\int_{\Delta}^{\epsilon}\frac{2\rho_F x}{\sqrt{x^2-\Delta^2}}dx=2\rho_F\sqrt{\epsilon^2-\Delta^2}.
\end{equation}
At the same time, $F(\epsilon_n)$ can also be written as
\begin{equation}\label{StatesDensity_2}
    F(\epsilon_n)=\frac{1}{V_{\text{SC}}}\sum_{m\le n}D_m
\end{equation}
where $D_m$ is the degeneracy of $\epsilon_m$ for $\epsilon_m\le\epsilon_n$. (Note that $\epsilon_1=\Delta+0^+$.) By equating Eqs. \eqref{StatesDensity_1} and \eqref{StatesDensity_2}, we can find the energy level (for SC states and metallic states) is at
\begin{equation}\label{SC_En}
    \epsilon_n=\sqrt{\Delta^2+(\frac{\sum_{m\le n}D_m}{2\rho_F V_{SC}})^2}.
\end{equation}

The benefit from the technical side is that we only need coarse profiles of DOS. The DOS does not need to have very sharp and precise peaks of each of the energy states. We can just presume there are a lot of degenerate states occupying one dominant peak. The details, which may not be precisely known for the experimental system, would not be crucial in such a situation.

\section{Numerical Results and Discussions}\label{sec:level1_5}
\begin{figure}[bhtp]
	\includegraphics[scale=0.35]{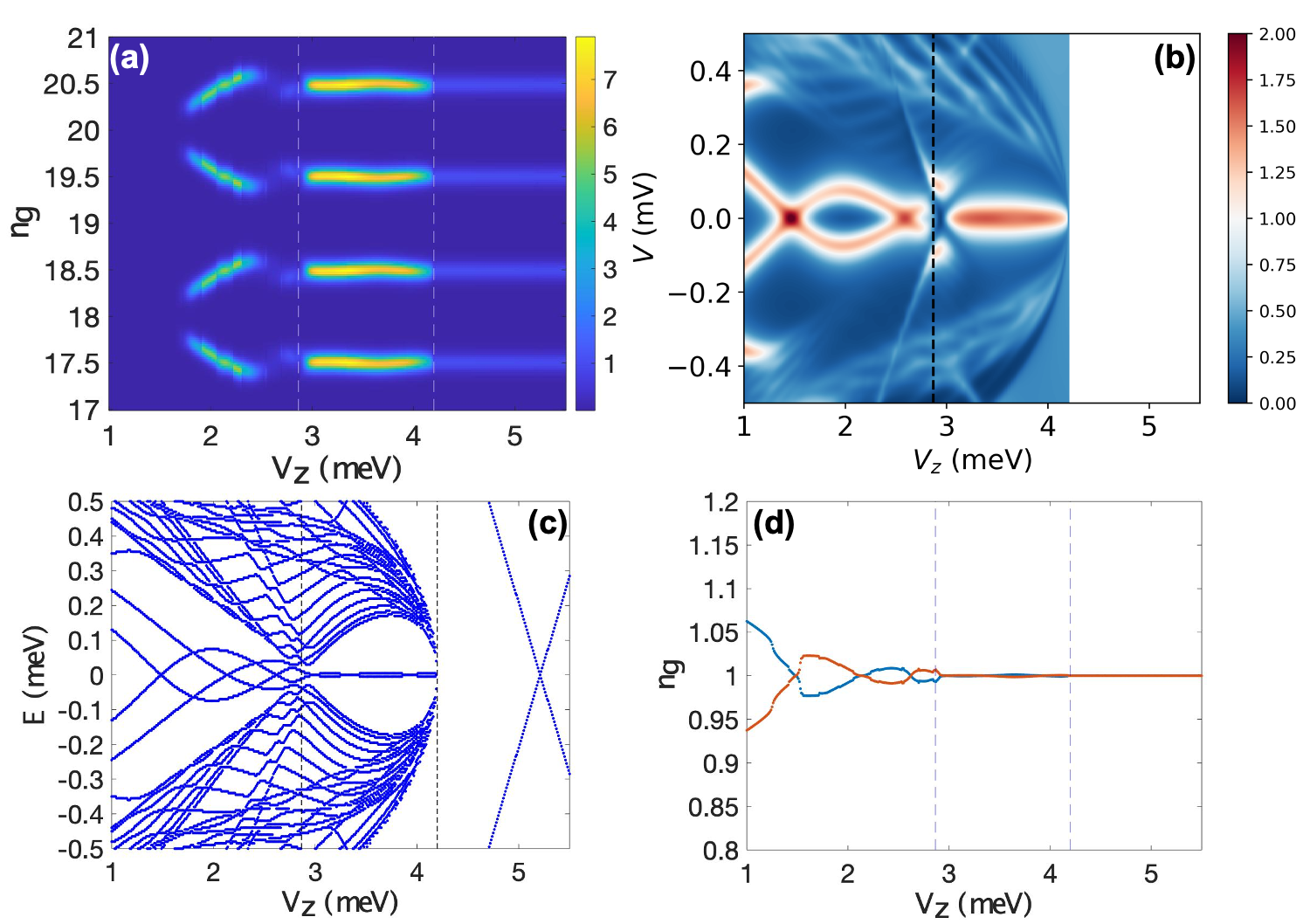}
	\caption{The reference case: We will use this set of results as reference for the later results in this section. The difference from Fig. \ref{fig:idealCase} is that this set of result does not have soft gap. The parameters are: the temperature $T=0.01$ meV, the wire length $L=1.5$ $\micro$m, the SC gap at zero Zeeman field $\Delta_0=0.9$ meV, the SC collapsing field $V_c=4.2$ meV. Other relevant parameters are given in Sec.~\ref{sec:level1_5}. (a) Coulomb blockaded conductance $G$ as a function of the gate-induced charge number $n_g$ and Zeeman field $V_z$ at $E_c=0.13$ meV. We only show two periods in the range of $n_g \in[17,21]$. (b) Non-Coulomb blockaded conductance $G$ from the left lead as a function of bias voltage $V$ and Zeeman field $V_z$. (c) Energy spectrum as a function of Zeeman field $V_z$. (d) OCPS as a function of Zeeman field $V_z$ that is extracted from the vertical peak spacing in panel (a).}
	\label{fig:refCase}
\end{figure}

In this section, we will use the results from Sec.~\ref{sec:level1_2_4} to compute the conductance from single-electron tunneling processes as shown in Fig. \ref{fig:idealCase}(a) and discussed in the Introduction. As already mentioned in the Introduction, transport from such processes, in the strong CB limit, can only occur near a degeneracy between the energy of the SC island with $N$ and $N+1$ electrons. This occurs for magnetic fields that are large enough to reduce the SC gap below the charging energy\cite{Albrecht2016Exponentiala,vanHeck2016Conductancea}. The range of Zeeman field over which the conductance is shown [similar to Fig. \ref{fig:idealCase}(a)], is thus limited to the range where the $2e$ periodic CB peaks seen at small Zeeman field are split. As mentioned in the context of the discussion of the results in Fig. \ref{fig:idealCase} in the Introduction, the CB conductance depends on many details of the Hamiltonian such as soft gap, self-energy, quantum dots, etc. The results in this section will systematically study the contribution of various mechanisms and ingredients used in Fig. \ref{fig:idealCase} by changing parameters from a reference case shown in Fig. \ref{fig:refCase}. The reference result shown in Fig. \ref{fig:refCase} is computed for the Hamiltonian described in Sec.~\ref{sec:level1_4} with the following parameters: the temperature $T=0.01$ meV, the nanowire length $L=1.5$ $\micro$m (150 sites, with the lattice space $a=10$ nm), the hopping strength $t=25$ meV, the SC gap at zero Zeeman field $\Delta_0=0.9$ meV, the SC collapsing field $V_c=4.2$ meV, the spin-orbit coupling constant $\alpha=2.5$ meV, the chemical potential $\mu=2.5$ meV, the self-energy coupling constant $\lambda=1.4$ meV, the TQPT field is theoretically at $V_{TQPT}=\sqrt{\lambda^2+\mu^2}=2.87$ meV for this parameter choice, the QD potential height at the left end $V_D=1.0$ meV, the QD potential height at the right end $V_D=4.0$ meV, and the two QD widths are both $l_D=0.26$ $\micro$m. The tunneling barrier created by the lead occupies $N_{barrier}=20$ nm (2 sites), with the energy height $E_{barrier}=10$ meV. DOS at Fermi level $\rho_F=0.1$ $(\micro m^3\cdot meV)^{-1}$, SC bulk volume $V_{SC}=10^5$ $\micro m^3$, and the upper bound of energy level is 2.5 meV (roughly 3 times larger than $\Delta_0$). The numerical results in this work will use variations around these standard parameters as described in the various sub-sections. Our choice of parameters is generic for the currently used experimental Majorana nanowire systems.

Figure \ref{fig:refCase}(a) shows the Coulomb-blockade conductance as a function of Zeeman field $V_z$ and the gate-induced charge number $n_g$. We show the calculated CB conductance for two periods in $n_g$ space, which are seen to have identical conductance pattern, thus explicitly verifying the periodicity of the conductance shown in Sec.~\ref{sec:level1_2_4}. Panel (b) is the regular tunnel conductance probed locally from the left lead as a function of Zeeman field $V_z$ and the bias voltage $V$ without Coulomb blockade. Panel (c) is the energy spectrum as a function of Zeeman field $V_z$, which is aligned with the conductance in Fig. \ref{fig:refCase}(a). The spectrum in Fig. \ref{fig:refCase}(c) is identical to the one in Fig. \ref{fig:idealCase}(c) and shows a closure of the bulk spectrum at the TQPT marked by the dashed line at the lower Zeeman field $V_z$. The $1e$ periodic conductance features in Fig. \ref{fig:refCase}(a) that are observed to arise following the TQPT are found to be brighter than in the case of the Fig. \ref{fig:idealCase}(a) and shows an abrupt drop in intensity following the Zeeman field $V_c$ (marked by the dashed line at the higher Zeeman field $V_z$) where the superconductivity of the parent Al SC is destroyed. The drop in intensity of the $1e$ periodic conductance peak in Fig. \ref{fig:refCase}(a) is a result of a transition from conductance peak associated with transport through MBSs suggested in earlier works\cite{Albrecht2016Exponentiala,vanHeck2016Conductancea} to $1e$ periodic CB in a normal metal\cite{Kouwenhoven1997Electron}. The enhanced CB peak in Fig. \ref{fig:refCase}(a) is a result of resonant transport through MBS, similar to the case in quantum dots\cite{Kouwenhoven1997Electron} and has been discussed in more detail in Sec.~\ref{sec:level1_3_2}. 

Panel (d) is obtained by tracing the maximum conductance value $n_g$ for each $V_z$ value from Panel (a) for both even and odd $N$, and then calculating the absolute value of difference of these two tracking $n_g$ from both even and odd $N$. Comparing the positions of the dashed lines in Fig. \ref{fig:refCase}(d) with the spectrum shown in Fig. \ref{fig:refCase}(c), we see that the OCPS arises from the ABS states before the TQPT. The oscillatory splitting of ABSs at each end appears in the local tunneling spectrum at the respective end as seen in Fig. \ref{fig:refCase}(b), though the oscillation in this case does not show significant suppression with Zeeman potential. On the other hand, the decreasing OCPS as in Fig. \ref{fig:refCase}(d) demonstrates the lobes coming from the two ABSs at both ends exhibited in Fig. \ref{fig:refCase}(c). Therefore, OCPSs, which we obtain from the CB conductance in Panel (a), give information about the non-local states, rather than just local states. The non-local transport through a pair of localized levels at each end, which is what is responsible for the shift of the energy of the resonance between the two ABSs, is described in more detail by Eq. \eqref{G_twoLevel} in Sec.~\ref{sec:level1_3_3}. The main conclusion in our model where occupation of one ABS is allowed to relax to the other ABS is that the intensity of the transport peak is suppressed by the energy difference between the two ABSs relative to temperature. This is in contrast to the case where such relaxation is forbidden\cite{Albrecht2016Exponentiala} where non-local transport requires both ABSs to be near zero energy relative to temperature. Both these models would lead to suppression of conductance from ABS states for Zeeman field in the beginning of the range in Fig. \ref{fig:refCase}(a) [and Fig. \ref{fig:idealCase}(a)] where the ABS energy difference is much larger than temperature.  

Aside from demonstrating the ideal case (the most similar to the experimental data) as in Fig. \ref{fig:idealCase}, we will also discuss different effects by changing various parameters relative to the reference case shown in Fig. \ref{fig:refCase}, such as wire length, temperature, chemical potential, superconductor collapsing field, with and without self-energy, and with and without quantum dots in the following subsections of Sec.~\ref{sec:level1_5}.

\subsection{Soft-gap dependence}\label{sec:level1_5_1}
Most of the CB experiments\cite{Albrecht2016Exponentiala,Shen2018Parity,Vaitiekenas2020Fluxinduced} in nanowires do not show the abrupt drop in the $1e$ periodic conductance seen in the ideal case plotted in Fig. \ref{fig:refCase}(a). Additionally, as discussed in Sec.~\ref{sec:level1_3_4} as well as in previous work\cite{vanHeck2016Conductancea}, the conductance into the ideal MBS seen in Fig. \ref{fig:refCase}(a) is expected to have a significantly higher intensity than the $2e$ periodic conductance from Cooper-pair transport. Such an enhanced intensity is not seen in experiments\cite{Albrecht2016Exponentiala,Shen2018Parity,Vaitiekenas2020Fluxinduced}, leading to a contradiction between theory and experiment. These issues are resolved in the calculated conductance in Fig. \ref{fig:idealCase}, in which the soft SC gap is considered in the topological regime completely. The CB conductance difference of $1e$ transition between the topological regime and the normal-metal regime (i.e., regime above $V_c$) do not show any visible difference. Clearly, soft gap plays a key role in the experimental CB phenomenology, and brings agreement between theory and experiment.

Physically, the soft SC gap arises from impurity-induced bound states in the superconductor that is subject to a strong Zeeman field. Experimentally, Majorana nanowires always manifest soft gaps at finite magnetic fields even if the gap is hard at zero field. We model the soft gap phenomenologically by splitting the superconductivity of the proximity-inducing superconductivity into two parts with two different gaps $\Delta_1(V_Z)$ and $\Delta_2(V_Z)$ with different critical Zeeman fields such that 
\begin{subequations}\label{softgap}
	\begin{align}
	\Delta_1(V_z)=\Delta_0\sqrt{1-(V_z/V_{\text{TQPT}})^2}\label{softgap_a}\\
	\Delta_2(V_z)=\Delta_0\sqrt{1-(V_z/V_c)^2}\label{softgap_b}
	\end{align}
\end{subequations}
, where $V_{\text{TQPT}}$ is assumed to be the topological quantum critical point and $V_c$ in $\Delta_2(V_z)$ is the SC collapsing field. The SC gap collapse is thought to arise from the magnetic field entering the parent SC (i.e., Al in these CB experiments) destroying the parent superconductivity, and hence all Majorana physics. The soft gap regime ($V_{\text{TQPT}}<V_z<V_c$) is characterized by weakened SC, which we model using a self-energy that is as an average of the self-energy from a clean SC and that from a normal metal. The soft gap is the generic experimental situation in Majorana nanowires at finite magnetic field values even if the zero field system has a hard superconducting gap. With this choice, the non-vanishing sub-gap density of states is generated in the bulk superconductor above $V_z>V_{\text{TQPT}}$ from the closing of the first SC gap $\Delta_1$ which is given by 
\begin{equation}\label{rho_soft}
    \begin{aligned}
    \rho_{SC}(\epsilon)=&\frac{1}{2}\{\frac{2\rho_F}{\sqrt{\epsilon^2-\Delta_1(V_z)^2}}\theta(\epsilon-\Delta_1(V_z))\\
    &+\frac{2\rho_F}{\sqrt{\epsilon^2-\Delta_2(V_z)^2}}\theta(\epsilon-\Delta_2(V_z))\}
    \end{aligned}
\end{equation}
At the same time, the system remains superconducting below $V_z<V_c$ from the second superconducting part Eq.\eqref{softgap_b}. We will then follow the same procedure as Eqs. \eqref{StatesDensity_1}-\eqref{SC_En} to obtain the energy levels within the soft-gap regime, i.e., $\Delta_1(V_z)<\epsilon<\Delta_2(V_z)$, and the hard gap regime i.e., $\epsilon>\Delta_2(V_z)$, as before. The two SC gaps $\Delta_j(V_z)$ also correspondingly modify the SC self-energy in Eq. \eqref{selfE} to a form which averages between the two SC gaps as in Eq. \eqref{rho_soft}.

The effect of the soft SC gap on the CB conductance is a reduction of the peak height associated with the MBS in the topological regime of the Zeeman field ($V_{\text{TQPT}}<V_z<V_c$) from Fig. \ref{fig:refCase}(a) to Fig. \ref{fig:idealCase}(a). For the parameters chosen in Fig. \ref{fig:idealCase}(a), the MBS conductance is found to be almost identical to the normal state conductance above $V_z>V_c$, which appears to be the case in the experimental data\cite{Albrecht2016Exponentiala,Shen2018Parity}. Our results suggest that an MBS peak height comparable to the normal CB peak at higher Zeeman field would be indicative of a soft gap. While the precise matching of the MBS and normal state CB peaks might be a result of our parameter choices, the smooth intensity variations of the CB peak in experiments\cite{Albrecht2016Exponentiala,Shen2018Parity} suggest the emergence (possibly gradual) of a soft SC gap at some Zeeman field above the TQPT. This is of course the experimental phenomenology observed in the usual tunneling spectroscopy of all Majorana nanowires studied so far where the SC gap always becomes monotonically softer with increasing magnetic field.

\subsection{Temperature dependence}\label{sec:level1_5_2}
\begin{figure*}[t]
	\includegraphics[scale=0.62]{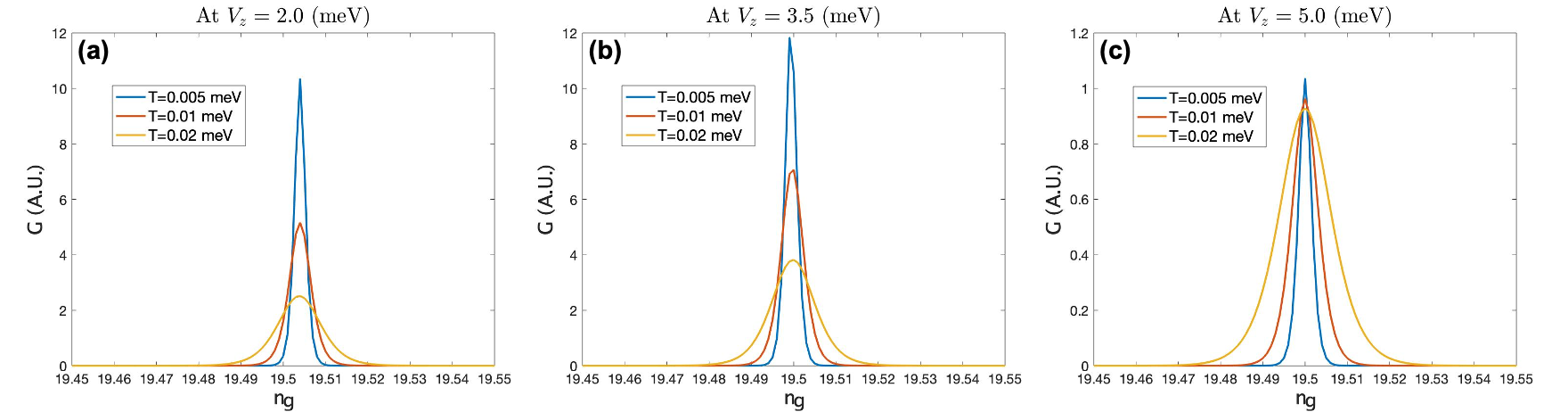}
	\caption{Line cuts of the conductance as a function of the gate-induced charge number $n_g$ (i.e., Fig. \ref{fig:refCase}), for fixed Zeeman fields $V_z$, with different temperature values $T=0.005$ meV, $T=0.01$ meV, and $T=0.02$ meV in a single panel. (a) At $V_z=2.0$ meV, the nanowire is in the Andreev-bound-state/trivial regime. (b) At $V_z=3.5$ meV, the nanowire is in the Majorana/topological regime. (c) At $V_z=5.0$ meV, the nanowire is in the normal-metal regime.}
	\label{fig:TempCase}
\end{figure*}
As discucssed in Sec.~\ref{sec:level1_3_2}, the $1e$ periodic CB peaks arising from MBS and normal metal behavior show, respectively, an inverse and vanishing temperature dependence of the CB peak height. This is a characteristic difference of a CB peak arising from a resonant bound state or a continuum of states\cite{Aleiner2002Quantum,Kouwenhoven1997Electron}. In Fig. \ref{fig:TempCase}(a), $V_z=2.0$ meV, which is in the topologically-trivial regime below TQPT, the conductance peak heights for $T=0.005$, $0.01$, and $0.02$ meV are roughly 0.05, 0.025, 0.012 in arbitrary units. As explained in Sec.~\ref{sec:level1_3_1}, the overall scale of the conductance in our work is determined by the tunneling parameter $\tau$. Our results can be compared to experiment by setting the normal state CB conductance for $V_z>V_c$ to the normal state tunnel conductance. Since the CB conductance peaks arise from ABSs, which are isolated bound states, the peak heights vary inversely with temperature as discussed in Sec.~\ref{sec:level1_3_2}\cite{Aleiner2002Quantum,Kouwenhoven1997Electron}. A similar temperature dependence is seen for the resonance in the topological regime ($V_z=3.5$ meV) in Fig. \ref{fig:TempCase}(b) where the peak heights are approximately 0.06, 0.035, and 0.018 at $T=0.005$, $0.01$, and $0.02$ meV, respectively. This is in contrast to the CB conductance peak in the normal-metal regime at $V_z=5.0$ meV which is shown in Fig. \ref{fig:TempCase}(c), where we find the peak heights to be almost temperature independent as with normal metal CB (see Sec.~\ref{sec:level1_3_2})\cite{Aleiner2002Quantum,Kouwenhoven1997Electron}. The results in Fig. \ref{fig:TempCase} show that the temperature dependence of the CB peaks can be used to distinguish between $1e$ periodic CB peaks arising from MBSs or trivial non-superconducting CB effect in regular metallic grains.

\subsection{Length dependence}\label{sec:level1_5_3}
\begin{figure}[htbp]
	\includegraphics[scale=0.35]{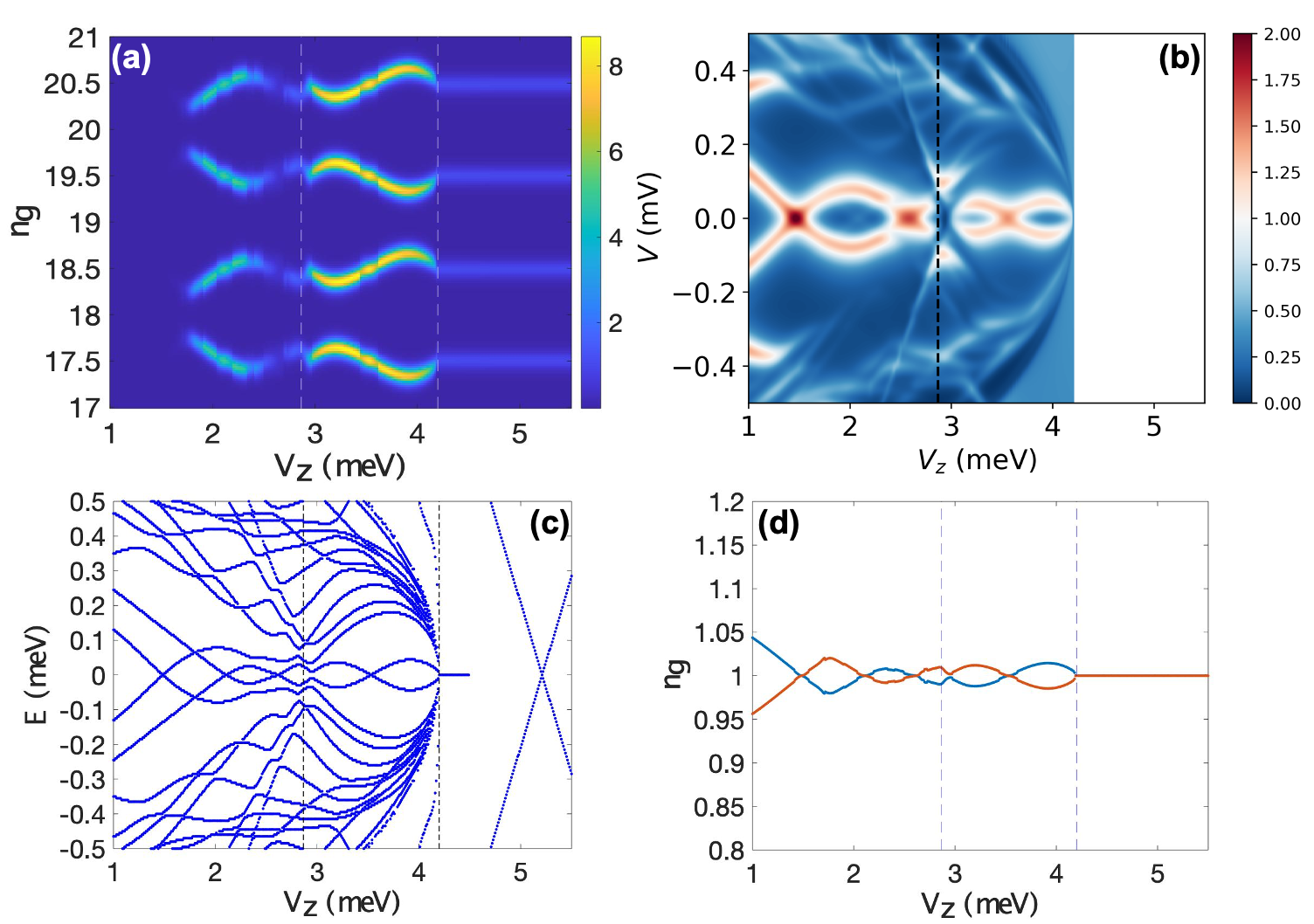}
	\caption{This set of results has the shorter wire length $L=1.0$ $\micro$m, while keeping all the other parameters the same as Fig. \ref{fig:refCase}. (a) Coulomb blockaded conductance $G$ as a function of the gate-induced charge number $n_g$ and Zeeman field $V_z$ at $E_c=0.13$ meV. (b) Non-Coulomb blockaded conductance $G$ as a function of bias voltage $V$ and Zeeman field $V_z$. (c) Energy spectrum as a function of Zeeman field $V_z$. (d) OCPS as a function of Zeeman field $V_z$.}
	\label{fig:LengthCase_10}
\end{figure}

\begin{figure}[htbp]
	\includegraphics[scale=0.35]{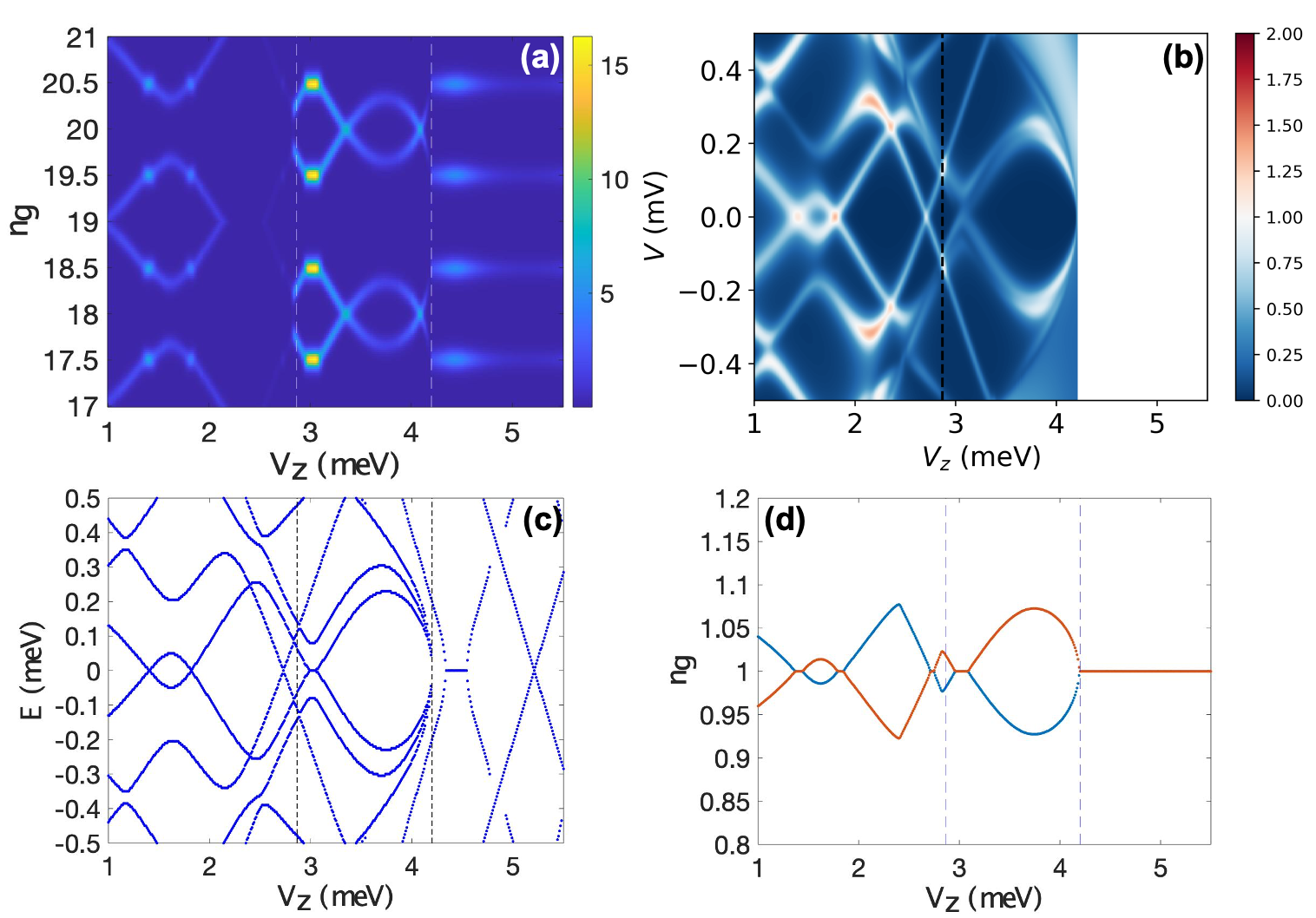}
	\caption{This set of results has the shorter wire length $L=0.6$ $\micro$m, while keeping all the other parameters the same as Fig. \ref{fig:refCase}. (a) Coulomb blockaded conductance $G$ as a function of the gate-induced charge number $n_g$ and Zeeman field $V_z$ at $E_c=0.13$ meV. (b) Non-Coulomb blockaded conductance $G$ as a function of bias voltage $V$ and Zeeman field $V_z$. (c) Energy spectrum as a function of Zeeman field $V_z$. (d) OCPS as a function of Zeeman field $V_z$.}
	\label{fig:LengthCase_06}
\end{figure}
In Fig. \ref{fig:LengthCase_10}, we decrease the nanowire length to $L=1.0$ $\micro$m, compared to Fig. \ref{fig:refCase} ($L=1.5$ $\micro$m). The Majorana oscillations between the two dashed lines in Fig. \ref{fig:LengthCase_10} become more obvious in the shorter wire, which satisfies the trend of the Majorana splitting $e^{-2L/\xi}$\cite{DasSarma2012Splittinga} which is necessarily enhanced in shorter wires indicating an exponential weakening of the topological protection. Due to the Majorana splitting, the lobes of the OCPS, which project the combination of the lowest-energy states on both ends, start to increase as soon as we enter the topological regime for the shorter wire. On the other hand, Majorana oscillation is suppressed for the longer wire, as in Fig. \ref{fig:refCase} for the case of $L=1.5$ $\micro$m. Therefore, we can see that lobes of the OCPS for $L=1.5$ $\micro$m decrease monotonously as the Zeeman field increases, while the counterparts of $L=1.0$ $\micro$m decrease only before the TQPT field. We also expect that the OCPS for the wire length longer than $L=1.5$ $\micro$m will look no different from the one of $L=1.5$ $\micro$m, considering that both ABS and MBS will be even more localized and stable, and thus the oscillations in the topological regime will be suppressed and negligible as in Fig. \ref{fig:refCase}. Based on Figs. \ref{fig:refCase}(c) and \ref{fig:LengthCase_10}(c), both of which show the decreasing lowest-lying energies as a function of Zeeman field below the TQPT field (first dashed line), we can also observe that the decreasing lobes of OCPS in Fig. \ref{fig:LengthCase_10}(d) mainly come from the ABSs induced by the two quantum dots at both ends. The decreasing OCPS coming from the lowest-lying ABSs on both ends will be destroyed when the nanowire is too short so that the ABSs on both ends interfere with each other, such as the $L=0.6$ $\micro$m case in Fig. \ref{fig:LengthCase_06}.

In Fig. \ref{fig:LengthCase_06}(a), with shorter nanowire length $L=0.6$ $\micro$m, the conductance peak is visible in the range of Zeeman potential at the lowest end in Fig. \ref{fig:LengthCase_06}(a), which is in contrast to Figs.\ref{fig:idealCase}(a), \ref{fig:refCase}(a), and \ref{fig:LengthCase_10}(a) where the conductance is suppressed based on Eq.\eqref{G_twoLevel} in Sec.~\ref{sec:level1_3_3} for the longer wire case. 

The suppression of the conductance based on Eq. \eqref{G_twoLevel} in the long wire case is eliminated for shorter wires with length comparable to the coherence (or the localization) length of the bound states. In this case, electrons from either end can tunnel into each of the ABSs so that electron tunneling between the ends of the wire through either of the ABSs described by Eq. \eqref{G_one-level} provides a measurable contribution to the CB conductance in shorter wires such as in Fig. \ref{fig:LengthCase_06}(a). These results lead to the conclusion that the observation of the dark region at the transition from $2e$ periodic conductance near the lowest part of the Zeeman range shown in the CB conductance plots [e.g. Figs. \ref{fig:idealCase}(a), \ref{fig:refCase}(a), \ref{fig:LengthCase_10}(a). etc.] can be understood to be a consequence of ABSs at the ends of the wire.  

\begin{figure}[htbp]
	\includegraphics[scale=0.47]{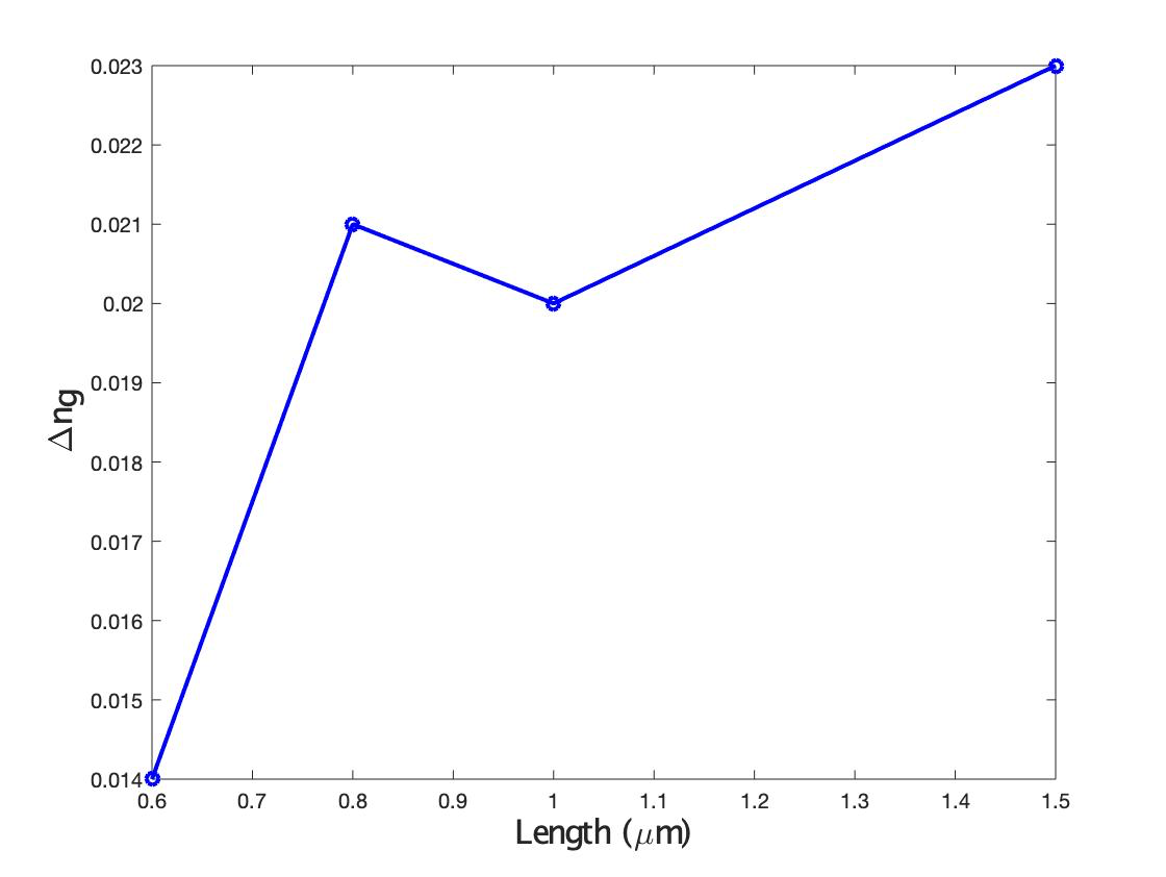}
	\caption{The oscillatory amplitude of the first lobe in the OCPS $(|S_o-S_e|/2)$ for different wire lengths. All the other parameters are kept the same as Fig. \ref{fig:refCase}, except for the nanowire length. The maximum oscillatory amplitude does not display strong length dependence as experimental data.}
	\label{fig:LengthDependence}
\end{figure}
The length dependence of the first lobe of the OCPS, which is plotted in Fig. \ref{fig:LengthDependence} shows a $40\%$ increase in magnitude of the OCPS with increasing length. This is in contrast to the exponential decrease in the magnitude of OCPS expected from MBS splitting oscillations, which has been claimed to be observed experimentally\cite{Albrecht2016Exponentiala,Vaitiekenas2020Fluxinduced}. Additionally, the change in the OCPS observed by Albrecht \textit{et al.} \cite{Albrecht2016Exponentiala,Vaitiekenas2020Fluxinduced} is two orders of magnitude in contrast to the $40\%$ we see in Fig. \ref{fig:LengthDependence} over the same range of lengths. A similar ($\sim 40\%$) length dependence of OCPS has been obtained in earlier theoretical work\cite{Chiu2017Conductance} using a master-equation approach. The origin of the large discrepancy of the length dependence of our OCPS with experiment is the fact that the OCPS in our models arise from ABSs rather than MBSs as expected in the experiments. Since the ABS energy is dominated by the profile of the confining potential at the end, we do not expect them to have the exponential length dependence induced by MBS. While this might be a motivation to restrict to models where OCPS from ABSs are absent, one should note that OCPS from MBSs are found numerically to increase or show no significant decrease with increasing applied Zeeman field\cite{Chiu2017Conductance} as seen in Fig. \ref{fig:LengthCase_10}(d). Thus the OCPS pattern from a single device in the experiment\cite{Albrecht2016Exponentiala,Shen2018Parity,Vaitiekenas2020Fluxinduced} is significantly more consistent with those arising from ABSs [such as Fig. \ref{fig:idealCase}(d)] than from MBSs. Therefore, while our results show qualitative consistency of the OCPS with single device, the range of models we study cannot reproduce both the decreasing OCPS with field as well as with length seen in the experiments\cite{Albrecht2016Exponentiala,Vaitiekenas2020Fluxinduced}. Again, this is consistent with the earlier theoretical conclusion based on the master equation approach\cite{Chiu2017Conductance}, and we believe that the conclusion of an ``exponential protection" made in Ref.\cite{Albrecht2016Exponentiala} is incorrect and is an artifact of using very few samples with each sample having its own different sets of ABS, etc. (i.e., the sample length was not varied \textit{in situ}, but only by going from sample to sample where obviously many things, not just the sample length, are changing in an uncontrolled manner).

\subsection{SC collapsing field dependence}\label{sec:level1_5_4}
\begin{figure}[htbp]
	\includegraphics[scale=0.35]{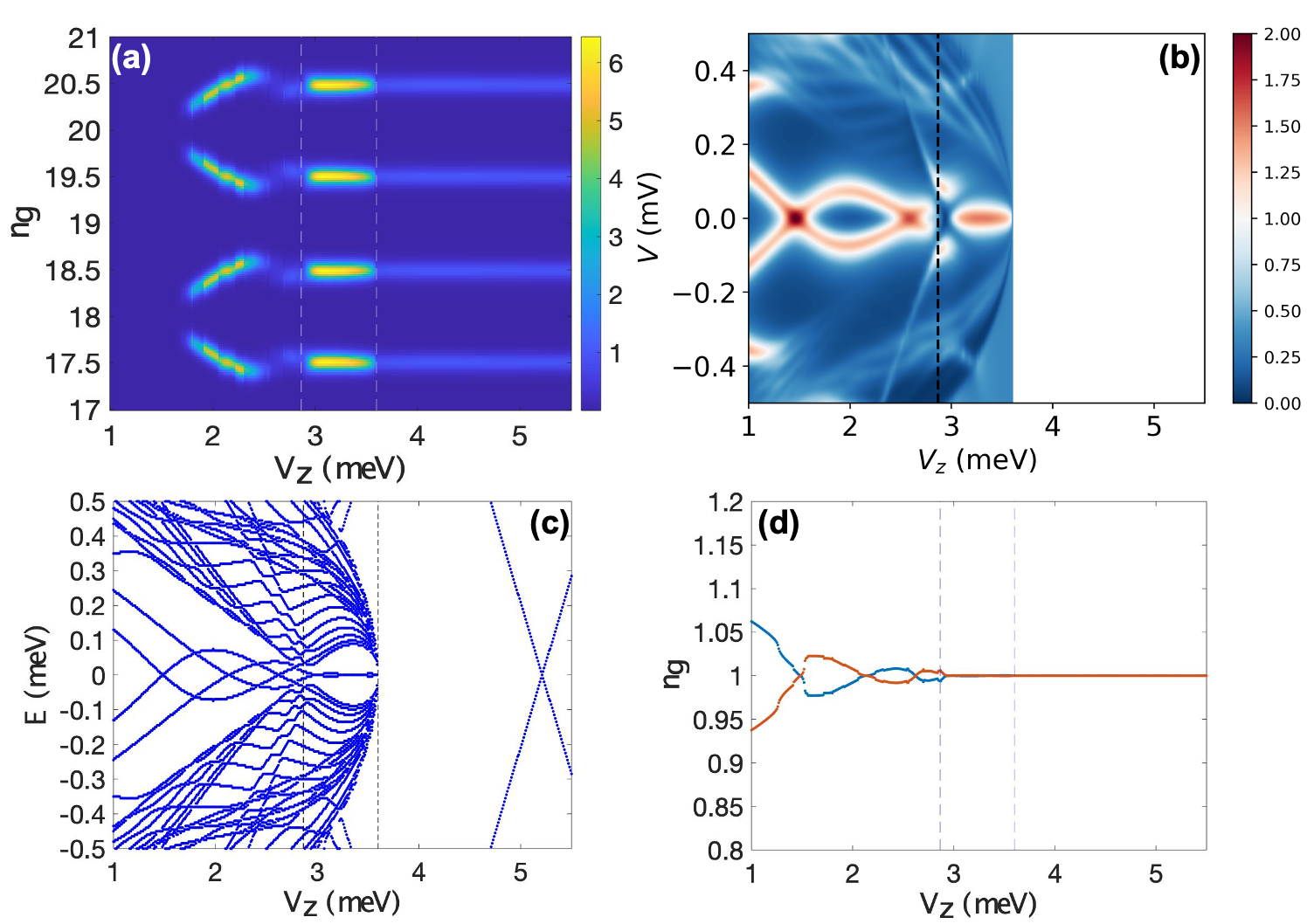}
	\caption{This set of results has the lower SC collapsing field $V_c=3.6$ meV, while keeping all the other parameters the same as Fig. \ref{fig:refCase}. (a) Coulomb blockaded conductance $G$ as a function of the gate-induced charge number $n_g$ and Zeeman field $V_z$ at $E_c=0.13$ meV. (b) Non-Coulomb blockaded conductance $G$ as a function of bias voltage $V$ and Zeeman field $V_z$. (c) Energy spectrum as a function of Zeeman field $V_z$. (d) OCPS as a function of Zeeman field $V_z$.}
	\label{fig:VcCase_3.6}
\end{figure}
\begin{figure}[htbp]
	\includegraphics[scale=0.35]{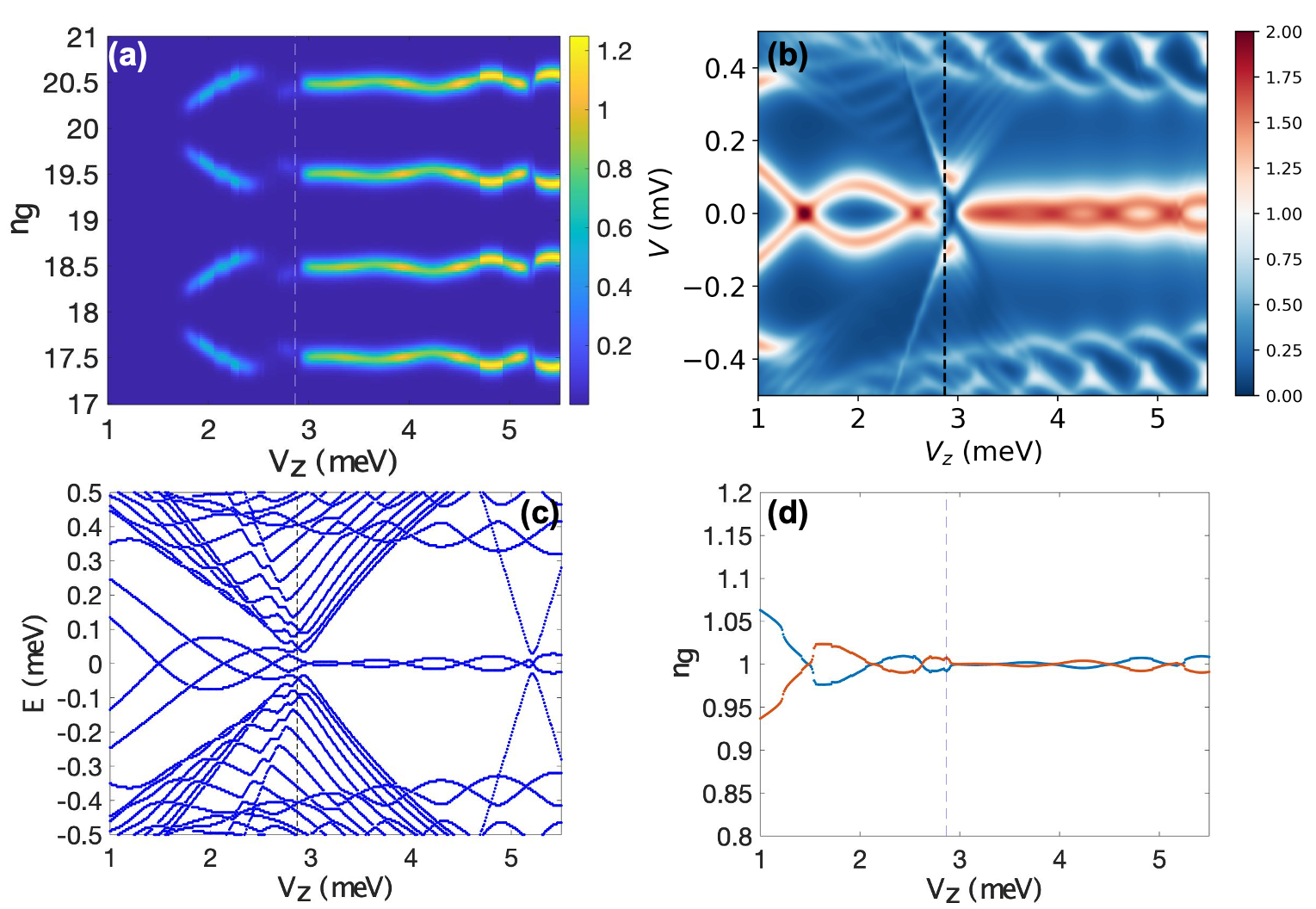}
	\caption{This set of results has the infinite SC collapsing field, i.e., $\Delta(V_z)=\Delta_0$, while keeping all the other parameters the same as Fig. \ref{fig:refCase}. (a) Coulomb blockaded conductance $G$ as a function of the gate-induced charge number $n_g$ and Zeeman field $V_z$ at $E_c=0.13$ meV. (b) Non-Coulomb blockaded conductance $G$ as a function of bias voltage $V$ and Zeeman field $V_z$. (c) Energy spectrum as a function of Zeeman field $V_z$. (d) OCPS as a function of Zeeman field $V_z$.}
	\label{fig:VcCase_infinity}
\end{figure}

In this subsection, we change the SC collapsing field in Fig. \ref{fig:refCase} from $V_c=4.2$ meV to lower value $V_c=3.6$ meV, as in Fig. \ref{fig:VcCase_3.6}, and to higher value $V_c=\infty$, i.e., $\Delta(V_z)=\Delta_0$, as in Fig. \ref{fig:VcCase_infinity}. The value of $V_c$ determines the size of the topological regime relative to the normal metal regime where the SC gap is destroyed. Since the Coulomb blockade conductance for a normal metal is exactly $1e$ periodic\cite{Scott-Thomas1989Conductance,vanHouten1989Comment,Glazman1989Coulomb,Meirav1989Onedimensional,Meirav1990Singleelectron,Beenakker1991Theory}, a low value of $V_c$ can appear as a suppression of the Majorana splitting oscillations expected in the topological superconducting regime. The range of the topological regime in Figs. \ref{fig:refCase} and \ref{fig:VcCase_3.6}, which starts at the TQPT field and ends at the SC collapsing field $V_c$ (i.e., region between the two dashed lines), does not give enough parameter space for the Majorana bound states to be delocalized by the strong external magnetic field. On the contrary, if the parent SC gap is highly robust to the applied magnetic field (i.e., $V_c\gg V_{\text{TQPT}}$) as is the case in Fig. \ref{fig:VcCase_infinity}, the range of topological SC becomes large enough to accommodate an observable range of Majorana splitting oscillations with an amplitude that increases with increasing Zeeman field. Thus, the oscillations in the topological superconducting phase are found to have an amplitude increasing with field, which is in contradiction with the experiments. As discussed in the previous subsections, the experimental observations of OCPSs\cite{Albrecht2016Exponentiala,Shen2018Parity,Vaitiekenas2020Fluxinduced} are more consistent with ABSs which dominate when $V_c$ is not too large relative to $V_{\text{TQPT}}$ as in Fig. \ref{fig:VcCase_3.6} or Figs. \ref{fig:idealCase} or \ref{fig:refCase}.

\subsection{Chemical potential dependence}\label{sec:level1_5_5}
\begin{figure}[tbp]
	\includegraphics[scale=0.35]{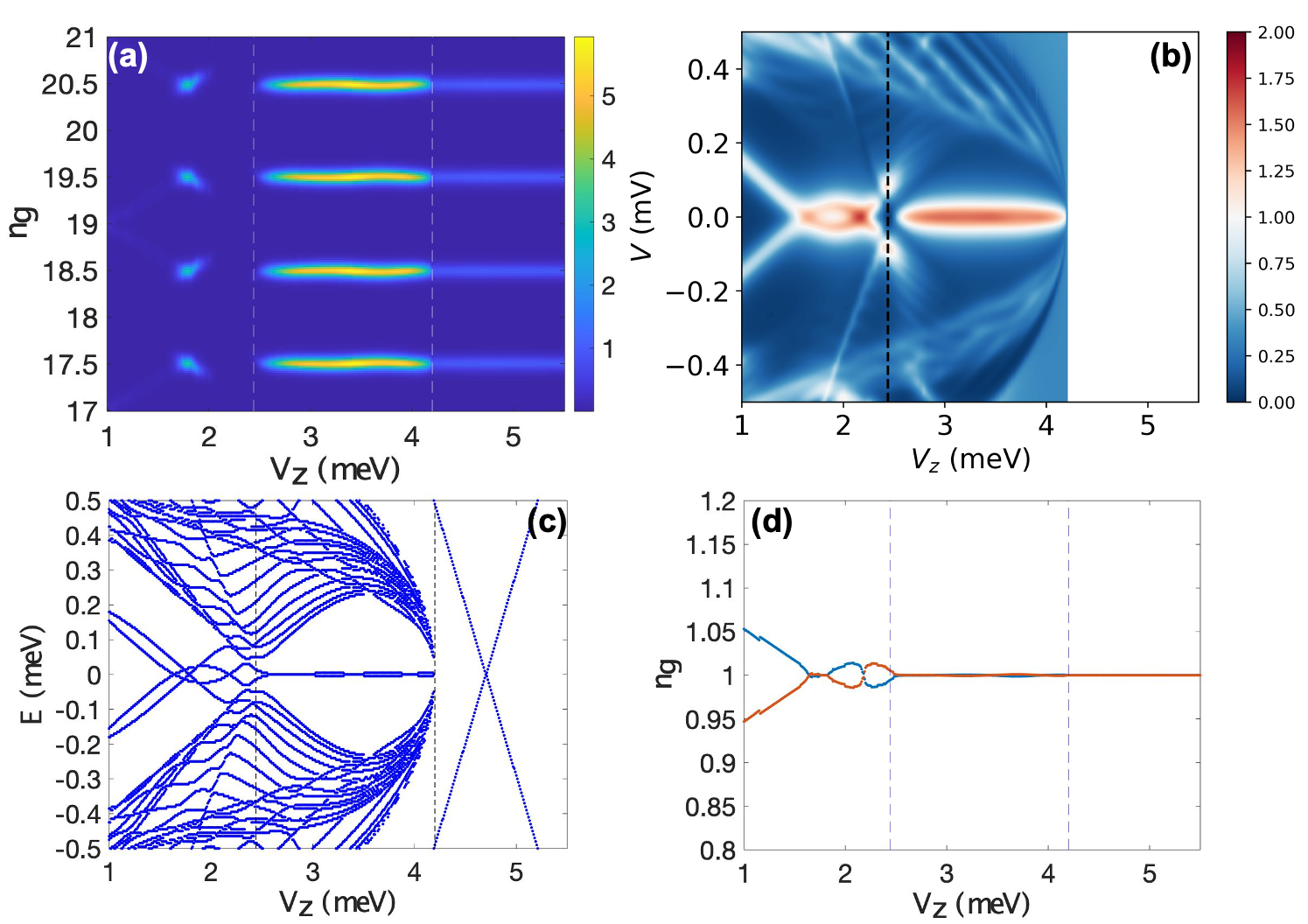}
	\caption{This set of results has the lower chemical potential $\mu=2.0$ meV, while keeping all the other parameters the same as Fig. \ref{fig:refCase}. (a) Coulomb blockaded conductance $G$ as a function of the gate-induced charge number $n_g$ and Zeeman field $V_z$ at $E_c=0.16$ meV. (b) Non-Coulomb blockaded conductance $G$ as a function of bias voltage $V$ and Zeeman field $V_z$. (c) Energy spectrum as a function of Zeeman field $V_z$. (d) OCPS as a function of Zeeman field $V_z$.}
	\label{fig:muCase_2.0}
\end{figure}

\begin{figure}[tbp]
	\includegraphics[scale=0.35]{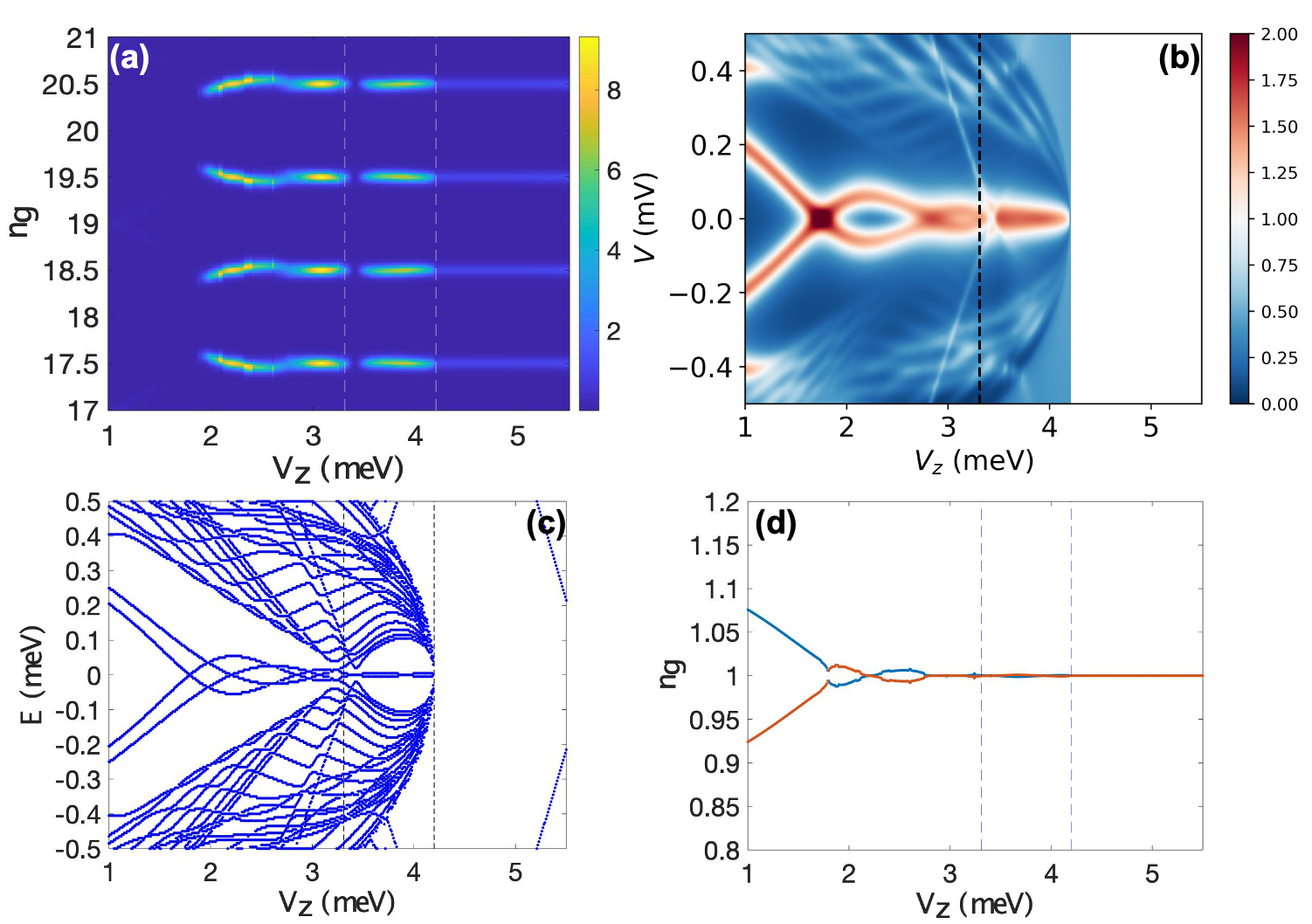}
	\caption{This set of results has the higher chemical potential $\mu=3.0$ meV, while keeping all the other parameters the same as Fig. \ref{fig:refCase}. (a) Coulomb blockaded conductance $G$ as a function of the gate-induced charge number $n_g$ and Zeeman field $V_z$ at $E_c=0.21$ meV. (b) Non-Coulomb blockaded conductance $G$ as a function of bias voltage $V$ and Zeeman field $V_z$. (c) Energy spectrum as a function of Zeeman field $V_z$. (d) OCPS as a function of Zeeman field $V_z$.}
	\label{fig:muCase_3.0}
\end{figure}

Unlike the dependencies discussed so far, changing the chemical potential can substantially change the spectrum of the nanowire even below the TQPT in a way that has been studied in the context of tunneling transport\cite{Kells2012Nearzeroenergya,Liu2017Andreeva,Liu2018Distinguishinga,Moore2018Twoterminala,Vuik2019Reproducing,Pan2020Physical}. Aside from changing the value of the TQPT field (the first dashed line), changing the chemical potential $\mu$ also modifies the spectrum of subgap ABS energies\cite{Kells2012Nearzeroenergya,Liu2017Andreeva,Liu2018Distinguishinga,Moore2018Twoterminala,Vuik2019Reproducing}. In addition, the ABSs induced from both ends of the nanowire do not follow a monotonic trend as the chemical potential is varied, due to the fact that the potential heights of QDs are not the same on both ends. Since the OCPS roughly projects the combination of the lowest energies on both ends, it is not guaranteed to generate the decreasing trend of OCPS by simply tuning the chemical potential. We can compare Fig. \ref{fig:refCase} with Figs. \ref{fig:muCase_2.0} and \ref{fig:muCase_3.0}, where the chemical potential changes from $\mu=2.5$ meV to $\mu=2.0$ meV and $\mu=3.0$ meV, respectively. The energy spectrum below TQPT field in Fig. \ref{fig:muCase_2.0}(c) has no similarity to the one in Fig. \ref{fig:refCase}(c), while the structure in Fig. \ref{fig:muCase_3.0}(c) is similar to Fig. \ref{fig:refCase}(c), even the chemical potentials in Figs. \ref{fig:muCase_2.0} and \ref{fig:muCase_3.0} are both just away from $\mu=2.5$ meV by $0.5$ meV. Nevertheless, the behaviors in the topological regime for these three plot sets do not seem to show a significant difference when the nanowire is long enough to suppress the Majorana splitting oscillation. Thus, as seen in Figs.\ref{fig:refCase}(d), \ref{fig:muCase_2.0}(d), and \ref{fig:muCase_3.0}(d), the first OCPS lobes arise from end ABS. Comparing the different chemical potential cases in Figs. \ref{fig:refCase}, \ref{fig:muCase_2.0}, and \ref{fig:muCase_3.0}, we see that while the OCPSs do not show any strong increase with Zeeman potential, observing a strong decrease with Zeeman field as in Fig. \ref{fig:refCase} and seen in experiments\cite{Albrecht2016Exponentiala,Shen2018Parity,Vaitiekenas2020Fluxinduced} depends on the choice of chemical potential.  

While the results in Figs. \ref{fig:idealCase}, \ref{fig:refCase}, \ref{fig:VcCase_3.6}, \ref{fig:VcCase_infinity}, and \ref{fig:oneQD_case} suggest the decreasing oscillations of CB conductance peaks are indicative of conductance in the ABS regime, the results in Figs. \ref{fig:muCase_2.0} and \ref{fig:muCase_3.0} suggest otherwise. These latter results show small decreasing oscillations across the $V_z=V_{\text{TQPT}}$ making it difficult to distinguish the ABS regime from the MBS regime. This is especially so in Fig. \ref{fig:muCase_3.0}(a), where the CB conductance patterns between the ABS regime and MBS regime look similar (i.e., both display almost $1e$ periodic CB transitions) because the ABSs stay close to zero energy, giving rise to the brightness pattern, which barely has oscillations. This result means that the CB conductance profiles by themselves cannot provide enough information about the difference between the ABS regime and the MBS regime. However, it should be noted that in Fig. \ref{fig:muCase_3.0}(a), a little patch of darkness near TQPT separates the ABS regime and MBS regime though it is not so visible in the general case. This will be discussed in more detail in Sec.~\ref{sec:level1_6_1}. In general, however, our detailed theoretical results indicate that particular caution is warranted in interpreting experimental CB conductance peaks as arising from Majorana zero modes since the distinction between the manifestations of ABS and MBS in CB conductance is rather subtle and small.

\subsection{Quantum dot dependence}\label{sec:level1_5_6}
\begin{figure}[tbp]
	\includegraphics[scale=0.35]{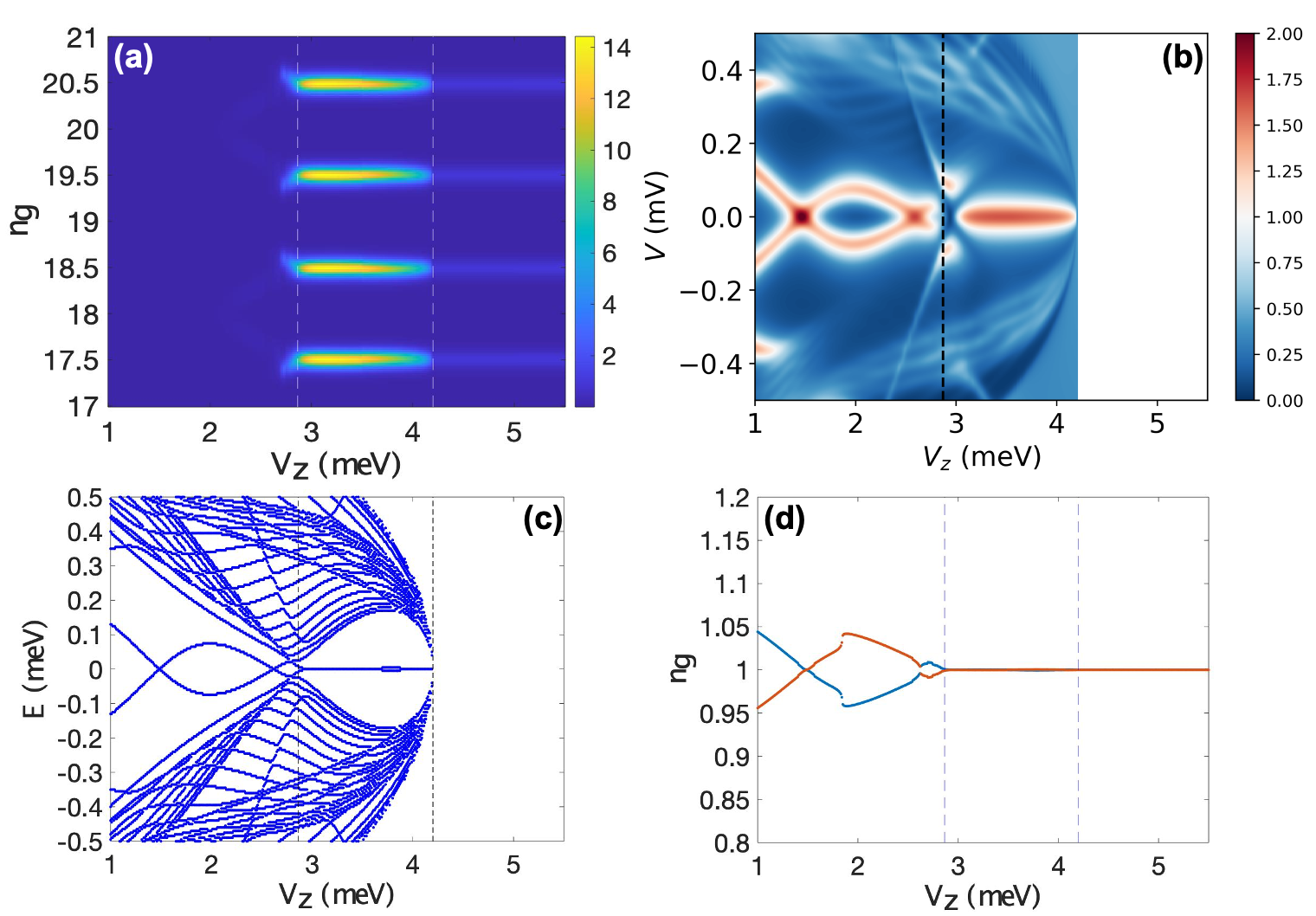}
	\caption{This set of results has only one QD on the left end of the wire, while keeping all the other parameters the same as Fig. \ref{fig:refCase}. (a) Coulomb blockaded conductance $G$ as a function of the gate-induced charge number $n_g$ and Zeeman field $V_z$ at $E_c=0.13$ meV. (b) Non-Coulomb blockaded conductance $G$ as a function of bias voltage $V$ and Zeeman field $V_z$. (c) Energy spectrum as a function of Zeeman field $V_z$. (d) OCPS as a function of Zeeman field $V_z$.}
	\label{fig:oneQD_case}
\end{figure}

\begin{figure}[tbp]
	\includegraphics[scale=0.35]{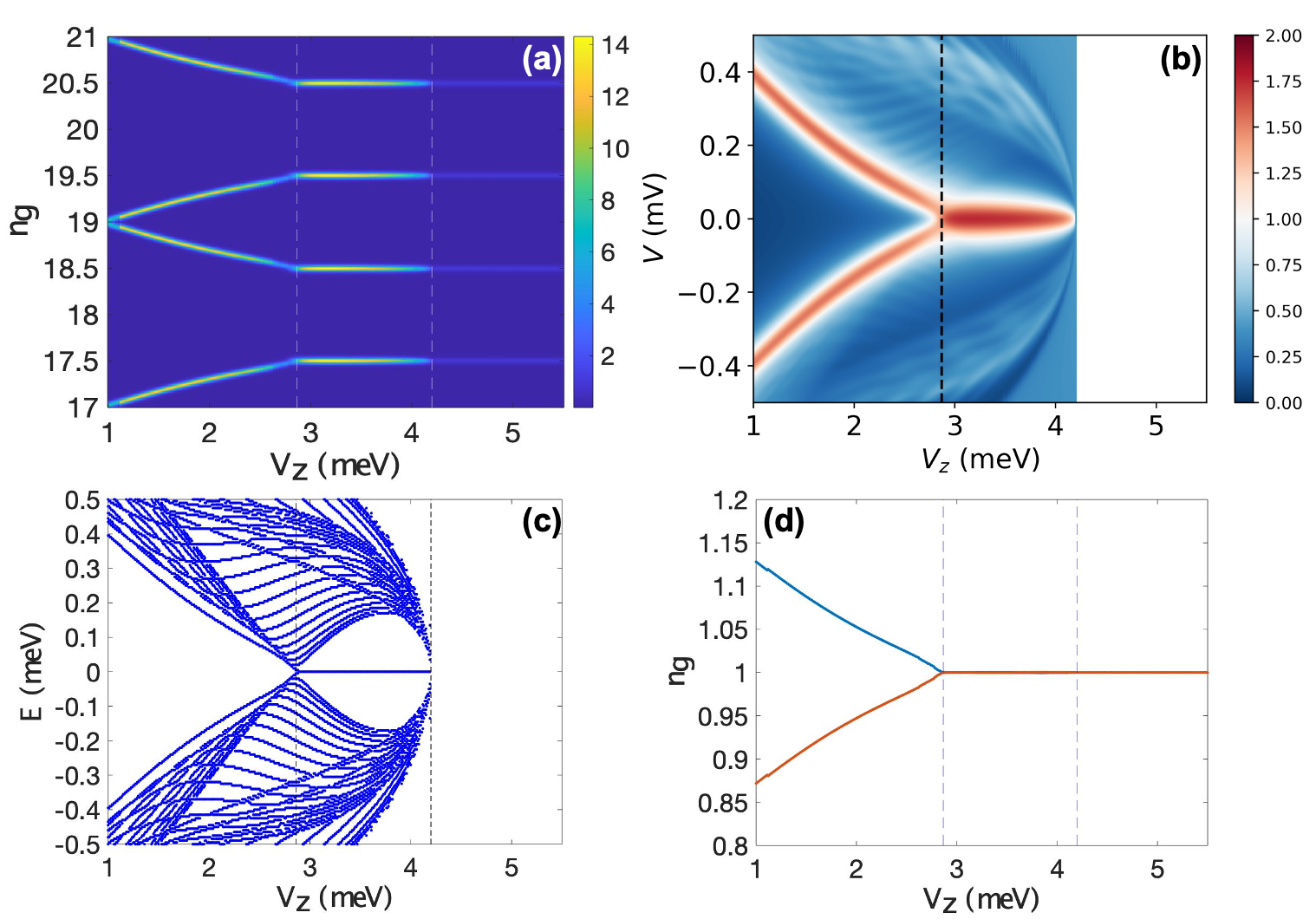}
	\caption{This set of results has none QD on both ends of wire, while keeping all the other parameters the same as Fig. \ref{fig:refCase}. (a) Coulomb blockaded conductance $G$ as a function of the gate-induced charge number $n_g$ and Zeeman field $V_z$ at $E_c=0.40$ meV. (b) Non-Coulomb blockaded conductance $G$ as a function of bias voltage $V$ and Zeeman field $V_z$. (c) Energy spectrum as a function of Zeeman field $V_z$. (d) OCPS as a function of Zeeman field $V_z$.}
	\label{fig:noQD_case}
\end{figure}

In Fig. \ref{fig:refCase}, we show the calculated results with two QDs on both ends of the nanowire. In this subsection, we show the results of one QD in Fig. \ref{fig:oneQD_case} by removing the QD on the right end but keeping the left one. We also show the results without any QD in Fig. \ref{fig:noQD_case}. Compared to Fig. \ref{fig:oneQD_case}(c), the non-Coulomb-blockade conductance in Fig. \ref{fig:oneQD_case}(b) reflects the ABS located at the left end as it is measured from the left lead. We also observe that there is no difference between Figs. \ref{fig:refCase}(b) and \ref{fig:oneQD_case}(b), even though there are two low-energy ABSs in Fig. \ref{fig:refCase}. Both Fig. \ref{fig:refCase}(b) and Fig. \ref{fig:oneQD_case}(b) only demonstrate the ABS induced from the QD at the left end. So the regular tunnel conductance (without Coulomb blockade) can only provide local information. On the other hand, Fig. \ref{fig:refCase}(d) projects the combination of the two ABSs located at both ends, by comparing it with Fig. \ref{fig:refCase}(c). Therefore, the OCPS obtained from the Coulomb blockade conductance gives us non-local information that the regular (non-Coulomb-blockade) conductance cannot provide.

In Fig. \ref{fig:oneQD_case}(a), the Coulomb blockade conductance peak is suppressed below the TQPT field (first dashed line), relatively lower than the Majorana peak, which lies between the two dashed lines. The diminished conductance associated with the ABSs in Fig. \ref{fig:oneQD_case} arises from the difference in the ABS energies (seen in Fig. \ref{fig:oneQD_case}(c)) at the two ends. Such a difference in energies leads to suppression in conductance as seen in Eq.\eqref{G_twoLevel}. The CB conductance intensity increases once the ABS energies approach zero energy and continues to remain high past the TQPT until the superconducting gap closes.

Comparing the OCPS with QDs [Fig. \ref{fig:refCase}(d)] with the case without QDs [Fig. \ref{fig:noQD_case}(d)], we conclude that the Zeeman field oscillations of OCPS depend on the presence of QDs. This is consistent with the absence of oscillations in the ABS spectrum in Fig. \ref{fig:noQD_case}. While the MBS spectrum typically shows oscillations with Zeeman energy\cite{Cheng2009Splittinga}, the splitting amplitude is significantly below the thermal resolution\cite{Liu2017Andreeva} at the temperatures we consider. Indeed, neither the non-Coulomb blockade conductance [Fig. \ref{fig:noQD_case}(b)] nor the OCPS from CB conductance [Fig. \ref{fig:noQD_case}(d)] show any oscillations, which is different from the results in Figs.\ref{fig:refCase} or \ref{fig:oneQD_case}, both of which have at least one QD. Therefore, the oscillatory lobes of OCPSs in the experimental data\cite{Albrecht2016Exponentiala,Shen2018Parity,Vaitiekenas2020Fluxinduced} most likely come from the ABSs induced by the QDs.

\subsection{Self-energy dependence}\label{sec:level1_5_7}
\begin{figure}[htbp]
	\includegraphics[scale=0.35]{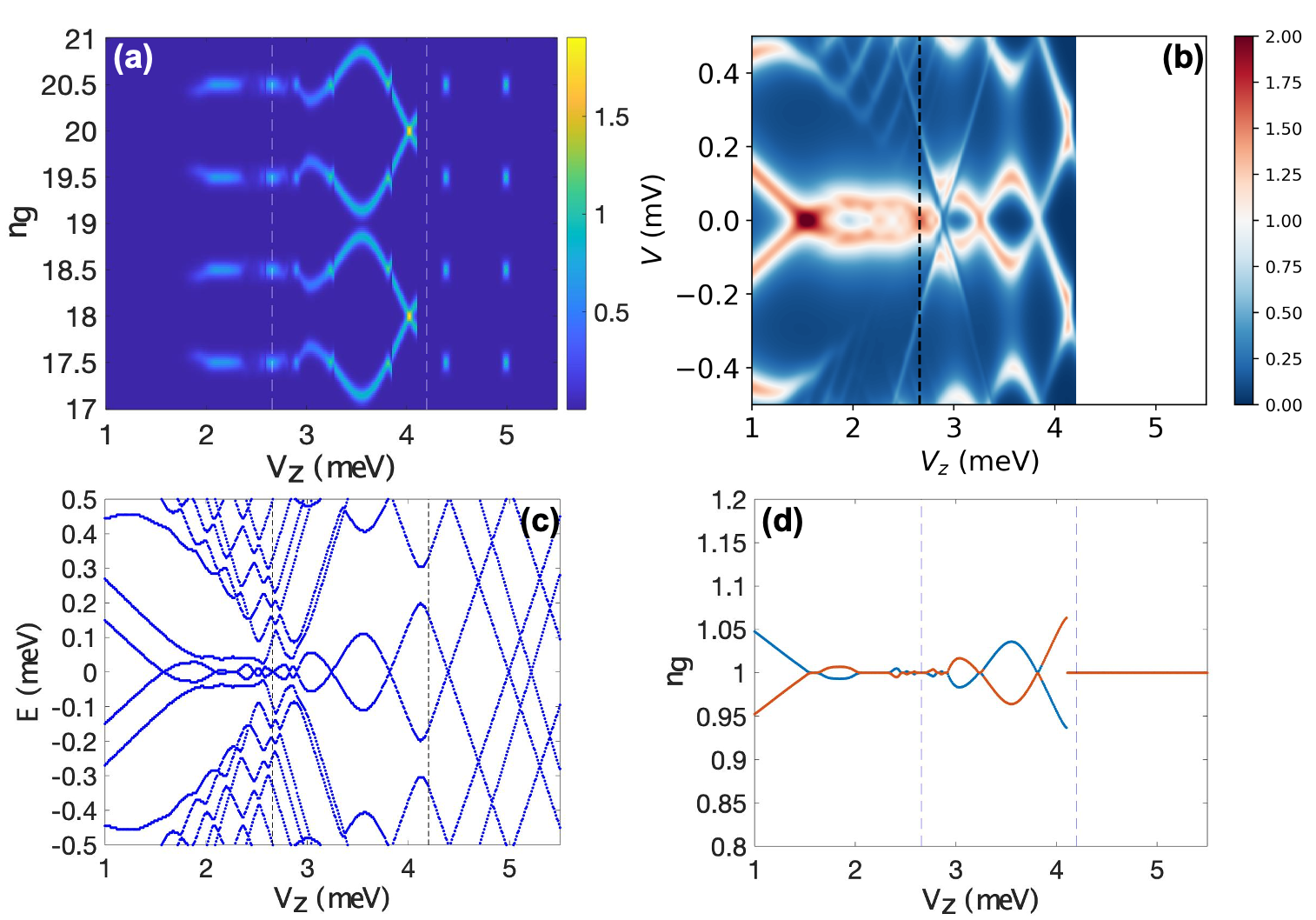}
	\caption{This set of results does not include the self-energy, while keeping all the other parameters the same as Fig. \ref{fig:refCase}. (a) Coulomb blockaded conductance $G$ as a function of the gate-induced charge number $n_g$ and Zeeman field $V_z$ at $E_c=0.13$ meV. (b) Non-Coulomb blockaded conductance $G$ as a function of bias voltage $V$ and Zeeman field $V_z$. (c) Energy spectrum as a function of Zeeman field $V_z$. (d) OCPS as a function of Zeeman field $V_z$.}
	\label{fig:NonSelfCase}
\end{figure}
In Fig. \ref{fig:NonSelfCase}, we remove the self-energy term induced by the parent superconductor. The energy splittings from both ABSs and MBSs are enlarged without self-energy, compared to Fig. \ref{fig:refCase}, which incorporates self-energy. The self-energy suppresses the energy splitting by a factor of $1/(1+\lambda/\Delta)$\cite{Stanescu2019Robust}. Aside from the effective SC gap $\Delta$ being replaced by the self-energy coupling $\lambda$, the TQPT field is shifted to $V_{\text{TQPT}}=\sqrt{\lambda^2+\mu^2}$ for the self-energy case, rather than $V_{\text{TQPT}}=\sqrt{\Delta_0^2+\mu^2}$ for the non-self-energy case. According to weak-coupling BCS theory, coherence length is $\xi=\hbar v_F/(\pi\Delta)$, so the coherence length becomes shorter when $\lambda>\Delta$. Therefore, the Majorana splitting, which follows $e^{-L/\xi}$, becomes smaller with parent SC coupling\cite{Peng2015Strong,Hui2015Majorana}. The subgap states become more localized as we increase the coupling strength $\lambda$ too. The self-energy generically reduces the Majorana oscillation amplitude above TQPT, but the quantitative suppression depends on many details of the parameters.

The inclusion of self-ernergy may be a key to understanding why MBS oscillations increasing in amplitude with Zeeman field are not observed in OCPS experiments or conventional conductance experiments without CB. For the non-self-energy results in Fig. \ref{fig:NonSelfCase}, we do not observe the OCPS decreasing with the Zeeman field monotonously. In fact, the splitting in the absence of self-energy seen in Fig. \ref{fig:NonSelfCase} is large enough to eliminate any $1e$ periodic features that are seen in the experiments and in Fig. \ref{fig:refCase} beyond a magnetic field which is not large enough to kill SC completely. Whether the almost universal experimental absence of Majorana oscillations with increasing Zeeman field is a consequence of the self-energy effect or simply a manifestation of the dominance of ABS in the experiments is an important question. Obviously, if the experiments are manifesting only ABS and no MBS, there will not be any Majorana oscillations whether self-energy effects are included or not.  The current nanowire experiments, including the CB experiments, are certainly more consistent with the physics of ABS being dominant, which provides a natural explanation for why the MBS oscillations are never seen.

\section{Discussion of key features}\label{sec:level1_6}
In this section, we summarize how the generic features of the experiment can be understood from a compilation of our results in Sec.~\ref{sec:level1_5} together with the analytical arguments in Sec.~\ref{sec:level1_3}.

\subsection{Bright-dark-bright pattern}\label{sec:level1_6_1}
One of the commonly seen features in the CB experiments\cite{Albrecht2016Exponentiala,Shen2018Parity,Vaitiekenas2020Fluxinduced} is the bright-dark-bright intensity pattern that is seen as the Zeeman field increases. Specifically, the $2e$ periodic CB peaks at small Zeeman field are seen to be bright followed by a dark region, which then becomes bright as the CB resonances approach the purely $1e$ periodic regime.  While the numerical results shown in Fig. \ref{fig:refCase}-\ref{fig:NonSelfCase} do not show the $2e$ periodic region, they do show a dark region of suppressed conductance at the lower end of the Zeeman scale corresponding to immediately after the $2e$ periodic region, consistent with experiments. The suppressed conductance seen in our numerical plots in this so-called dark region can be understood from Eq.\eqref{G_twoLevel}, which shows that the conductance intensity will become suppressed when the eigen-energy difference from both ends is larger than the temperature. From the spectrum in Figs. \ref{fig:idealCase}(c), \ref{fig:refCase}(c) and \ref{fig:LengthCase_10}(c), there are two sub-gap ABSs which are located on both wire ends respectively. The higher-energy subgap ABS suppresses the resonant conductance as the arguments below Eq.\eqref{G_twoLevel} suggest. Therefore, we are unable to observe the conductance below $V_z=1.7$ meV. The brightness appears above $V_z=1.7$ meV, when the energy difference from $L$ and $R$ ends is within the temperature range, i.e., $|\epsilon_L-\epsilon_R|<T$. These numerical results also prove that Eq.\eqref{G_twoLevel} applies in the long nanowire, where the ABSs on both ends are localized. On the other hand, when the wire is short, the ABSs on both ends are delocalized and one can use Eq.\eqref{G_one-level} instead of Eq.\eqref{G_twoLevel} to describe conductance through each of these states. The resulting conductance is no longer suppressed by the energy difference between the ABSs, though it is instead suppressed by the short length. The topological MBS peaks are associated with a single ABS that is therefore described by Eq.\eqref{G_one-level} and shows a bright CB resonance even when split\cite{vanHeck2016Conductancea}. This Majorana peak (in the absence of a soft gap) described by Eq.\eqref{G_one-level}, is the brightest in the CB conductance plot. Thus, the presence of exponential suppression based on Eq.\eqref{G_twoLevel} can in principle explain the bright-dark-bright feature seen in experiments on long Majorana wires\cite{Albrecht2016Exponentiala,Shen2018Parity,Vaitiekenas2020Fluxinduced} although the experimental wires are unlikely to be in the long wire regime given the rather small induced SC gap.

The mechanism for the dark feature discussed in the previous paragraph is different from that proposed previously\cite{vanHeck2016Conductancea} where the dark feature resulted from a continuum of delocalized states. The suppression in the latter mechanism arises from the delocalization of the continuum states, as opposed to the localization of the ABSs considered in our work. The numerical results in Figs.\ref{fig:refCase}-\ref{fig:NonSelfCase}(a) also contain contribution from continuum states, but only near the TQPT when the bulk gap closes. This is seen as a separate small dark region in the vicinity of the TQPT (dashed line at the lower $V_z$) in Figs. \ref{fig:refCase}-\ref{fig:NonSelfCase}. Since this region is characterized by a large number of states, the numerical treatment of this region requires the generalized Meir-Wingreen formalism described in Sec.~\ref{sec:level1_2} of this work. One can use this small dark region as a signature to distinguish the ABS regime and the MBS regime, as seen in Fig. \ref{fig:muCase_3.0}(a). But we can barely pinpoint this small dark patch generically because this signature may be confused with a larger range of dark conductance arising from localized ABSs at the two ends as seen in other CB conductance numerical results, i.e., Panel (a) of Figs. \ref{fig:refCase}, \ref{fig:LengthCase_10}, \ref{fig:LengthCase_06}, \ref{fig:VcCase_3.6}, \ref{fig:VcCase_infinity}, \ref{fig:muCase_2.0}, and \ref{fig:NonSelfCase}. Furthermore, the soft-gap effect in the MBS regime together with certain ABS wave-function profiles could obscure the darkness associated with the TQPT patch, making the presence of Majorana challenging to distinguish, as in Fig. \ref{fig:idealCase}(a). This further reinforces our earlier comment that CB conductance studies may not be a good experimental technique to discern MBS from ABS.

A separate puzzle that is not immediately resolved by our numerical treatment is the relative intensity between the two bright regions. The first $2e$ periodic bright region results from elastic co-tunneling of Cooper pairs and the second one arises from tunneling of electrons. Thus, one expects the second bright region to be brighter than the first region\cite{vanHeck2016Conductancea}, which is quite different from what is seen in experiments\cite{Albrecht2016Exponentiala,Shen2018Parity,Vaitiekenas2020Fluxinduced}. A potential resolution of this puzzle is provided in this work by details of the proximity-induced semiconductor structure. Specifically as described in Sec.~\ref{sec:level1_3_1}, the normal state CB conductance at high fields, which is equal to the normal state conductance, is suppressed in semiconductor structures if the local density of states is suppressed away from an ABS resonance in the semiconductor at low densities. Such a suppression does not affect the virtual elastic co-tunneling process since a non-resonant ABS can contribute to Cooper-pair tunneling. Thus, the inclusion of the transmission resonance associated with an ABS can explain the enhanced $2e$ periodic Cooper-pair tunneling. Resolution of this puzzling brightness paradox in the CB experiments is one of the major conceptual achievements of our theory. Our explanation, however, further reinforces the dominance of trivial ABS over topological MBS in the existing Majorana nanowire experiments.

\subsection{Suppression of normal Coulomb-blockade peak relative to ABS/MBS}\label{sec:level1_6_2}
The results for the CB conductance in Figs. \ref{fig:refCase}, \ref{fig:LengthCase_10}-\ref{fig:VcCase_3.6} in Sec.~\ref{sec:level1_5} show $1e$ periodic conductance both from MBSs in the topological region (i.e., $V_{\text{TQPT}}< V_z < V_c$) as well as normal metallic CB for $V_z>V_c$. In all these cases, the MBS conductance peak, being resonant follows Eq. \eqref{G_one-level} and is higher than the normal state CB conductance described by Eq. \eqref{G_delocalised}. In principle, this makes the bright $1e$ periodic conductance from MBS even brighter relative to the $2e$ periodic conductance mentioned in the previous subsection. Additionally, such a difference in brightness between the MBS and normal conductance is not seen in experiments\cite{Albrecht2016Exponentiala,Shen2018Parity}. Both these issues get resolved in Fig. \ref{fig:idealCase} where we have included the effect of a soft-gap. Such a soft-gap arises from the interplay of disorder and magnetic field on the parent superconductor\cite{Takei2013Soft}. Interestingly, the presence of large oscillations from the ABS states prior to the TQPT suggests the absence of disorder induced sub-gap states below the TQPT.   

\section{Conclusion}\label{sec:level1_7}
We have developed a theory for and numerically calculated the two-terminal conductance of a semiconductor-superconductor nanowire in the Coulomb blockade regime, including all the important realistic effects, such as the soft-gap, SC proximity effect, temperature, nanowire length, SC collapsing field, chemical potential, QDs, ABS, self-energy, SC states, and metallic states. The realistic model for the wire used in our work, in certain parameter regimes such as the TQPT or beyond the critical Zeeman field $V_c$, contains a large number of low energy states. In order to compute CB conductance in this region, we have derived a generalized Meir-Wingreen formula, which is based on assuming the tunneling rate to be lower than that of equilibration rate in the nanowire. This assumption reduces the complexity of the rate equation formalism from the exponential\cite{Chiu2017Conductance} to linear in the number of low-energy levels. However, the assumption requires an equilibration process that might not be very efficient in the limit of a few levels. We have discussed the resulting differences in Sec.~\ref{sec:level1_3_2}. Our calculation also entirely focuses on the $1e$ tunneling regime and we have only provided analytic estimates for the Cooper-pair tunneling regime relative to the normal state conductance. The normal state conductance seen in the high Zeeman field regime of our numerical results thus provides a calibration scale to compare the results to experiments.

Our results are best summarized by Fig. \ref{fig:idealCase}(a), which shows an example of the electron-tunneling part of the CB conductance as a function of Zeeman field. While this plot excludes the $2e$ periodic Cooper-pair tunneling part of the transport seen at low Zeeman fields as the brightest feature in the experiments\cite{Albrecht2016Exponentiala,Shen2018Parity,Vaitiekenas2020Fluxinduced}, our CB conductance plot shows a range of Zeeman field which is dark followed by a bright region similar to experiments\cite{Albrecht2016Exponentiala,Shen2018Parity,Vaitiekenas2020Fluxinduced}. By comparing to the spectrum shown in Fig. \ref{fig:idealCase}(c), the dark region in our simulations arises from ABSs, in contrast to the mechanisms studied in earlier works\cite{vanHeck2016Conductancea}. The CB conductance becomes visible where the ABSs approach zero energy and remains bright in the topological region where the conductance is due to MBSs, except for a dark patch near the TQPT. At large enough Zeeman field, the SC is driven normal and the CB conductance peak from MBSs crosses over to the $1e$ periodic CB peak of a normal metal. While the normal metal CB peak is higher than the dark region from ABSs, it is substantially (in results apart from Fig. \ref{fig:idealCase}) weaker than the resonant CB peaks for MBSs. We have verified that resonant CB peaks such as those from MBSs and ABSs can be distinguished from normal metal and TQPT related CB peaks by their temperature dependence. The difference in intensities between the MBS and normal CB regime, which is not seen in experiments\cite{Albrecht2016Exponentiala,Shen2018Parity}, is suppressed in Fig. \ref{fig:idealCase} by the introduction of a soft gap i.e., subgap density of states in the superconductor that is introduced by semiconductor disorder. The complete bright-dark-bright feature is not naturally included in our numerical simulations since the first bright feature in the experiments results from $2e$ periodic Cooper-pair tunneling. However, we have argued in Sec.~\ref{sec:level1_3_1} that resonant transmission features associated with the barrier potential can suppress the normal CB conductance (at high Zeeman field) relative to the $2e$ periodic Cooper-pair conductance (at low Zeeman field). Thus, we find that the intensity pattern seen in experiments can be matched by a specific nanowire model in our generalized Meir-Wingreen formalism.

Apart from the intensity fluctuations, the positions of the CB peaks that deviate from $1e$ periodicity provide interesting spectroscopic information such as MBS splitting\cite{Albrecht2016Exponentiala,vanHeck2016Conductancea}. In fact, the observation of such breaking of $1e$ periodicity allows one to verify the absence of quasiparticle poisoning by low energy sub-gap states\cite{Higginbotham2015Parity}. Following Albrecht \textit{et al.}\cite{Albrecht2016Exponentiala}, we have characterized the positions of the peaks through the OCPS [plotted in panel (d) of Figs.\ref{fig:idealCase}-\ref{fig:NonSelfCase}, except Figs. \ref{fig:TempCase} and \ref{fig:LengthDependence}]. We find that in the case of ABSs at both ends of the wire, the first lobes of the OCPS result from a combination of both ABSs. This is distinct from tunneling conductance at a single end, which is sensitive to the spectrum only at one end. We find that this model of ABSs at both ends produces OCPS that decreases with increasing Zeeman field. The OCPS arising from MBSs\cite{Albrecht2016Exponentiala,vanHeck2016Conductancea,Chiu2017Conductance}, which increases in amplitude with increasing Zeeman field, is suppressed by the inclusion of self-energy from the proximity-inducing superconductor. By comparing OCPS from models with and without QDs, we find that QDs are necessary to obtain oscillations that decrease with increasing Zeeman field.

We emphasize that the experimentally claimed ``exponential protection" in Refs.\cite{Albrecht2016Exponentiala,Vaitiekenas2020Fluxinduced} may be a misleading artifact of the data being taken at  very few samples with each sample having its own set of ABS dominating the CB transport. Our work shows that because of the non-universal nature of ABS dominating CB tunneling transport in the currently available SC-SM-QD nanowire samples, there is no universal length dependence in the CB physics. We believe that the strong length dependence in the claimed experimental data can be an artifact of the few samples considered in each of the systems considered\cite{Albrecht2016Exponentiala,Vaitiekenas2020Fluxinduced}. Since each sample has a totally different parameter set (and not just different length), such experiments can tell us absolutely nothing about the intrinsic length dependence of Majorana physics since all system parameters are varied along with the wire length in such experiments.

In summary, the qualitative features, i.e., bright-dark-bright intensity patterns as well as decreasing OCPS of CB oscillations in semiconductor nanowires can be understood in terms of our semi-realistic model for the superconductor/semiconductor structure. We found that ABSs at both ends, self-energy and soft gap are all necessary ingredients to explain the features of the experiment. We find that the CB conductance has certain distinct advantages over direct tunneling conductance. Specifically, the CB conductance is sensitive to the lowest-energy states, unlike single-end tunneling conductance which picks up signal from all states. Furthermore, the CB conductance is sensitive to the delocalization of states. Away from zero energy, conductance through localized states is suppressed. An experimental characterization of the temperature dependence of the CB peak intensities would be important to verify the characters of the states contributing to the conductance. The rate-equation formalism in this work can be extended to compute the finite-bias differential conductance as opposed to the zero-bias conductance presented here. An interesting future direction would be to explore whether the finite-bias signatures could provide more information about the Majorana wire systems, i.e., be able to distinguish the ABS and MZM regimes. An important point to keep in mind in this context is, however, the fact that our current zero-bias conductance theory indicates that CB conductance measurements may not be particularly useful in distinguishing topological MBS signatures from trivial ABS signatures.

Acknowledgement: This work is supported by Laboratory for Physical Sciences and Microsoft Corporation. The authors acknowledge the support of the University of Maryland High Performance Computing Center for the use of Deep Thought II cluster for carrying out the numerical work. 

\bibliography{CoulombBK_ref.bib}

\clearpage
\onecolumngrid
\vspace{1cm}
\begin{center}
	{\bf\large Appendix}
\end{center}
\vspace{0.5cm}
\setcounter{section}{0}
\setcounter{secnumdepth}{3}
\setcounter{equation}{0}
\setcounter{figure}{0}
\renewcommand{\thefigure}{A\arabic{figure}}
\newcommand\Scite[1]{[S\citealp{#1}]}
\makeatletter \renewcommand\@biblabel[1]{[S#1]} \makeatother

\section{Microscopic tunneling rates}\label{sec:levelA_1}
The rate of absorption of electrons from lead $\alpha=L,R$ is given by
\begin{equation}\label{Gamma}
	\begin{aligned}
	\Gamma_N^\alpha&=\tau_\alpha\sum_{i,j}P_N(E_i)\int d\epsilon f(\epsilon-\mu_\alpha)\delta(E_j-E_i-\epsilon)\left|\langle\psi_j|d_\alpha^\dagger|\psi_i\rangle\right|^2\\
	&=\tau_\alpha\sum_{i,j}P_N(E_i)f(E_j-E_i-\mu_\alpha)\left|\langle\psi_j|d_\alpha^\dagger|\psi_i\rangle\right|^2
	\end{aligned}
\end{equation}
, where $f(\epsilon)=(1+e^{\beta\epsilon})^{-1}$ is the Fermi function.

Similarly, the rate of emission of electrons into lead $\alpha$ is given by
\begin{equation}\label{Lambda}
	\begin{aligned}
	\Lambda_N^\alpha&=\tau_\alpha\sum_{i,j}P_N(E_i)\int d\epsilon\left[1-f(\epsilon-\mu_\alpha)\right]\delta(E_j+\epsilon-E_i)\left|\langle\psi_j|d_\alpha|\psi_i\rangle\right|^2\\
	&=\tau_\alpha\sum_{i,j}P_N(E_i)\left[1-f(E_i-E_j-\mu_\alpha)\right]\left|\langle\psi_j|d_\alpha|\psi_i\rangle\right|^2.
	\end{aligned}
\end{equation}
This equation can be simplified and related to the coefficient $\Gamma_N^\alpha$ by interchanging the indices $i,j$ and considering the charge state $(N+1)$ as
\begin{equation}\label{Lambda_Np1}
	\begin{aligned}
	\Lambda_N^\alpha&=\tau_\alpha\sum_{i,j}P_{N+1}(E_j)\left[1-f(E_j-E_i-\mu_\alpha)\right]\left|\langle\psi_j|d_\alpha^\dagger|\psi_i\rangle\right|^2\\
	&=\tau_\alpha\sum_{i,j}P_{N+1}(E_j)f(E_i-E_j+\mu_\alpha)\left|\langle\psi_j|d_\alpha^\dagger|\psi_i\rangle\right|^2\\
	&=\tau_\alpha e^{-\beta\mu_\alpha}\frac{Z_N}{Z_{N+1}}\sum_{i,j}P_N(E_i)e^{-\beta(E_j-E_i-\mu_\alpha)}f(E_i-E_j+\mu_\alpha)\left|\langle\psi_j|d_\alpha^\dagger|\psi_i\rangle\right|^2\\
	&=\tau_\alpha e^{-\beta\mu_\alpha}\frac{Z_N}{Z_{N+1}}\sum_{i,j}P_N(E_i)f(-E_i+E_j-\mu_\alpha)\left|\langle\psi_j|d_\alpha^\dagger|\psi_i\rangle\right|^2\\
	&=e^{-\beta\mu_\alpha}\frac{Z_N}{Z_{N+1}}\Gamma_N^\alpha.
	\end{aligned}
\end{equation}

\section{Steady-state probabilities}\label{sec:levelA_2}
The rate of the system leaving the state $N$ must match the rate that is entering into the state $N$. This leads to the equation
\begin{equation}\label{detailedBalance_ap1}
	P_{0,N}\sum_{\alpha}(\Gamma_N^\alpha+\Lambda_N^\alpha)=\sum_{\alpha}(P_{0,N-1}\Gamma_{N-1}^\alpha+P_{0,N+1}\Lambda_{N+1}^\alpha).
\end{equation}
Substituting $\Lambda$ in terms of $\Gamma$, the equilibrium condition becomes,
\begin{equation}\label{detailedBalance_ap2}
	P_{0,N}\sum_{\alpha}(\Gamma_N^\alpha+e^{-\beta\mu_\alpha}\frac{Z_{N-1}}{Z_N}\Gamma_{N-1}^\alpha)=\sum_{\alpha}(P_{0,N-1}\Gamma_{N-1}^\alpha+P_{0,N+1}e^{-\beta\mu_\alpha}\frac{Z_N}{Z_{N+1}}\Gamma_N^\alpha).
\end{equation}
Collecting the rates $\Gamma_N^\alpha$, the equilibrium condition becomes
\begin{equation}\label{detailedBalance_ap3}
	\sum_{\alpha}\Gamma_N^\alpha\left[P_{0,N}-P_{0,N+1}e^{-\beta\mu_\alpha}\frac{Z_N}{Z_{N+1}}\right]
	=\sum_{\alpha}\Gamma_{N-1}^\alpha\left[P_{0,N-1}-P_{0,N}e^{-\beta\mu_\alpha}\frac{Z_{N-1}}{Z_N}\right].
\end{equation}
Defining $\Gamma_N^\alpha Z_N=\tilde{\Gamma}_N^\alpha$ and $P_{0,N}/Z_N=\tilde{P}_{0,N}$, the steady-state condition simplifies to
\begin{equation}\label{detailedBalance_ap4}
	\sum_{\alpha}\tilde{\Gamma}_N^\alpha\left[\tilde{P}_{0,N}-\tilde{P}_{0,N+1}e^{-\beta\mu_\alpha}\right]
	=\sum_{\alpha}\tilde{\Gamma}_{N-1}^\alpha\left[\tilde{P}_{0,N-1}-\tilde{P}_{0,N}e^{-\beta\mu_\alpha}\right].
\end{equation}
The above equation is solved by the detailed balance condition, where both sides of the above equation vanish so that
\begin{equation}\label{tP_Np1}
	\tilde{P}_{0,N+1}=\tilde{P}_{0,N}\left(\frac{\sum_{\alpha}\tilde{\Gamma}_N^\alpha e^{-\beta\mu_\alpha}}{\sum_{\alpha}\tilde{\Gamma}_N^\alpha}\right)^{-1}
	=\tilde{P}_{0,N}e^{\beta\mu}\left(1+\frac{\sum_{\alpha}\tilde{\Gamma}_N^\alpha\left[e^{-\beta(\mu_\alpha-\mu)}-1\right]}{\sum_{\alpha}\tilde{\Gamma}_N^\alpha}\right)^{-1}.
\end{equation}
In the limit of a small applied voltage $\mu_\alpha=\mu+V_\alpha$ and expanding to linear order in $V_\alpha$,
\begin{equation}\label{linear_tP_Np1}
	\tilde{P}_{0,N+1}=\tilde{P}_{0,N}e^{\beta\mu}\left(1+\beta\frac{\sum_{\alpha}\tilde{\Gamma}_N^\alpha V_\alpha}{\sum_{\alpha}\tilde{\Gamma}_N^\alpha}\right).
\end{equation}
Solving the recursion, we get
\begin{equation}\label{tP_N}
	\tilde{P}_{0,N}=Z_{tot}^{-1}e^{\beta N\mu}\left(1+\beta\sum_{j<N}\frac{\sum_{\alpha}\tilde{\Gamma}_j^\alpha V_\alpha}{\sum_{\alpha}\tilde{\Gamma}_j^\alpha}\right).
\end{equation}
In equilibrium, if $V_\alpha=0$, the above equations become
\begin{equation}\label{tP_N_eq}
	\tilde{P}_{0,N}^{(eq)}=Z_{tot}^{-1}e^{N\beta\mu}
\end{equation}
, where $Z_{tot}=\sum_N Z_N e^{-N\beta\mu}$ (noting that $\sum_N Z_N\tilde{P}_{0,N}^{(eq)}=1$).

\section{Current and conductance}\label{sec:levelA_3}
The current at the left lead is given by
\begin{equation}\label{currentDerived}
	\begin{aligned}
	I&=\sum_N P_{0,N}(\Gamma_N^L-\Lambda_N^L)\\
	&=\sum_N P_{0,N}\left(\Gamma_N^L-e^{-\beta \mu_L}\frac{Z_{N-1}}{Z_N}\Gamma_{N-1}^L\right)\\
	&=\sum_N\tilde{P}_{0,N}\left(\tilde{\Gamma}_N^L-e^{-\beta \mu_L}\tilde{\Gamma}_{N-1}^L\right)\\
	&=\sum_N\tilde{\Gamma}_N^L\left(\tilde{P}_{0,N}-e^{-\beta\mu_L}\tilde{P}_{0,N+1}\right).
	\end{aligned}
\end{equation}
This current vanished at $V_\alpha=0$ because the combination of probability factors vanishes. Since the probability factor vanishes to linear order, the conductance can be extracted by expanding this factor to linear order in $V_\alpha$:
\begin{equation}\label{linearPdiff}
	\begin{aligned}
	\tilde{P}_{0,N}-e^{-\beta\mu_L}\tilde{P}_{0,N+1}&=\tilde{P}_{0,N}-(1-\beta V_L)e^{-\beta\mu}\tilde{P}_{0,N+1}\\
	&=\tilde{P}_{0,N}\left[1-(1-\beta V_L)\left(1+\beta\frac{\sum_{\alpha}\tilde{\Gamma}_N^\alpha V_\alpha}{\sum_{\alpha}\tilde{\Gamma}_N^\alpha}\right)\right]\\
	&\approx\tilde{P}_{0,N}\left(V_L-\frac{\sum_{\alpha}\tilde{\Gamma}_N^\alpha V_\alpha}{\sum_{\alpha}\tilde{\Gamma}_N^\alpha}\right)\\
	&-\beta\tilde{P}_{0,N}\left(\frac{\sum_{\alpha}\tilde{\Gamma}_N^\alpha(V_\alpha-V_L)}{\sum_{\alpha}\tilde{\Gamma}_N^\alpha}\right)\\
	&=-\beta\tilde{P}_{0,N}\left(\frac{\tilde{\Gamma}_N^R(V_R-V_L)}{\sum_{\alpha}\tilde{\Gamma}_N^\alpha}\right)
	\end{aligned}
\end{equation}
Writing $V=V_R-V_L$, the conductance is written as
\begin{equation}\label{G}
	\begin{aligned}
	G=\frac{dI}{dV}|_{V=0}&=-\beta\sum_N\tilde{P}_{0,N}\frac{\tilde{\Gamma}_N^R\tilde{\Gamma}_N^L}{\tilde{\Gamma}_N^R+\tilde{\Gamma}_N^L}\\
	&=-\beta\sum_N P_{0,N}\frac{\Gamma_N^R\Gamma_N^L}{\Gamma_N^R+\Gamma_N^L}\\
	&=\sum_N\frac{\gamma_N^R\gamma_N^L}{\gamma_N^R+\gamma_N^L}
	\end{aligned}
\end{equation}
where $\gamma_N^\alpha=-\beta P_{0,N}\Gamma_N^\alpha$.

In the above expression, we use the equilibrium tunneling rate
\begin{equation}\label{gamma_eq}
	\begin{aligned}
	\gamma_N^\alpha&=-\beta\tau_\alpha Z_{tot}^{-1}\sum_{i,j}e^{\beta N\mu}Z_N P_N(E_i)f(E_j-E_i-\mu)\left|\langle\psi_j|d_\alpha^\dagger|\psi_i\rangle\right|^2\\
	&=-\beta\tau_\alpha\sum_{i,j}P(E_i)f(E_j-E_i-\mu)\left|\langle\psi_j|d_\alpha^\dagger|\psi_i\rangle\right|^2.
	\end{aligned}
\end{equation}
Here we have used the identity we showed earlier, i.e.,
\begin{equation}\label{P_Ei}
	P(E_i)=P_N(E_i)P_{0,N}=P_N(E_i)Z_N e^{\beta N\mu}/Z_{tot}=e^{-\beta(E_i-N\mu)}/Z_{tot}.
\end{equation}
Using the identity
\begin{equation}\label{identity}
	\begin{aligned}
	e^{-\beta N\mu}P(E_i)f(E_j-E_i-\mu)&=\frac{e^{-\beta E_i}}{1+e^{\beta(E_j-E_i-\mu)}}\\
	&=\frac{1}{e^{\beta E_i}+e^{\beta(E_j-\mu)}}\\
	&=-\beta^{-1}\left[e^{-\beta E_i}+e^{-\beta(E_j-\mu)}\right]f'(E_j-E_i-\mu)\\
	&=-\beta^{-1}e^{-\beta N\mu}\left[P(E_i)+P(E_j)\right]f'(E_j-E_i-\mu)
	\end{aligned}
\end{equation}
, the above conductance can be written in the Meir-Wingreen form
\begin{equation}\label{MW_gamma}
	\gamma_N^\alpha=\tau_\alpha\sum_{i,j}\left[P(E_i)+P(E_j)\right]f'(E_j-E_i-\mu)\left|\langle\psi_j|d_\alpha^\dagger|\psi_i\rangle\right|^2.
\end{equation}

\section{Meir-Wingreen formula for one-terminal Coulomb blockaded Majorana nanowire}\label{sec:levelA_4}
The original Meir-Wingreen's formula is the Landauer formula, when we consider the transport through an interacting region, with one terminal connecting the system and the environment.\cite{Meir1992Landauer} The general Hamiltonian is of this form
\begin{equation}\label{MW_hami}
    H=\sum_{k,\alpha\in L,R}\epsilon_{k\alpha}a_{k\alpha}^\dagger a_{k\alpha}+H_{int}(\{c_n^\dagger\};\{c_n\}) + \sum_{k,n,\alpha\in L,R}(V_{k\alpha,n}a_{k\alpha}^\dagger c_n + \text{H.c.})
\end{equation}
where $a_{k\alpha}^\dagger$ ($a_{k\alpha}$)  creates (destroys) an electron with momentum $k$ in channel $\alpha$ from either left (\textit{L}) or the right (\textit{R}) lead, and $\{c_n^\dagger\}$ and $\{c_n\}$ form a complete and orthonormal set of single-electron creation and annihilation operators in the interacting region. The channel index includes spin and all other quantum numbers which, in addition to $k$, are necessary to define uniquely a state in the leads.

Through the Keldysh formalism, one can get the linear-response conductance $G$ in the form
\begin{equation}\label{originalMW}
    G=\frac{e^2}{\hbar}\sum_{m,n}\Gamma_{n,m}(E_j - E_i)\sum_{i,j}(P_i + P_j)\left[-\frac{\partial f_{eq}(E_j - E_i)}{\partial\epsilon}\right]\langle\psi_j|c_n^\dagger|\psi_i\rangle\langle\psi_i|c_m|\psi_j\rangle
\end{equation}
where the $\psi_i$ are eigenstates, with energies $E_i$, of the uncoupled interacting region, and $P_i$ is the equilibrium probability of state $\psi_i$. For non-interacting electrons, one can choose the $c$'s to correspond to single-particle eigenstates of the uncoupled system, and the overlap factor in each term, $\langle\psi_j|c_n^\dagger|\psi_i\rangle\langle\psi_i|c_m|\psi_j\rangle$, is trivially 0 or 1.

Now we apply our system to the Meir-Wingreen's formula in Eq. \eqref{originalMW}. In our case, we consider only $m=n=\{L,R\}\otimes\{\uparrow,\downarrow\}$, where the transmission only occurs on the left and right ends of the nanowire, and the tunneling factor is independent of the energy (setting $\Gamma_{n,m}=1$), then Eq. \eqref{originalMW} becomes 
\begin{equation}\label{nanowireMW}
    G=\frac{e^2}{\hbar}\sum_{x=L,R}\sum_{\sigma=\uparrow,\downarrow}\sum_{i,j}(P_i+P_j)\left[-\frac{\partial f_{eq}(E_j - E_i)}{\partial\epsilon}\right]|\langle\psi_j|c_{x,\sigma}^\dagger|\psi_i\rangle|^2
\end{equation} 
where $|\psi_i\rangle=|\{n_i\}\rangle$ is the state with some quasi-electron configuration over $N_l$ energy levels, and
\begin{equation}\label{E_i}
	E_i=\sum_s n_s\epsilon_s +U(N)
\end{equation}
is the energy of configuration $\{n_i\}$ with electrostatic energy $U(N)=E_c(N-n_g)^2$. $n_s$ is the occupation number of $s$ state of the quasiparticle, i.e., $n_s$ is the eigenvalue of the operator $d_s^\dagger d_s$ while $d_s^\dagger$ and $d_s$ are creation and annihilation operators for the quasi-particles. (Note that $N=\sum_s n_s + 2N_c$ is the total electron number and $N_c$ is the additional number of the Cooper pairs away from the charge-neutral point as the gate voltage of the nanowire is zero, and $n_g$ is the number of electrons corresponding to the gate voltage.) $\epsilon_s$ is the eigen-energy of $s$ state of the Hamiltonian as Eq.\eqref{H_BdG}. The probability for quasi-particle distribution configuration $\{n_i\}$ is described by the Gibbs distribution
\begin{equation}\label{Gibbs_P}
    P_i = e^{-\beta E_i}/Z = Z^{-1}\exp\left[-\beta\left(\sum_s n_s \epsilon_s + U(N)\right)\right]
\end{equation}
, where the canonical partition function is given by
\begin{equation}\label{Gibbs_Z}
    Z=\sum_i e^{-\beta E_i}=\sum_i \exp\left[-\beta\left(\sum_s n_s\epsilon_s + U(N)\right)\right].
\end{equation}

Due to fermion parity conservation, we only need to separate the total electron number $N$ into even and odd groups. The superconductor ground state favors the even number of electrons due to the condensate. If one more electron adds in the superconductor and overcomes the superconducting gap, then there will be one extra quasiparticle besides the condensate. Otherwise, the condensate ground state will remain and only allow $2e$ transport. Therefore, the behaviors of even parity and odd parity for this superconductor-induced nanowire are different. We only need to discuss the parity of the total electron number $N$, i.e., we discuss only the two values from ($N$ mod 2). To simplify the formula later, we can rewrite Eqs. \eqref{Gibbs_P} and \eqref{Gibbs_Z} as
\begin{equation}\label{Gibbs_P_shift}
    P_i=Z^{-1}\exp\left[-\beta\left(\sum_s n_s \epsilon_s + \frac{1}{2}Q_i\cdot \Delta U\right)\right]
\end{equation}
and
\begin{equation}\label{Gibbs_Z_shift}
    Z=\sum_i\exp\left[-\beta\left(\sum_s n_s\epsilon_s + \frac{1}{2}Q_i\cdot \Delta U\right)\right]
\end{equation}
where $Q_i=Q_0\cdot(-1)^{\sum_s n_s}$ is the fermion parity (note that $Q_0=\pm 1$ is the ground-state fermion parity, and the subscript $i$ means the quasi-particle distribution configuration.), and $\Delta U=U(N)-U(N-1)$. The parity of $N$ and $(N-1)$ is opposite. Therefore, we can shift $U(N)$ and $U(N-1)$ by the mean value $\left[U(N-1)+U(N)\right]/2$ and express them with the parity $Q_i$ and the electrostatic energy difference $\Delta U$. Equation \eqref{Gibbs_Z_shift} can be further simplified as
\begin{equation}\label{Z_derived}
    \begin{aligned}
    Z&=\sum_{\{n_i\}}\exp\left[-\beta\left(\sum_s n_s\epsilon_s+\frac{1}{2}Q_i\Delta U\right)\right]\\
    &=\sum_Q\sum_{\{n_i\}}\delta(Q-Q_i)\exp\left[-\beta\left(\sum_s n_s\epsilon_s+\frac{1}{2}Q\Delta U\right)\right]\\
    &=\sum_{Q=\pm 1}\sum_{\{n_i\}}\frac{1}{2}\left[1+Q\cdot Q_0(-1)^{\sum_s n_s}\right]\exp\left[-\beta\left(\sum_s n_s\epsilon_s+\frac{1}{2}Q\Delta U\right)\right]\\
    &=\sum_{Q=\pm 1}\frac{1}{2}\exp\left(-\frac{\beta Q\cdot\Delta U}{2}\right)\left[\prod_s\left(1+e^{-\beta\epsilon_s}\right)+Q\cdot Q_0\cdot\prod_s\left(1-e^{-\beta\epsilon_s}\right)\right]\\
    &=\sum_{Q=\pm 1}\frac{1}{2}e^{-\beta(Q\cdot\Delta U/2)}\cdot\prod_s(1+e^{-\beta\epsilon_s})\left[1+Q\cdot Q_0\cdot\prod_s\tanh\left(\frac{\beta\epsilon_s}{2}\right)\right]
    \end{aligned}
\end{equation}
The energy of configuration $\{n_i\}$, i.e., $E_i$ [the original definition is in Eq.\eqref{E_i}], can also be shifted and redefined as
\begin{equation}\label{E_i_shift}
    E_i=\sum_s n_s\epsilon_s+\frac{Q_i\Delta U}{2}
\end{equation}

Since the electron operator is of the form as Eqs. \eqref{e_create} and \eqref{e_annihilation}, the quasi-particle creation and annihilation operators can only change the orbital occupation number by one, say in the $p$th orbital. Hence, the eligible final transition state can only be the same configuration with one orbital occupation changed, i.e.,
\begin{equation}\label{psi_j}
    |\psi_j\rangle=|\{n_{j\neq p},\bar{n}_p\}\rangle=(1-n_p)(d_p^\dagger|\{n_i\}\rangle) + n_p(d_p|\{n_i\}\rangle)
\end{equation}
with $n_p=1$ ($n_p=0$) if the $p$th orbital is occupied (empty) in the configuration $\{n_i\}$. Then the energy difference between these two states is 
\begin{equation}\label{E_p}
    E_p=E_j - E_i = (1-2n_p)\epsilon_p -Q_i\Delta U.
\end{equation}

The transition matrix element is therefore $\langle\psi_j|c_{x\sigma}^\dagger|\psi_i\rangle\equiv\langle\{n_j\}|c_{x\sigma}^\dagger|\{n_i\}\rangle=\langle\{n_{i\neq p},\bar{n}_p\}|c_{x\sigma}^\dagger|\{n_i\}\rangle=\left((1-n_p)\langle\{n_i\}|d_p + n_p\langle\{n_i\}|d_p^\dagger\right)c_{x\sigma}^\dagger|\{n_i\}\rangle$. With Eqs. \eqref{e_create} and \eqref{e_annihilation}, 
\begin{equation}\label{TMelement}
    \langle\psi_j|c_{x\sigma}^\dagger|\psi_i\rangle = (1-n_p)u_{p,x\alpha}^* + n_p v_{p,x\alpha}
\end{equation}
Hence,
\begin{equation}\label{Abs2_TMelement}
    |\langle\psi_i|c_{x\sigma}^\dagger|\psi_j\rangle|^2=(1-n_p)\Gamma_p^{x,\sigma} + n_p \Lambda_p^{x,\sigma}
\end{equation}
where 
\begin{equation}\label{tunnel_Gamma_spin}
    \Gamma_p^{x,\sigma}=\left|u_{p,x\sigma}\right|^2
\end{equation}
is the tunneling rate for the electron to tunnel from the lead at $x$ to the nanowire (same for the opposite direction), and 
\begin{equation}\label{tunnel_Lambda_spin}
    \Lambda_p^{x,\sigma}=\left|v_{p,x\sigma}\right|^2
\end{equation}
is the tunneling rate for the hole to tunnel from the lead at $x$ to the nanowire (same for the opposite direction). Note that $\Gamma_p^x=\sum_{\sigma=\uparrow,\downarrow}\Gamma_p^{x,\sigma}$ and $\Lambda_p^x=\sum_{\sigma=\uparrow,\downarrow}\Lambda_p^{x,\sigma}$ as defined in Eqs. \eqref{tunnel_Gamma} and \eqref{tunnel_Lambda}.

With the ingredients above, we can rewrite Eq. \eqref{nanowireMW} as
\begin{equation}\label{G_derived}
    \begin{aligned}
    G & =\frac{e^2}{\hbar}\sum_{x=L,R}\sum_{\sigma=\uparrow,\downarrow}\sum_i P_i\sum_p\left(1+P_j/P_i\right)\left[-\frac{\partial f_{eq}\left((1-2n_p)\epsilon_p-Q_i\Delta U\right)}{\partial\epsilon}\right]\cdot\left[(1-n_p)\Gamma_p^{x,\sigma} + n_p \Lambda_p^{x,\sigma}\right]\\
    & = \frac{e^2}{\hbar}\sum_{x=L,R}\sum_i P_i\sum_p\left(1+\exp\left[-\beta\left((1-2n_p)\epsilon_p - Q_i\Delta U\right)\right]\right)\\
    &\qquad\times\left\{-f'_{eq}\left((1-2n_p)\epsilon_p - Q_i\Delta U\right)\right\}\left[(1-n_p)\Gamma_p^{x} + n_p \Lambda_p^{x}\right]\\
    & = \frac{e^2}{\hbar}\sum_{x=L,R}\sum_p\sum_{n=0,1}\sum_{Q=-1,1}\underbrace{\left(Z^{-1}\cdot\sum_{\substack{i:n=n_p\\Q=Q_0\cdot(-1)^{\sum n_s}}}\exp\left[-\beta\left(\sum_s n_s\epsilon_s + \frac{1}{2}Q\Delta U\right)\right]\right)}_{F_p(n,Q)}\\
    &\qquad \times\underbrace{\left(1+\exp\left[-\beta\left((1-2n)\epsilon_p - Q\Delta U\right)\right]\right)\cdot\left[-f'_{eq}\left((1-2n)\epsilon_p-Q\Delta U\right)\right]}_{=\beta\cdot f_{eq}((1-2n)\epsilon_p-Q\Delta U)}\cdot\left[(1-n)\Gamma_p^{x} + n \Lambda_p^{x}\right]\\
    & = \beta\frac{e^2}{\hbar}\sum_{x=L,R}\sum_p\sum_{n=0,1}\sum_{Q=-1,1}\underbrace{f_{eq}((1-2n)\epsilon_p-Q\Delta U)\cdot F_p(n,Q)}_{\tilde{F}_p(n,Q)}\cdot\left[(1-n)\Gamma_p^{x} + n \Lambda_p^{x}\right]\\
    \end{aligned}
\end{equation}
$F_p(n,Q)$ sums over all the possible configurations, but with the constriction that $n=n_p$ and $Q=Q_0\cdot(-1)^{\sum_s n_s}$ being selected properly, i.e.,
\begin{equation}\label{F_derived}
    \begin{aligned}
    F_p(n,Q) & = Z^{-1}e^{-\beta Q\Delta U/2}\sum_{\{n_s\}}\delta(n-n_p)\delta\left(Q-Q_0\cdot(-1)^{\sum n_s}\right)\exp\left(-\beta\sum_s n_s\epsilon_s\right)\\
    & = Z^{-1}e^{-\beta Q\Delta U/2}e^{-\beta n \epsilon_p}\sum_{\{n_{s\neq p}\}}\underbrace{\frac{1}{2}\left[1+Q\cdot Q_0\cdot(-1)^{\sum_{s\neq p}n_s}(-1)^{n_p}\right]}_{\star}\exp\left(-\beta\sum_{s\neq p}n_s\epsilon_s\right)\\
    & = \frac{1}{2Z}e^{-\beta Q\Delta U/2}e^{-\beta n \epsilon_p}\left[\prod_{s\neq p}\left(1+e^{-\beta\epsilon_s}\right) + Q\cdot Q_0\cdot(-1)^{n_p}\prod_{s\neq p}\left(1-e^{-\beta\epsilon_s}\right)\right]\\
    & = \frac{1}{2Z}e^{-\beta(Q\Delta U/2+n\epsilon_p)}\left[\frac{\prod_s(1+e^{-\beta\epsilon_s})}{(1+e^{-\beta\epsilon_p})}+Q\cdot Q_0\cdot(-1)^n\cdot\frac{\prod_s(1-e^{-\beta\epsilon_s})}{(1-e^{-\beta\epsilon_p})}\right]\\
    & = \frac{1}{2Z}e^{-\beta(Q\Delta U/2+n\epsilon_p)}\prod_s(1+e^{-\beta\epsilon_s})\left[\frac{1}{1+e^{-\beta\epsilon_p}}+Q\cdot Q_0\cdot(-1)^n\cdot\frac{\prod_s\tanh\left(\beta\epsilon_s/2\right)}{1-e^{-\beta\epsilon_p}}\right]\\
    & = \frac{e^{-\beta(Q\Delta U/2+n\epsilon_p)}\left[\frac{1}{(1+e^{-\beta\epsilon_p})}+Q\cdot Q_0\cdot(-1)^n\cdot\frac{\prod_s\tanh\left(\beta\epsilon_s/2\right)}{(1-e^{-\beta\epsilon_p})}\right]}{\sum_{Q=\pm 1}e^{-\beta(Q\Delta U/2)}\left[1+Q\cdot Q_0\cdot\prod_s\tanh\left(\frac{\beta\epsilon_s}{2}\right)\right]}
    \end{aligned}
\end{equation}
Note that the star $\star$ part in Eq. \eqref{F_derived} is technically a delta function for $Q$ when we sum over $(-1)$ and $1$.
\begin{equation}\label{f}
    f\equiv\frac{1}{2}\left[1+Q\cdot Q_0\cdot(-1)^{\sum_{s\neq p}n_s}(-1)^{n_p}\right]
\end{equation}
If $Q$ is the correct parity, i.e.,
\begin{equation}\label{Q}
    Q=Q_0\cdot(-1)^{\sum_s n_s}=Q_0\cdot(-1)^{\sum_{s\neq p}n_s}(-1)^{n_p}
\end{equation}
, then $f=(1+1)/2=1$. On the contrary, if $Q=-Q_0\cdot(-1)^{\sum_s n_s}$ (not correct parity), then $f=(1-1)=0$. So only $Q$ that represents the correct fermion parity is picked and evaluated.

We can actually simplify Eq.\eqref{F_derived} further.
\begin{equation}\label{prodTanh_derived}
    \begin{aligned}
    \frac{\prod_s\tanh\left(\frac{\beta\epsilon_s}{2}\right)}{1-e^{-\beta\epsilon_p}}
    &=\frac{\tanh\left(\beta\epsilon_p/2\right)}{e^{-\beta\epsilon_p/2}\left(e^{\beta\epsilon_p/2}-e^{-\beta\epsilon_p/2}\right)}\cdot\prod_{s\neq p}\tanh\left(\frac{\beta\epsilon_s}{2}\right)\\
    &=e^{\beta\epsilon_p/2}\frac{\tanh\left(\beta\epsilon_p/2\right)}{2\sinh\left(\beta\epsilon_p/2\right)}\cdot\prod_{s\neq p}\tanh\left(\frac{\beta\epsilon_s}{2}\right)\\
    &=\frac{e^{\beta\epsilon_p/2}}{2\cosh\left(\beta\epsilon_p/2\right)}\cdot\prod_{s\neq p}\tanh\left(\frac{\beta\epsilon_s}{2}\right)\\
    &=\frac{e^{\beta\epsilon_p/2}}{e^{\beta\epsilon_p/2}+e^{-\beta\epsilon_p/2}}\cdot\prod_{s\neq p}\tanh\left(\frac{\beta\epsilon_s}{2}\right)\\
    &=\frac{1}{1+e^{-\beta\epsilon_p}}\cdot\prod_{s\neq p}\tanh\left(\frac{\beta\epsilon_s}{2}\right)
    \end{aligned}
\end{equation}
Then, the $F_p(n,Q)$ factor is simplified to be
\begin{equation}\label{F_derived2}
    \begin{aligned}
    F_p(n,Q)&=\frac{e^{-\beta(Q\Delta U/2+n\epsilon_p)}\left[\frac{1}{(1+e^{-\beta\epsilon_p})}+Q\cdot Q_0\cdot(-1)^n\cdot\frac{1}{(1+e^{-\beta\epsilon_p})}\cdot\prod_{s\neq p}\tanh\left(\frac{\beta\epsilon_s}{2}\right)\right]}{\sum_{Q=\pm 1}e^{-\beta(Q\Delta U/2)}\left[1+Q\cdot Q_0\cdot\prod_s\tanh\left(\frac{\beta\epsilon_s}{2}\right)\right]}\\
    &=\frac{\frac{e^{-\beta(Q\Delta U/2+n\epsilon_p)}}{(1+e^{-\beta\epsilon_p})}\left[1+Q\cdot Q_0\cdot (-1)^n\cdot\prod_{s\neq p}\tanh\left(\frac{\beta\epsilon_s}{2}\right)\right]}{\sum_{Q=\pm 1}e^{-\beta(Q\Delta U/2)}\left[1+Q\cdot Q_0\cdot\prod_s\tanh\left(\frac{\beta\epsilon_s}{2}\right)\right]}\\
    &=\frac{e^{-\beta(Q\Delta U/2+n\epsilon_p)}\left[1+Q\cdot Q_0\cdot(-1)^n\cdot\prod_{s\neq p}\tanh\left(\frac{\beta\epsilon_s}{2}\right)\right]}{\sum_{Q=\pm 1}\sum_{n=0,1}e^{-\beta(Q\Delta U/2+n\epsilon_p)}\left[1+Q\cdot Q_0\cdot\prod_s\tanh\left(\frac{\beta\epsilon_s}{2}\right)\right]}
    \end{aligned}
\end{equation}
For numerical evaluation [in MATLAB, $\tanh(x)$ gives 1, when $x$ is above some threshold, and this causes the conductance computation unable], it is necessary to replace $\prod_s\tanh(\beta\epsilon_s/2)$ with $(1-e^{-y})$, where
\begin{equation}\label{y}
    y=-\log\left(1-\prod_s\tanh(\frac{\beta\epsilon_s}{2})\right)
\end{equation}
is evaluated by the identity
\begin{equation}\label{tanh_Idensity}
    \tanh(x)=1-2e^{-\left[2x+\log(1+e^{-2x})\right]}.
\end{equation}

\section{Strong CB near three-fold $(N,N+1,N+2)$ transitions}\label{sec:levelA_5}
Let us continue to consider the strong CB limit, where three states $N=0,1,2$ are in play. The rate constants required would be $\gamma_0^{\alpha,1}$ and $\gamma_0^{\alpha,2}$. In this case, $\nu_2=-\rho_2=0$ (since $\gamma_2=0$), $\nu_1=-\rho_1=\frac{\gamma_1^{R,1}}{\gamma_1^{R,1}+\gamma_1^{L,1}}>0$ and $\nu_0=-\frac{\rho_1\sum_\alpha\gamma_0^{\alpha,1}}{\sum_\alpha\gamma_0^{\alpha,1}\gamma_0^{\alpha,2}}-\rho_0$ where $\rho_0=-\frac{(\gamma_0^{R,1}+2\gamma_0^{R,2})}{\sum_\alpha\gamma_0^{\alpha,1}+\gamma_0^{\alpha,2}}$. The difference $\nu_0-\nu_1=-\frac{\rho_1\sum_\alpha\gamma_0^{\alpha,2}}{\sum_\alpha\gamma_0^{\alpha,1}+\gamma_0^{\alpha,2}}-\rho_0>0$. The conductance in this case is a combination of these two positive quantities, i.e.,
\begin{equation}\label{G_3fold}
    \begin{aligned}
    G=&\gamma_0^{L,1}(\nu_0-\nu_1)+\gamma_1^{L,1}\nu_1+2\gamma_0^{L,2}\nu_0\\
    =&\frac{\gamma_1^{R,1}\gamma_1^{L,1}}{\gamma_1^{R,1}+\gamma_1^{L,1}}+\frac{\gamma_1^{R,1}\gamma_0^{L,1}}{\gamma_1^{R,1}+\gamma_1^{L,1}}\frac{\sum_{\alpha}\gamma_0^{\alpha,2}}{\sum_\alpha\gamma_0^{\alpha,1}+\gamma_0^{\alpha,2}}+\frac{\gamma_0^{L,1}(\gamma_0^{R,1}+2\gamma_0^{R,2})}{\sum_{\alpha}\gamma_0^{\alpha,1}+\gamma_0^{\alpha,2}}+2\gamma_0^{L,2}\nu_0.
    \end{aligned}
\end{equation} 

This equation clearly includes the previous case if two-electron transfer rate $\gamma_N^{\alpha,2}=0$. Another possibility is $\gamma_N^{\alpha,1}=0$, in which case, the conductance takes the obvious generalization
\begin{equation}\label{G_gamma12=0}
    G=\frac{\gamma_0^{R,2}\gamma_0^{L,2}}{\gamma_0^{R,2}+\gamma_0^{L,2}}.
\end{equation}
One expects this conductance to be weaker than the other case because it involves higher-order processes. 

Finally, a third case that is of interest for long wires is one where there are one-electron processes only on the left so that $\gamma_N^{R,1}\approx 0$. For simplicity, we also assume $\gamma_N^{L,2}=0$. This leaves a simple answer
\begin{equation}\label{localG}
    G=\frac{2\gamma_0^{L,1}\gamma_0^{R,2}}{\gamma_0^{L,1}+\gamma_0^{R,2}}.
\end{equation}
This would track the spectrum only on the left until the state at $L$ becomes delocalized enough and can explain the bright-dark-bright feature.

\section{Tunneling rate expressed in LDOS/DOS}\label{sec:levelA_6}
The purpose of this appendix is to construct the relation between the tunneling rate and LDOS and DOS. Since we cannot distinguish a discretized state above the superconducting gap (SC states) or after the gap collapses (metallic states), we need to utilize LDOS and DOS to calculate the tunneling rate. In order to make sure the results are consistent, we use this method even to calculate the tunneling rate for the bound states (below the gap) as well.

The local density of states (LDOS) at position $x$ with energy $\epsilon$ is defined as
\begin{equation}\label{LDOS}
   \rho_L^\sigma\left(x,\epsilon\right)=-\frac{1}{\pi}\text{Im}\{\sum_{s=\uparrow,\downarrow} \langle x,\sigma,s|G(\epsilon)|x,\sigma,s\rangle\},
\end{equation}
where $\sigma$ denotes electron or hole, and $s$ denotes spins. $G(\epsilon)$ is the Green's function:
\begin{equation}\label{Green}
    G(\epsilon)=\frac{1}{H_{BdG}(\epsilon)-\epsilon-i\delta},
\end{equation}
where $\delta$ is an infinitesimal number. Because the tunneling rate expressed by the wavefunction needs to meet the normalization condition, we need to coordinate the (total) Density of States (DOS). The DOS is defined as the sum of LDOS over all spatial space, i.e.,
\begin{equation}\label{DOS}
    \begin{aligned}
    \rho_{tot}^\sigma(\epsilon) & \equiv\sum_x\rho_L^\sigma(x,\epsilon)\\
    & = -\frac{1}{\pi}\text{Im}\{\sum_x\sum_{s=\uparrow,\downarrow} \langle x,\sigma,s|G(\epsilon)|x,\sigma,s\rangle\}.
    \end{aligned}
\end{equation}

With the wave function in the position basis and the assumption that the energy states are sharply distributed, LDOS at position $x$ with energy $\epsilon$ can be expressed as
\begin{equation}\label{LDOS_delta}
    \rho_L^\sigma(x,\epsilon)=\sum_n \langle\psi_n^\sigma|x\rangle\langle x|\psi_n^\sigma\rangle\delta(\epsilon-\epsilon_n),
\end{equation}
where $n$ are discretized bound states. For the SC states and metallic states, they are distributed dense enough to become continuum. We can approximate the metallic continuum by the discretized state distribution, i.e., the eigen-state linewidth is much smaller than the eigen-energy differences.

Suppose $\epsilon_p$ is the energy of state with infinitely small linewidth. We can apply the derivation below to either bound states, or discretized states in the continuum regime (either SC states or metallic states), as long as we can ignore the size of linewidth. Then,
\begin{equation}\label{int_LDOS}
    \begin{aligned}
    \int_{\epsilon_p-a}^{\epsilon_p+a}\rho_L^\sigma(x,\epsilon)d\epsilon & = \int_{\epsilon_p-a}^{\epsilon_p+a}\sum_n\left|\langle\psi_n^\sigma|x\rangle\right|^2\delta(\epsilon-\epsilon_n)d\epsilon\\
    & = \left|\langle\psi_{\epsilon_p}^\sigma|x\rangle\right|^2,
    \end{aligned}
\end{equation}
where $a$ is an infinitesimal energy spacing. That is to say, the tunneling rate at $x$ contributed by the energy level $\epsilon_p$ is the integral of LDOS at $x$ over the single state $\epsilon_p$. Theoretically, Eq. \eqref{int_LDOS} is correct; however, the energy spacing $a$ we choose will affect the results of $\left|\langle\psi_{\epsilon_p}^\sigma|x\rangle\right|^2$. Hence, we need to use DOS to make the normalization of the wave function satisfied. The DOS is the sum of LDOS over all spatial space [equivalent to Eq.\eqref{rho_tot}], i.e.,
\begin{equation}\label{rho_tot_intLDOS}
    \begin{aligned}
    \rho_{tot}^\sigma(\epsilon) & =\int\rho_L^\sigma(x,\epsilon)dx\\
    &=\int\sum_n\langle\psi_n^\sigma|x\rangle\langle x|\psi_n^\sigma\rangle\delta(\epsilon-\epsilon_n)dx\\
    &=\sum_n\delta(\epsilon-\epsilon_n).
    \end{aligned}
\end{equation}
Then, based on the normalization of the wavefunction,
\begin{equation}\label{int_DOS}
    \int_{\epsilon_p-a}^{\epsilon_p+a}\rho_{tot}^\sigma(\epsilon)d\epsilon = \int_x\left(\int_{\epsilon_p-a}^{\epsilon_p+a}\rho_L^\sigma(x,\epsilon)d\epsilon\right)dx = \int_x \left|\langle\psi_{\epsilon_p}^\sigma|x\rangle\right|^2 dx =1
\end{equation}
We can approximate Eq. \eqref{int_DOS} by
\begin{equation}\label{int_DOS_approx}
    \rho_{tot}^\sigma(\epsilon_p)\cdot(2a)=1
\end{equation}
if $a$ is infinitesimally small and the linewidth of $\rho_{tot}^\sigma(\epsilon)$ at $\epsilon_p$ can be ignored. Then the energy spacing we should choose in order to satisfy the normalization condition is
\begin{equation}\label{normSpacing}
    2a=\frac{1}{\rho_{tot}^\sigma(\epsilon_p)}.
\end{equation}
Therefore, the tunneling rate as expressed in Eqs. \eqref{tunnel_Gamma} and \eqref{tunnel_Lambda} can precisely be
\begin{equation}\label{tunneling_DOS}
    \left|\langle\psi_{\epsilon_p}^\sigma|x\rangle\right|^2=\int_{\epsilon_p-a}^{\epsilon_p+a}\rho_L^\sigma(x,\epsilon)d\epsilon = \rho_L^\sigma(x,\epsilon_p)\cdot(2a)=\frac{\rho_L^\sigma(x,\epsilon_p)}{\rho_{tot}^\sigma(\epsilon_p)}.
\end{equation}
Note that the eigenwave function already includes all the internal degree of freedom (spin up and spin down), so the LDOS and DOS also traces out these internal degrees of freedom as well.

For the SC states and metallic states, we can take the robust total DOS equations \eqref{rho_SC} and \eqref{rho_tot} into \eqref{tunneling_DOS}. On the contrary, since generally the LDOS $\rho_L^\sigma(x,\epsilon)$ has some finite linewidth below the gap, we cannot assume $a$ to be infinitesimal. We can just stick to Eq. \eqref{int_LDOS}, but extend the size of $a$ and integrate completely over one bound state, for the eigen-states below the superconducting gap.

There are some technical subtleties to have the perfect integral of Eq. \eqref{int_LDOS}:
\begin{enumerate}
	\item The infinitesimal imaginary part from Green's function method ($\delta$) must be much smaller than the energy spacing between  states. This means our probe resolution needs to be sharp enough to distinguish two states.
	\item The grid spacing for the integral needs to be much smaller than the imaginary part of the Green's function $\delta$ (the width of the LDOS peak), which is the scale on which the integrand is smooth.
	\item Equation \eqref{int_LDOS} is wrong for degenerate states, so it cannot be used at energy crossing. Therefore, we have to follow the procedure of Sec.~\ref{sec:level1_4_2} to calculate the tunneling rates.
\end{enumerate}

\clearpage

\end{document}